\documentclass[preprint, 12pt]{elsarticle}

\usepackage{amsmath,physics}
\usepackage{algorithm}
\usepackage{algorithmic}
\usepackage{mathtools}
\usepackage{amssymb}
\usepackage{tikz}
\usetikzlibrary{mindmap}
\usetikzlibrary{shapes}
\usepackage[center]{caption}
\usepackage{epstopdf}
\usepackage{subcaption}
\usepackage{booktabs}
\usepackage{float}
\usepackage[inline]{enumitem}
\usepackage{pdfcomment}
\usepackage{colortbl}
\usepackage{bm}

\newtheorem{definition}{Definition}
\newtheorem{theorem}{Theorem}

\newproof{IEEEproof}{Proof}

\DeclareMathOperator*{\argmax}{arg\,max}
\DeclareMathOperator*{\argmin}{arg\,min}
\newcommand*{\myComb}[2]{{}^{#1}C_{#2}}

\journal{Signal Processing}

\graphicspath{{graphics/}{graphics/sim_3_radar}{graphics/publisher}}

\begin{document}
	\begin{frontmatter}
		\title{Harmonic Mean Density Fusion in Distributed Tracking: Performance and Comparison }
		\author{Nikhil Sharma%
		}
		\ead{sharmn66@mcmaster.ca}
		\author{
			Ratnasingham ~Tharmarasa
		}
		\ead{thamas@mcmaster.ca}
		\author{
			Thiagalingam Kirubarajan
		}
		\ead{kiruba@mcmaster.ca}
		
		\begin{abstract}	
			
			A distributed sensor fusion architecture is preferred in a real target-tracking scenario as compared to a centralized scheme since it provides many practical advantages in terms of computation load, communication bandwidth, fault-tolerance, and scalability. In multi-sensor target-tracking literature, such systems are better known by the pseudonym—track fusion, since processed tracks are fused instead of raw measurements. A fundamental problem, however, in such systems is the presence of unknown correlations between the tracks, which renders a standard Kalman filter (naive fusion) useless.
			
			A widely-accepted solution is covariance intersection (CI) which provides near-optimal estimates but at the cost of a conservative covariance. Thus, the estimates are pessimistic, which might result in a delayed error convergence. Also, fusion of Gaussian mixture densities is an active area of research where standard methods of track fusion cannot be used. In this article, harmonic mean density (HMD) based fusion is discussed, which seems to handle both of these issues. We present insights on HMD fusion and prove that the method is a result of minimizing average Pearson divergence. This article also provides an alternative and easy implementation based on an importance-sampling-like method without the requirement of a proposal density. Similarity of HMD with inverse covariance intersection is an interesting find, and has been discussed in detail.
			
			Results based on a real-world multi-target multi-sensor scenario show that the proposed approach converges quickly than existing track fusion algorithms while also being consistent, as evident from the normalized estimation-error squared (NEES) plots.
		\end{abstract}	
		
		
	\end{frontmatter}	
	
	\section{Introduction}
	Even though sensor technologies are reaching new heights in terms of accuracy, size, processing time, and portability, it is still not prudent to depend on a single sensor for large-scale systems. A multi-sensor system provides unmatched benefits, including accuracy, fault tolerance, scalability, and, in some cases, observability. In the widely researched domain of multiple target tracking, the employed sensors have a limited field of view (FOV), which constrains their applicability. Thus, to scale the area of observation, multiple sensors are deployed, which transmit their observations either in the form of raw measurements (centralized) or process the raw data before sending the resulting processed observations (tracks) to a fusion facility. Such networked architectures are highly employed in systems such as simultaneous localization and mapping (SLAM), autonomous vehicles, air traffic control, adversary interception, underwater surveillance, and others.
	
	The architecture wherein raw observations are processed locally using standard techniques like a Kalman filter and sent to the nearest fusion center is the focus of this article. Such an architecture is better suited to cater multi-sensor target tracking problems since it offers numerous advantages in comparison to a centralized mechanism. These include, 
	\begin{enumerate}[label	= (\roman*)]
		\item Clutter management at the local node level, thus reducing link bandwidth requirements.
		\item Homogeneity of track state  space, which eases track association.
		\item Fault-tolerance due to the existence of multiple local fusion centers.
		\item Near optimal performance.
		\item Full communication rate is not required, as the local node has the ease to transmit tracks only when they are confirmed by the local track-management scheme.
	\end{enumerate}
	Since the local node is assumed to employ a Kalman filter like estimator, the processed observations that are transmitted to the fusion center are generally parameters of the posterior density of the target state conditioned upon measurements. Due to such reasons, distributed systems in the target-tracking domain are termed track-to-track fusion (T2TF) or simply track-fusion. Since a standard Kalman filter assumes independence between measurements and target-state errors, it cannot be used to fuse tracks as they are correlated, primarily due to the following three reasons,
	\begin{enumerate}[label = (\roman*)]
		\item Due to the presence of common process noise in tracks originating from the same target.
		\item Correlation due to the same statistics used in track initialization among local nodes.
		\item Correlation between a global (fused) and a local track due to the presence of a common history of measurements. This occurs in fusion centers with memory and can be removed by restarting global track periodically.
	\end{enumerate} 
	
	There are two major classes of solution preferred for the case of track-fusion -- \begin{enumerate*} \item Bayesian methods, \item Pooling methods \end{enumerate*}. The Bayesian methods refer to the solution where Bayes' theorem is employed in some sense either after decorrelating the local tracks, or by computing a globalized likelihood function using local node parameters. It is also possible to compute the exact value of cross-correlation between any two local tracks using the recursion provided in \cite{bar1981track}, but this would require knowledge of the local Kalman gain at each fusion instant and also memory for storing $\myComb{n}{2}$ number of cross-correlations at the fusion center, for $n$ nodes. Such methods includes the optimal T2TF algorithm, information matrix fusion, and, tracklet-fusion. Using a standard Kalman filter by ignoring cross-correlation is usually termed as naive fusion. 
	
	The pooling methods combine probability density functions directly using mixing weights such that some criteria is minimized. Under the target-tracking domain, such methods include the famous covariance intersection and its variants, and also its generalized form known as the geometric mean density (GMD) fusion. These methods are preferred due to their relative ease of implementation in the Gaussian case. The ease of implementation stems from the fact that these methods do not inquire into the nature of cross-correlation. Thus, extra information from the local node viz. Kalman gain, local prior estimates since last fusion or sensor characteristics are not required for fusion (at least for the ideal case). Such methods include the covariance intersection (CI) and its variants. A new class of pooling method known as harmonic mean density (HMD) based track fusion was introduced by the authors in \cite{nikhil_journal_1}, which is also the focus of this article.
	
	\subsection{Covariance Intersection and Variants}
	
	For a known cross-covariance, the resulting fused covariance ellipsoid can be shown to lie in the intersection between the local covariance ellipsoids \cite{julier1997non, chen2002fusion}. Thus, a convex combination of the covariance and mean of the estimates should yield a consistent fused estimate \cite{julier1997non}, which was utilized in the formulation of the original covariance intersection algorithm. In the presence of unknown cross-correlations, CI constitutes an upper bound on the cross-correlation, which makes it conservative. Later, many variants have been proposed to increase the tightness of this upper bound. 
	
	One such variant is the split-covariance intersection filter (SCIF), which assumes that a single local track is a package consisting of independent and dependent estimates separately, such that the local estimate is a weighted sum of independent and dependent estimates. The algorithm then operates on these components in a way similar to the covariance intersection algorithm. The result is the fused estimates with their dependent and correlated components. The existence of a correlation such that independent and dependent components sum up to form a track is, however, vague. Also, the independent parts of a track are impossible to identify. If both components are available explicitly at the fusion center, then an optimal algorithm could be possible, according to the author's knowledge.
	
	The ellipsoidal intersection (EI) in \cite{sijs2010state} aims to separate the mutual and exclusive information contents from a local estimate and fuse them such that the mutual information is only counted once. The update equation resembles naive fusion before subtracting the mutual-information component. An interesting feature is that the algorithm computes the maximum possible cross-correlation ellipsoid, which results in a lower fused covariance relative to CI. However, it was shown in \cite{noack2017decentralized,noack2017inverse} that EI can result in non-consistent estimates. The paper also proposed an additional CI variant, termed inverse-covariance intersection (ICI), which claims consistency of the fused estimate in addition to providing a less-conservative fused covariance relative to CI. However, both papers on ICI have not provided any effect on the root-mean-square error. Also, both CI and ICI need to compute a one-dimension minima search for the fusion parameter $\omega$. Though the cost function in both algorithms is convex, the one in ICI requires inverting 3 matrices compared to 2 in CI, thus being slightly slower. Also, no alternatives to ICI, EI, and SCIF for densities other than Gaussian exist in the current literature.
	
	\subsection{Fusion of Gaussian Mixtures}
	The appeal of using Gaussian mixtures is increasing with advancing computation power. Mixture densities are typically used among local nodes primarily for multiple model tracking, modeling unknown noise statistics (Gaussian-sum filtering), parameterization of non-observable measurements, and among extensions of probability hypothesis density and multiple hypothesis trackers. Track fusion of Gaussian-mixture densities is still an active area of research, where exact closed-form solutions cease to exist. 
	
	A generalized version of covariance intersection is employed for fusing densities other than Gaussian. This was first proposed in \cite{mahler2000optimal} by Mahler as the generalized Uhlmann-Julier covariance intersection. Also, in \cite{hurley2002information}, Hurley succinctly expressed the relation between Chernoff information and covariance intersection. In \cite{battistelli2014kullback}, the Chernoff fusion was studied for a non-Gaussian distribution and was termed the geometric mean density (GMD). Since the covariance intersection (CI) assumes a closed form for Gaussian densities, component-wise CI can be used while fusing Gaussian mixtures using Chernoff information. This was the approach used in \cite{upcroft2005rich}, where the GMM were fused for a bearing-only tracking scenario. A better result was discussed in \cite{julier2006empirical}, where a first-order approximation of the power of a Gaussian mixture was employed. This was the first paper that dealt with approximating the power of a Gaussian mixture model with another Gaussian mixture. The result was succeeded by Gunay et al., which employed sigma-points for generalizing the power of a Gaussian mixture \cite{gunay2016chernoff} though it was computationally demanding. Moreover, such a fusion strategy is affected adversely when the individual modes of a mixture are very close, which is usually the case with an IMM tracker. A review of various methods for approximating the non-integer power of a Gaussian mixture was presented in \cite{ajgl2015approximation} along with a comparison.
	
		Harmonic mean density (HMD) based track fusion is a viable solution to the fusion of uni-modal and multi-modal Gaussian densities. In this context, the contributions of this paper are as follows :	
	\begin{itemize}
		\item We prove that, like the GMD, which is the result of minimizing average Kullback-Leibler divergence (KLD), the HMD minimizes average Pearson $\chi^2$ divergence and, thus, the reverse Neyman $\chi^2$ divergence.
		\item The article presents important developments on the consistency and unbiasedness of fused estimates. It has been proved that consistency of HMD depends on the consistency of local posterior densities.
		\item Similarity of HMD implementation has been shown towards inverse covariance intersection (ICI).
		\item We provide an alternate implementation of the HMD based on importance sampling without the requirement of a proposal density. 
		\item  A simple optimization approach to calculating the fusion weight has also been provided. The approach requires one matrix inversion and hence, performs faster. 
		\item It has been shown that the existing strategy for calculation of fusion weights in CI and ICI fails in the case of scalar estimates. Through an example, the superiority of HMD in the fusion of such cases has been shown. 
	\end{itemize}

	The rest of the paper is organized as follows: in Section \ref{probForm}, a brief summary of the issue of cross-correlation in a distributed sensor fusion scenario is presented. In Section \ref{review}, the properties of HMD fusion are reviewed along with standard track fusion techniques. Here, the existing implementation for HMD, where the Gaussian mixture denominator is approximated by a Gaussian density, is presented. We present a consistency analysis of harmonic mean density in Section \ref{sec_consistency}. The relationship between ICI and HMD has also been provided in this section, along with some comments on ICI. Section \ref{implementation} provides a new sample-based implementation of HMD and GMD using importance sampling-like techniques. The comparison and results on multiple track-fusion scenarios are presented in Section \ref{sim}. Finally, the article is concluded in Section \ref{conclusion}.

	\section{Problem Formulation} \label{probForm}
	Considering a scenario with two sensor platforms -- $\mathcal{S}^1$ and $\mathcal{S}^2$ generating processed track density conditioned on measurements upto time-step $k$ -- $ p(\mathbf{x}_k|\mathbf{z}_{1:k}^1)$ and $p(\mathbf{x}_k|\mathbf{z}_{1:k}^2)$. Assume that both tracks are conditioned on a common prior at time $k$ which is quantified as $p_c(\mathbf{x}_k)$. Then,
	\begin{align}
		\mathit{p}(\mathbf{x}_k|\mathbf{z}_{1:k}^1) \propto \mathit{p}(\mathbf{z}^1_k|\mathbf{x}_k) p_c(\mathbf{x}_k).
	\end{align}
	and similarly for the other node. Using naive fusion with such local densities results in,
	\begin{align}
		p_N(\mathbf{x}_k|\mathbf{z}_{1:k}^1 \cup \mathbf{z}_{1:k}^2) \propto p(\mathbf{z}^1_k|\mathbf{x}_k)p_c(\mathbf{x}_k)p(\mathbf{z}^2_k|\mathbf{x}_k)p_c(\mathbf{x}_k),\label{eq_naive}
	\end{align}
	where $p_N(\mathbf{x}_k|z_k^1 \cup z_k^2)$ is the naively fused density. Thus, the common information is accounted twice which will lead to an optimistic estimate (e.g. multiplying two Gaussian densities results in a Gaussian density with a lower or same covariance). A heuristic approach would be to use naive fusion while dividing once with the common information,
	\begin{align}
		p(\mathbf{x}_k|\mathbf{z}_k^1 \cup \mathbf{z}_k^2) &\propto \frac{p(\mathbf{z}^1_k|\mathbf{x}_k)p_c(\mathbf{x}_k)p(\mathbf{z}^2_k|\mathbf{x}_k)p_c(\mathbf{x}_k)}{p_c(\mathbf{x}_k)},\notag\\
		& \propto \frac{p(\mathbf{x}_k|\mathbf{z}_{1:k}^1)p(\mathbf{x}_k|\mathbf{z}_{1:k}^2)}{p(\mathbf{x}_k|\mathbf{z}^1_k\cap \mathbf{z}^2_k)}.\label{exactBayesian}
	\end{align}
	
	This is also the exact Bayesian formulation for distributed data fusion \cite{lu2019distributed}, where $p(\mathbf{x}_k|z^1_k\cap z^2_k)$ accounts for mutual density due to common prior. For Gaussian densities, an exact formulation of common prior was presented in \cite{bar1981track} which will require extra information from local trackers. This article focuses on a class of solutions where the common information can be accounted for (in a suboptimal sense) by only using the information available. Thus,
	\begin{align}
		p(\mathbf{x}_k|\mathbf{z}^1_k\cap \mathbf{z}^2_k) \propto \mathcal{F}\left(p(\mathbf{x}_k|\mathbf{z}_k^1), p(\mathbf{x}_k|\mathbf{z}_k^2), \omega\right),
	\end{align}
	where,  $\mathcal{F}(.)$ is an appropriate functional operating on probability densities, and $\omega$ $\in$ $[0,1]$ is a scalar parameter that can be evaluated such that the resulting density is optimal in some sense.
	
	If the local platform employs, for e.g., an IMM tracker, then the posterior density $p(\mathbf{x}_k|\mathbf{z}_{1:k}^i)$ consists of estimates from multiple modes packed in a Gaussian mixture of the form,
	\begin{align}
		p(\mathbf{x}_k|\mathbf{z}_{1:k}^i) = \sum_{p = 1}^{M} \mu_p \mathcal{N}\left(\mathbf{x}_k; \hat{\mathbf{x}}^p_k, \Gamma^p_k\right),
	\end{align}
	where $M$ is the total number of modes, $\mu_p$ is the mode probability of the $p^{\text{th}}$ Gaussian component. An exact solution to the fusion of such tracks is infeasible due to,
	\begin{enumerate}[label = (\roman*)]
		\item Communication constraint as $M\times \mathcal{S}$ number of filter gains and measurement Jacobian matrices need to be communicated at each fusion instant.
		\item Division by a Gaussian mixture \cite{ahmed2015s}.
	\end{enumerate}	
	The second point leads to sub-optimality even if exact procedures are followed. This necessitates the search for a robust and suboptimal class of strategies. 

	\section{Harmonic Mean Density Fusion} \label{review}
	Given any two probability density functions, $p_1(\mathbf{x})$ and $p_2(\mathbf{x})$, an interpolation between them can be constructed using an $\omega$-weighted generalized mean, $\mathcal{M}_\omega(p_1(\mathbf{x}), p_2(\mathbf{x}))$, also known as the Fr\'echet mean. Such generalized means can also be constructed as a result of optimizing a unique \textit{f}-divergence measure \cite{koliander2022fusion},
	\begin{align}
		\mathcal{M}_\omega(p_1(\mathbf{x}), p_2(\mathbf{x})) = \argmin_{m(\mathbf{x})} \sum_{i=1}^2 \omega_i D_f\left[m(\mathbf{x}) || p_i(\mathbf{x}) \right]
	\end{align}  
	where $D_f(.)$ is a \textit{f}-divergence \cite{yury}, defined by,
	\begin{align}
		D_f[p(\mathbf{x})||q(\mathbf{x})] = \int q(\mathbf{x})f\left(\frac{p(\mathbf{x})}{q(\mathbf{x})}\right)dx
	\end{align}
	with probability densities $p(\mathbf{x})$ and $q(\mathbf{x})$. $f(t)$ is any convex function that is defined for $t>0$, with $f(1) = 0$ \cite{csiszar2004information}. The resulting Fr\'echet mean is in general, a mixture probability distribution, which when normalized results in a valid probability density function known as the $\omega$-weighted M-mixture, 
	
	\begin{align}
		\mathbf{M}_\omega\big(p_1(\mathbf{x}_k), p_2(\mathbf{x}_k)\big) = \frac{\mathcal{M}_\omega\big(p_1(\mathbf{x}_k), p_2(\mathbf{x}_k)\big)}{\zeta_{\mathcal{M}_\omega}\big(p_1(\mathbf{x}_k), p_2(\mathbf{x}_k)\big)}, \label{M-mixture}
	\end{align}
	where $\zeta_{\mathcal{M}_\omega}\big(p_1(\mathbf{x}_k), p_2(\mathbf{x}_k)\big)$ is the normalization constant resulting from interpolation of $p_1(x)$ and $p_2(x)$ to form an abstract mean $\mathcal{M}_\omega(.)$,
	\begin{align}
		\zeta_{\mathcal{M}_\omega}\big(p_1(\mathbf{x}_k), p_2(\mathbf{x}_k)\big) = \int_{\mathbb{R}^n} \mathcal{M}_\omega\big(p_1(\mathbf{x}_k), p_2(\mathbf{x}_k)\big) d\mathbf{x}
	\end{align} 
	
	In the case of HMD, we take the abstract mean as the harmonic mean of participating local posterior track densities. The resulting fused M-mixture $p_f(\mathbf{x})$ is,
	\begin{align}
		\frac{1}{p_f(\mathbf{x})} &\propto \frac{1}{\mathcal{M}^h_\omega\big(p_1(\mathbf{x}), p_2(\mathbf{x})\big)} = \frac{\omega}{p_1(\mathbf{x})} + \frac{1-\omega}{p_2(\mathbf{x})} \\
		\implies 		p_f(\mathbf{x})		&\propto \frac{p_1(\mathbf{x}) p_2(\mathbf{x})}{(1-\omega)p_1(\mathbf{x}) + \omega p_2(\mathbf{x})} \label{eq_HMD_gen}
	\end{align}
	
	Where the proportionality constant is the normalization factor $1/\zeta_{\mathcal{M}^h_\omega}(.)$. Note that the mixture density in the denominator prohibits closed-form solutions, even for the Gaussian case. Even with a simple approximation, the HMD-based track fusion was found to perform exceptionally well compared to other track fusion strategies. Many important properties of HMD akin to track fusion have been proven by the authors in \cite{nikhil_journal_1}. Some of these are,
	\begin{enumerate}[label=(\roman*)]
		\item HMD avoids double counting of information, unlike Naive fusion in \eqref{eq_naive}. 
		
		Assuming conditional dependence, such that 
		\begin{align}
			p(\mathbf{z}^1|\mathbf{x}) = p(\mathbf{z}^{1/2}|\mathbf{x})p(\mathbf{z}^1\cap \mathbf{z}^2|\mathbf{x}),
		\end{align}		
		The notation $p(\mathbf{z}^{1/2}|\mathbf{x})$ denotes the exclusive information present with sensor track $1$ relative to $2$ such that $p(\mathbf{z}^{1/2}|\mathbf{x})\cap{p(\mathbf{z}^{2/1}|\mathbf{x})} = \o$. It is easy to observe that using HMD,
		\begin{align}
			p(\mathbf{x}|\mathbf{z}^1 \cup \mathbf{z}^2) &\propto \frac{p(\mathbf{z}^{1/2}|\mathbf{x}) p(\mathbf{z}^{2/1}|\mathbf{x})}{(1-\omega) p(\mathbf{z}^{1/2}|\mathbf{x}) + \omega p(\mathbf{z}^{2/1}|\mathbf{x})}  \notag \\
			&\quad \times p(\mathbf{z}^1\cap \mathbf{z}^2|\mathbf{x})p(\mathbf{x}).
		\end{align} 
		Thus, the common information $p(\mathbf{z}^1\cap \mathbf{z}^2|\mathbf{x})$ is only accounted once.
		
		\item HMD has the potential to be a recursive fusion strategy since it is an abstract mean. To illustrate, let $p_1$, $p_2$ and $p_3$ be three local densities. Then, it can be proved that
		\begin{align}
			\mathcal{M}^h_\omega\big(p_1, p_2, p_3\big) = \mathcal{M}^h_{\omega_2}\bigg(\mathcal{M}^h_{\omega_1}\big(p_1, p_2 \big), p_3\bigg),
		\end{align}
		which makes it easier to fuse local tracks pairwise.
		
		\item The second derivative of the normalization constant with respect to $\omega$ in case of HMD 
		can be proved to be always positive for valid local densities $p_1(\mathbf{x})$ and $p_2(\mathbf{x})$.  This suggests a convex nature of the normalization constant, like in the case of GMD. It was also found that the normalization constant in the case of HMD is less than that of GMD.
		\begin{align}
			\int_{\mathbb{R}^n} \frac{p_1(\mathbf{x})p_2(\mathbf{x})}{\omega_2p_1(\mathbf{x}) + \omega_1 p_2(\mathbf{x})} \leq \int_{\mathbb{R}^n} \frac{p_1(\mathbf{x})p_2(\mathbf{x})}{p^{\omega_1}_1(\mathbf{x})p^{\omega_2}_2(\mathbf{x})}
		\end{align}
		The bounds on normalization constant to be between 0 and 1 creates a lower bound on the normalized HMD, as seen in the later section.
		
		\item Since HMD is an abstract mean, it is bounded from above and below by participating densities. It ensures that the fusion can be performed unlimited number of times, and the worse result we can get is the infimum of the set of participating densities. We will use this result later in Section \ref{sec_consistency} while analyzing consistency of HMD estimator.
	\end{enumerate}
	
	A lot of work on conservative fusion \cite{julier2008fusion} in the contemporary literature is focused on how to reduce the conservativeness of covariance intersection and its generalization. It was shown in simulations of \cite{nikhil_journal_1} that HMD does not suffer from covariance inflation while also staying consistent in terms of normalized estimation error squared (NEES).
	
	\subsection{Average Divergence Minimization}
	Conservative fusion techniques like arithmetic and geometric averaging construct a fused density by creating a mutual agreement between the participating local densities. This mutual agreement is based on minimizing a metric between the output density and each of the local density functions. Therefore, the techniques existing in the literature are found to be a result of minimizing some average divergence measure. 
	
	\begin{itemize}
		\item The covariance intersection and its generalization are a result of minimizing $\omega$-weighted average KL divergence \cite{battistelli2014kullback}.
		\item The arithmetic average density is a result of minimizing $\omega$-weighted reverse KL divergence \cite{li2020arithmetic}.
	\end{itemize}
	
	While in \cite{nikhil_journal_1}, fusion using HMD was shown from an information theoretic viewpoint using K-L divergence, here we prove that it actually results in the minimization of the average Pearson $\chi^2$ divergence (an alternate proof is also presented in \cite{koliander2022fusion}). Though the divergence minimization in any of the pooling techniques does not affect the fusion accuracy in any sense, it provides an intuition about the geometry of the information process that the local densities undergo during fusion.
	
	\begin{theorem}
		The harmonic mean density between $p_1(\mathbf{x})$ and $p_2(\mathbf{x})$ minimizes the $\omega$-weighted average Pearson $\chi^2$ divergence, and thus, the reverse Neyman $\chi^2$ divergence.
	\end{theorem}
	
	\begin{IEEEproof}
		Denote the $\omega$-weighted average Pearson $\chi^2$ divergence by $J$, and the HMD by $p_h(\mathbf{x})$, then,
		\begin{align}
			J\big(p_h(\mathbf{x})\big) &= \sum^2_{i=1}\omega_i D_{\chi^2_P}\big(p_h(\mathbf{x})||p_i(\mathbf{x})\big)   \notag \\ 
			 & = \sum^2_{i=1}\omega_i\frac{1}{2}\int_{-\infty}^{\infty} \frac{\big(p_h(\mathbf{x}) - p_i(\mathbf{x})\big)^2}{p_i(\mathbf{x})} d\mathbf{x} \label{eq_divMin_obj}
		\end{align}
		 where $D_{\chi^2_P}$ stands for Pearson $\chi^2$ divergence. Note that there are two constraints on $p_h(\mathbf{x})$, which are --- \begin{enumerate*} \item $p_h(\mathbf{x}) \geq 0$, and \item $\int_{\mathbb{R}^n} p_h(\mathbf{x}) = 1$ \end{enumerate*}. In order to proceed with minimization using calculus of variations, we ignore the constraints and simply minimize the functional. The minimized function will then be checked for validity of a proper probability density function. 
		
		The condition of minimization is upon the first variation as,
		\begin{align}
			\delta J\big(p_h(\mathbf{x})\big) = \frac{d}{dt}\bigg(J\big(p_h(\mathbf{x}) + tg_h(\mathbf{x})\big)  \bigg)_{t=0} = 0 \label{eq_firstVar}
		\end{align}
		Reversing the summation and integral in eqn. \eqref{eq_divMin_obj}, and substituting in eqn. \eqref{eq_firstVar},
		\begin{align}
			&\delta J\big(p_h(\mathbf{x})\big) = \int_{-\infty}^\infty\frac{1}{2} \sum^2_{i=1}\omega_i \frac{\partial}{\partial t}  \frac{\big(p_h(\mathbf{x}) + tg_h(\mathbf{x}) - p_i(\mathbf{x})\big)^2}{p_i(\mathbf{x})} d\mathbf{x} \notag \\
			&\qquad= \int_{-\infty}^\infty\frac{1}{2} \sum^2_{i=1}\omega_i \frac{2\big(p_h(\mathbf{x}) + tg_h(\mathbf{x})\big)g_h(\mathbf{x}) - 2p_i(\mathbf{x})g_h(\mathbf{x}) }{p_i(\mathbf{x})}d\mathbf{x} \notag \\
			&\qquad = \int_{-\infty}^\infty \sum^2_{i=1}\omega_i  \left[\frac{\big(p_h(\mathbf{x}) + tg_h(\mathbf{x})\big)}{p_i(\mathbf{x})} - 1 \right]g_h(\mathbf{x})d\mathbf{x} = 0
		\end{align}
		Substituting $t = 0$ for the first variation, we get the following equation for $p_h(\mathbf{x})$
		\begin{align}
			p_h(\mathbf{x}) \sum^2_{i=1} \frac{\omega_i}{p_i(\mathbf{x})} = \sum^2_{i=1}\omega_i = 1
		\end{align}
		which implies,
		\begin{align}
			p_h(\mathbf{x}) = \frac{1}{\sum\limits^2_{i=1} \frac{\omega_i}{p_i(\mathbf{x})}} = \frac{p_1(\mathbf{x})p_2(\mathbf{x})}{\omega_2p_1(\mathbf{x}) + \omega_1p_2(\mathbf{x})}
		\end{align}
		In general, it can be proved that,
		\begin{align}
			p_h(\mathbf{x}) = \frac{1}{\sum\limits^N_{i=1} \frac{\omega_i}{p_i(\mathbf{x})}}
		\end{align}
		where $N$ is the number of local densities.
		
		Since the participating local densities are valid pdfs, the first constraint for positivity is satisfied. The second constraint can be enforced by normalizing the resulting density with $\zeta^h = \int p_h(\mathbf{x})d(\mathbf{x})$. The second variation can be shown to be positive, which proves that the result is actually a minimum. 
		\begin{align}
			\delta^2 J\big(p_h(\mathbf{x})\big) &= \frac{d^2}{dt^2}\bigg(J\big(p_h(\mathbf{x}) + tg_h(\mathbf{x})\big)  \bigg)_{t=0} \notag \\
												&= \int_\infty^\infty \sum_{i=1}^{2} \omega_i \frac{g_h(\mathbf{x})^2}{p_i(\mathbf{x})} d\mathbf{x} \geq 0.
		\end{align}		
		For proving the case for $D_{\chi^2_N}(.)$, the reverse Neyman divergence, it can be noted that both $D_{\chi^2_N}(.)$ and $D_{\chi^2_P}(.)$ are special cases of $\alpha$-divergence $D_{\alpha}(.)$ \cite{gul2016robust}, which satisfy the property,
		\begin{align}
		D_{\chi^2_P}(.) &= D_{(\alpha = 2)}\big(p_i(\mathbf{x})||p_j(\mathbf{x})\big) \notag \\ 
							& = D_{(\alpha = -1)}\big(p_j(\mathbf{x})||p_i(\mathbf{x})\big) = D'_{\chi^2_N}(.)
		\end{align}
		where $D'(.)$ stands for divergence obtained by reversing the arguments. 
		
	\end{IEEEproof}

%
%
%
%

	\section{Consistency Analysis} \label{sec_consistency}
	Consistency of HMD strongly depends on the consistency of participating local posterior densities. To prove this, first we adopt the following definition of consistency as provided in \cite{ghosal1997review, ghosal1995convergence},

	\begin{definition} \label{def_cons}
		A posterior density $p(\mathbf{x})$ is considered strongly consistent if, for any neighbourhood of $\mathbf{V}$ of true value $\mathbf{x}_0$, the following holds:
		\begin{align}
			\lim_{k\rightarrow\infty}	\text{Pr}\left(\mathbf{x} \notin \mathbf{V} | \mathbf{Z}_{1:k} \right)_{p(.)} = 0
		\end{align}
	\end{definition}
	
	where the subscript denotes that the probability is computed with respect to the posterior distribution $p(.)$, and $Z_{1:k}$ is the set of $k$ observations. The definition can be alternatively stated as,
	\begin{align}
		\lim_{k\rightarrow\infty} \text{Pr}\left(\mathbf{x} \notin \mathbf{V} | \mathbf{Z}_{1:k} \right)_{p(.)} = \epsilon,
	\end{align}
	and the posterior is strongly consistent if $\epsilon = 0$ and weakly consistent if $\epsilon$ converges to 0 in probability. In any case, it can be proved that consistency of HMD depends on the consistency of the local posterior densities.
	
	\begin{theorem}
		HMD is consistent as long as the infimum of the set of local posterior densities is consistent.
	\end{theorem}
	\begin{IEEEproof}
		Since the unnormalized HMD $\mathcal{M}^h_\omega(p_1(\mathbf{x}),  p_2(\mathbf{x}))$ is a generalized mean of local densities, it follows that,
		\begin{align}
			\inf\{p_1(\mathbf{x}),  p_2(\mathbf{x}) \} \leq \mathcal{M}^h_\omega(p_1(\mathbf{x}),  p_2(\mathbf{x})) \leq \sup\{p_1(\mathbf{x}),  p_2(\mathbf{x})\}.
		\end{align}
		Using the fact that normalized constant in the case of HMD is less than one \cite{nikhil_journal_1}, the normalized HMD follows,
		\begin{align}
			\mathbf{M}^h_\omega(.) = \frac{\mathcal{M}^h_\omega(.)}{\zeta_{\mathcal{M}^h_\omega}} \geq \inf\{p_1(\mathbf{x}),  p_2(\mathbf{x}) \}, \label{eq_bound_HMD}
		\end{align}
		where the equality is attained when $p_1(\mathbf{x}) = p_2(\mathbf{x})$. For the infimum of local posterior to be consistent, it should follow that for any interval $\mathbf{V}$
		\begin{align}
			\lim_{k\rightarrow \infty } \text{Pr}\left( \mathbf{x} \notin \mathbf{V} | \mathbf{Z}_{1:k} \right)_{\inf\{p_1(.), p_2(.)\}} = 1 - \int_\mathbf{V} \inf\{p_1(\bm{\lambda}), p_2(\bm{\lambda}) \} d\bm{\lambda} = \epsilon
		\end{align}
		where same notation as in Definition \ref{def_cons} have been used. Substituting eqn. \eqref{eq_bound_HMD} above, the consistency of HMD can be related as
		\begin{align}
			\lim_{k\rightarrow \infty } \text{Pr}\left( \mathbf{x} \notin \mathbf{V} | \mathbf{Z}_{1:k} \right)_{\mathbf{M}^h_\omega(.)} \leq  1 - \int_\mathbf{V} \inf\{p_1(\bm{\lambda}), p_2(\bm{\lambda}) \} d\bm{\lambda} = \epsilon \label{eq_HMD_cons}
		\end{align}
		Hence, consistency of the infimum is a lower bound for the consistency of HMD. If $p_1(.) \neq p_2(.)$, eqn. \eqref{eq_HMD_cons} becomes a strong inequality and $\epsilon$ converges to $0$ faster than the local posterior densities. This results in the fact that HMD is always more accurate than the infimum. 
	\end{IEEEproof}
	
	\subsection{Implementation using Gaussian Approximation}
	For Gaussian local posteriors, the simplest implementation of HMD is when the mixture in denominator is replaced by a Gaussian equivalent \cite{nikhil_journal_1}. Let's call this implementation HMD-GA (for Gaussian Approximation), which is given by (dropping time-step $k$ for brevity),
	\begin{align}
		\frac{1}{\zeta_{\mathcal{M}^h_\omega}} \frac{p_1(\mathbf{x})p_2(\mathbf{x})}{\omega_2p_1(\mathbf{x}) + \omega_1p_2(\mathbf{x})} \propto  \frac{p_1(\mathbf{x})p_2(\mathbf{x})}{\mathcal{N}\left(\mathbf{x}; \boldsymbol{\gamma}^h_m, \boldsymbol{\Gamma}^h_m\right)}
	\end{align}
	which gives the fused mean and covariance as, 
	\begin{subequations}\label{eq_hmdNaive}
		\begin{align} 
			\mathbf{\Gamma}^f &= \left( \mathbf{\Gamma}^{{-1}}_1 + \mathbf{\Gamma}^{{-1}}_2  - \mathbf{\Gamma}_{m}^{h^{-1}}\right)^{-1}\label{fusedCov},\\
			\hat{\mathbf{x}}^f &= \mathbf{\Gamma}^f \left[\mathbf{\Gamma}^{{-1}}_1\hat{\mathbf{x}}_{1} +  \mathbf{\Gamma}^{{-1}}_2\hat{\mathbf{x}}_{2}   -   \mathbf{\Gamma}_{m}^{h^{-1}}\boldsymbol{\gamma}^h_{m}\right],\label{fusedMean}
		\end{align}
	\end{subequations}
	where $\boldsymbol{\gamma}^h_{m}$ and $\mathbf{\Gamma}_{k}^{h}$ are obtained using the Gaussian approximation of the mixture. These quantities represent the mutual information due to cross-correlations, which is subtracted once in \eqref{eq_hmdNaive}. The implementation is simple, intuitive, and requires three matrix inversions, making it of the order $\mathcal{O}(n^3)$, where $n$ is the dimension of state estimates. This is equivalent to all other fusion algorithms.
	
	\subsection{Condition of Consistency of HMD-GA.}
	
	 Even though the simulations suggest HMD as a consistent fusion method, the spread-of-means term in Gaussian approximation might pose a problem in convergence. In this section, we present the conditions of consistency of HMD-GA in a general case. 
	
	Comparing the construction of HMD-GA in eqn. \eqref{eq_hmdNaive} with that of the optimal fusion equation in eqn. \eqref{exactBayesian}, it is easy to see that the mutual information component in HMD $\left(\boldsymbol{\gamma}_m^h, \mathbf{\Gamma}_m^h\right)$ is the Gaussian equivalent of the mixture,
	\begin{align}
		\boldsymbol{\gamma}_m^h &=  \omega_2 \mathbf{\hat{x}}_1 + \omega_1\mathbf{\hat{x}}_2 \\
		\mathbf{\Gamma}_m^h &= \omega_2 \mathbf{\Gamma_1} + \omega_1\mathbf{\Gamma_2} + \tilde{\mathbf{\Gamma}}
	\end{align}
	where $\tilde{\mathbf{\Gamma}}$ is the spread-of-means term, given by,
	\begin{align}
		\tilde{\mathbf{\Gamma}} = \sum_{i=1}^2 (1-\omega_i)\left[\mathbf{\hat{x}}_i - \boldsymbol{\gamma}_m^h\right]\left[\mathbf{\hat{x}}_i - \boldsymbol{\gamma}_m^h\right]^T,
	\end{align}
	which can be manipulated and alternatively written as,
	\begin{align}
		\tilde{\mathbf{\Gamma}} = \omega_1\omega_2 \left[\mathbf{\hat{x}}_1 - \mathbf{\hat{x}}_2\right]\left[\mathbf{\hat{x}}_1 - \mathbf{\hat{x}}_2\right]^T \label{eq_spread_of_means_alter}
	\end{align}
	Note that $\tilde{\mathbf{\Gamma}}$ is a rank-one positive semi-definite (PSD) matrix, with an eigen-value $\omega_1\omega_2\left[\mathbf{\hat{x}}_1 - \mathbf{\hat{x}}_2\right]^T\left[\mathbf{\hat{x}}_1 - \mathbf{\hat{x}}_2\right]$ with multiplicity 1, and the rest of the eigen-values zero with multiplicity $n-1$; $n$ being the dimension of the state. 
	
	As per the definition of finite sample consistency, the estimate error with respect to the true state of the target should match the covariance presented by the estimator \cite{bar2001estimation}. Mathematically, the mean squared error (MSE) should converge to its covariance matrix. If $\left(\hat{\mathbf{x}}, \mathbf{\Gamma}\right)$ is an unbiased estimate pair, then it is needed that,
	\begin{align}
		\mathbb{E}\left[\mathbf{x} - \hat{\mathbf{x}} \right]\left[\mathbf{x} - \hat{\mathbf{x}} \right]^T =  \mathbf{\Gamma}. \label{eq_consistency}
	\end{align}

	To analyze consistency, we first prove that HMD-GA provides an unbiased estimate. Taking expectation on both sides of eqn. \eqref{fusedMean}, we get,
	\begin{align}
		\mathbb{E}\left[\hat{\mathbf{x}}^f\right] = \mathbf{\Gamma}^f \left[\mathbf{\Gamma}^{{-1}}_1\mathbb{E}\left[\hat{\mathbf{x}}_{1}\right] +  \mathbf{\Gamma}^{{-1}}_2\mathbb{E}\left[\hat{\mathbf{x}}_{2}\right]   -   \mathbf{\Gamma}_m^{h^{-1}}\mathbb{E}\left[\boldsymbol{\gamma}^h_m\right]\right], \label{eq_expect_hmdMean}
	\end{align}
	The local estimates are assumed unbiased which means $\mathbb{E}[\hat{\mathbf{x}}^{1}] = \mathbb{E}[\hat{\mathbf{x}}^{2}] = \mathbb{E}[\boldsymbol{\gamma}^h_m] = \mathbf{x}$. Thus,
	\begin{align}
		\mathbb{E}\left[\hat{\mathbf{x}}^f\right] =  \mathbf{\Gamma}^f \left( \mathbf{\Gamma}^f\right)^{\mathrlap{-1}}\mathbf{x} = \mathbf{x}
	\end{align}
	wherein eqn. \eqref{fusedCov} has been employed. Next, we calculate the estimation error by subtracting eqn. \eqref{fusedMean} from eqn. \eqref{eq_expect_hmdMean} (using the notation $\tilde{\mathbf{x}} = \mathbf{x} - \hat{\mathbf{x}}$),
	\begin{align}
		\tilde{\mathbf{x}}^f = \mathbf{\Gamma}^f \left[\mathbf{\Gamma}^{{-1}}_1\tilde{\mathbf{x}}_{1} +  \mathbf{\Gamma}^{{-1}}_2\tilde{\mathbf{x}}_{2}  -   \mathbf{\Gamma}^{h^{-1}}_m\tilde{\boldsymbol{\gamma}}^h_m\right]
	\end{align}
	Our requirement becomes,
	\begin{align}
		\mathbf{\Gamma}^f - \mathbb{E}\left[\tilde{\mathbf{x}}^f(\tilde{\mathbf{x}}^f)^T\right] = \mathbf{0}
	\end{align}
	which leads us to,
	\begin{align}
		\left(\mathbf{\Gamma}^f\right) &- \mathbf{\Gamma}^f\bigg[\mathbf{\Gamma}^{{-1}}_1 + \mathbf{\Gamma}^{{-1}}_2 +\mathbf{\Gamma}^{{-1}}_1\mathbf{\Gamma}_{12}\mathbf{\Gamma}_2^{{-1}} + \mathbf{\Gamma}_2^{{-1}}\mathbf{\Gamma}_{21}\mathbf{\Gamma}_1^{{-1}} \notag \\ &- \left(\omega_2\mathbf{\Gamma}_1^{{-1}}\mathbf{\Gamma}_{12} + \omega_1\mathbf{\Gamma}_2^{{-1}}\mathbf{\Gamma}_{21} \right)\mathbf{\Gamma}_m^{h^{-1}} + \mathbf{\Gamma}^{h^{-1}}_m \notag \\
			&+   \mathbf{\Gamma}^{h^{-1}}_m\left(\omega_2 \mathbf{\Gamma}_{21}\mathbf{\Gamma}_1^{{-1}} + \omega_1\mathbf{\Gamma}_{12}\mathbf{\Gamma}_2^{{-1}} \right) \bigg]\mathbf{\Gamma}^f = \mathbf{0}
	\end{align}
	where $\mathbf{\Gamma}_{12}$ is the cross-correlation matrix between the estimation error of $\hat{\mathbf{x}}_{1}$ and $\hat{\mathbf{x}}_{2}$. It has also been assumed that $\boldsymbol{\gamma}^h_m,\mathbf{\Gamma}^h_m$ form a consistent pair, such that $\mathbb{E}\left[\tilde{\boldsymbol{\gamma}}^h_m\left(\tilde{\boldsymbol{\gamma}}^h_m\right)^T \right] = \mathbf{\Gamma}^h_m$. Multiplying both sides by $\mathbf{\Gamma}^{f^{-1}}$ and using eqn. \eqref{fusedCov}, we get,
	\begin{align}
		&\left(\omega_2\mathbf{\Gamma}_1^{{-1}}\mathbf{\Gamma}_{12}\mathbf{\Gamma}_m^{h^{-1}} + \omega_1\mathbf{\Gamma}_m^{h^{-1}}\mathbf{\Gamma}_{12}\mathbf{\Gamma}_2^{{-1}} - \mathbf{\Gamma}_1^{{-1}}\mathbf{\Gamma}_{12}\mathbf{\Gamma}_2^{{-1}}\right) + \notag \\
			& \left( \omega_1\mathbf{\Gamma}_2^{{-1}}\mathbf{\Gamma}_{21}\mathbf{\Gamma}_m^{h^{-1}} + \omega_2\mathbf{\Gamma}_m^{h^{-1}}\mathbf{\Gamma}_{21}\mathbf{\Gamma}_1^{{-1}} - \mathbf{\Gamma}_2^{{-1}}\mathbf{\Gamma}_{21}\mathbf{\Gamma}_1^{{-1}}\right) = \mathbf{0} \label{eq_cond_consist_1}
	\end{align}
	Since $\Gamma_{21} = \Gamma_{12}$ and all matrices are symmetric, the above equation is in the form of $A+A^T = \mathbf{0}$, where
		\begin{align}
		A = \left(\omega_2\mathbf{\Gamma}_1^{{-1}}\mathbf{\Gamma}_{12}\mathbf{\Gamma}_m^{h^{-1}} + \omega_1\mathbf{\Gamma}_m^{h^{-1}}\mathbf{\Gamma}_{12}\mathbf{\Gamma}_2^{{-1}} - \mathbf{\Gamma}_1^{{-1}}\mathbf{\Gamma}_{12}\mathbf{\Gamma}_2^{{-1}}\right).
	\end{align}
	Hence the condition for consistency can be simplified to,
	\begin{align}
		\left(\omega_2\mathbf{\Gamma}_1^{{-1}}\mathbf{\Gamma}_{12}\mathbf{\Gamma}_m^{h^{-1}} + \omega_1\mathbf{\Gamma}_m^{h^{-1}}\mathbf{\Gamma}_{12}\mathbf{\Gamma}_2^{{-1}} - \mathbf{\Gamma}_1^{{-1}}\mathbf{\Gamma}_{12}\mathbf{\Gamma}_2^{{-1}}\right) = \mathbf{0} \label{eqn_consis_cond}
	\end{align}
	
	Note that this is in the form of Sylvester equation $\mathbf{P}\mathbf{\Gamma}_m^{h^{-1}} + \mathbf{\Gamma}_m^{h^{-1}}\mathbf{Q} = \mathbf{R}$, where $\mathbf{P} = \omega_2\mathbf{\Gamma}_1^{{-1}}\mathbf{\Gamma}_{12} $, $\mathbf{Q} = \omega_1\mathbf{\Gamma}_{12}\mathbf{\Gamma}_2^{{-1}} $ and $\mathbf{R} = \mathbf{\Gamma}_1^{{-1}}\mathbf{\Gamma}_{12}\mathbf{\Gamma}_2^{{-1}} $. A unique solution is ensured when the spectra of matrices $\mathbf{P}$ and $\mathbf{Q}$ do not overlap \cite{bhatia1997and}. This can be ensured by enforcing $\omega_1 \neq \omega_2$.
	
	Note that in the case of naive fusion, there is no upper bound on $\mathbf{\Gamma}_m^{h}$ since $\mathbf{\Gamma}_m^{h^{-1}} = \mathbf{0}$, which makes the estimates inconsistent for unsymmetric positive definite cross-correlation matrices $\mathbf{\Gamma}_{12}$.  
	

	 Using a linear correlation coefficient $\rho$, which takes positive values in the case of homogeneous (same state-space) track fusion, it is easy to prove that HMD-GA is consistent for its extreme values.
	
	\noindent When $\rho = 0$, it is obvious that the estimates are independent and HMD-GA is therefore conservative. The case of perfect correlation, $\rho = 1$ can be understood using the following assumption for the local track estimate,
	\begin{align}
		\hat{\mathbf{x}}_i = \mathbf{x} + \mathbf{v}_i, \quad i = 1,2,\dots \label{eq_local_sampleFrom_truth}
	\end{align} 
	where $\mathbf{x}$ is the true value and the corresponding noise term, $\mathbf{v}_i$ is sampled from the distribution $\mathcal{N}(\mathbf{0}, \mathbf{\Gamma}_i)$. This simply follows from the unbiased-ness and consistency of the local tracker. In the case of perfect correlation, the local estimates $\hat{\mathbf{x}}_1$ and $\hat{\mathbf{x}}_2$, both following the relation in eqn. \eqref{eq_local_sampleFrom_truth}, should exhibit a linear relationship,
	\begin{align}
		\hat{\mathbf{x}}_1 = \mathbf{A}\hat{\mathbf{x}}_2 + \mathbf{B}
	\end{align}
	 It follows from unbiasedness that $\mathbf{A} = \mathbf{I}$ and $\mathbf{B} = \mathbf{0}$, where $\mathbf{I}$ is the identity matrix of appropriate order. Thus, $\hat{\mathbf{x}}_1 = \hat{\mathbf{x}}_2$ should follow and HMD-GA effectively reduces to inverse covariance intersection which is proven as a consistent method in \cite{noack2017inverse}.   

It can also be proved that without $\tilde{\mathbf{\Gamma}}$, $\mathbf{\Gamma}_m^{h}$ is consistent and exactly becomes inverse covariance intersection, which is discussed next.
	
	\subsection{Relation with ICI}
	Observing the implementation of HMD-GA in eqn. \eqref{eq_hmdNaive}, its similarity with inverse covariance intersection (ICI) is striking since the only change is with respect to $\Gamma_m^h$. The main motivation behind using ICI is its guaranteed consistency and a non-inflated covariance compared to covariance intersection. However, it doesn't enjoy the fruits of being a natural and generalized fusion algorithm like HMD and GMD. The fusion equation for ICI are,
	\begin{align}
		\mathbf{\Gamma}^\text{ICI} &= \left[\left( \mathbf{\Gamma}_1\right)^{-1} + \left( \mathbf{\Gamma}_2\right)^{-1} - \left(\omega_1\mathbf{\Gamma}_1 +\omega_2\mathbf{\Gamma}_2 \right)^{-1}\right]^{-1} \notag \\
		\mathbf{\hat{x}}^\text{ICI} &= \bigg[\left( \mathbf{\Gamma}_1\right)^{-1}\mathbf{\hat{x}}_1 + \left( \mathbf{\Gamma}_2\right)^{-1}\mathbf{\hat{x}}_2 - \notag \\
										& \quad \left(\omega_1\mathbf{\Gamma}_1 +\omega_2\mathbf{\Gamma}_2 \right)^{-1}\left(\omega_1\mathbf{\hat{x}}_1 + \omega_2\mathbf{\hat{x}}_2\right)	\bigg]
	\end{align}
	with $\omega_1 = (1-\omega_2)$ as before. Note that the fusion weights have been interchanged in comparison to HMD-GA. This doesn't result in any loss of generalization since there's no unique interpretation of $\omega$ in the literature.
	
	\subsubsection{Correlation Structure Employed for ICI}

	The correlation structure employed for ICI is presented in \cite{noack2017decentralized,noack2017inverse, sijs2010state} with the assumption of Gaussianity of all components. If the common information component $P(\mathbf{x}|\mathbf{z}_1\cap \mathbf{z}_2) = P(\mathbf{x}|\mathbf{z}_2\cap \mathbf{z}_1)$ are assumed to be Gaussian distributed as  $\mathcal{N}\left(\mathbf{x};\boldsymbol{\gamma}_m, \mathbf{\Gamma}_m,\right)$, and the mutually independent component in a track $P(\mathbf{x|z}_{1/2})$ is Gaussian as $\mathcal{N}\left(\mathbf{x};\boldsymbol{\gamma}_1^\text{ind}, \mathbf{\Gamma}_1^\text{ind} \right)$, then the local track density is,
	\begin{align}
		P(\mathbf{x}|\mathbf{z}_1) &= P(\mathbf{x|z}_{1/2})P(\mathbf{x}|\mathbf{z}_1\cap \mathbf{z}_2) \notag \\
		&\propto \mathcal{N}\left(\mathbf{x}; \mathbf{\hat{x}}_1,\mathbf{\Gamma_1}\right) 
	\end{align}
	where
	\begin{align}
		\left(\mathbf{\Gamma}_1 \right)^{-1} &= \left(\mathbf{\Gamma}^{-1}_m +  \left(\mathbf{\Gamma}_1^\text{ind}\right)^{-1} \right) \\
		\mathbf{\hat{x}}_1 &= \mathbf{\Gamma}_1 \left( \left(\mathbf{\Gamma}_1^\text{ind}\right)^{-1}  \boldsymbol{\gamma}_1^\text{ind} + \mathbf{\Gamma}^{-1}_m\boldsymbol{\gamma}_m \right) \label{eq_corr_struct}
	\end{align}
	which are the local track density parameters for Gaussian case. Similarly for the node $j$,
	\begin{align}
		\left(\mathbf{\Gamma}_2 \right)^{-1} &= \left(\mathbf{\Gamma}^{-1}_m +  \left(\mathbf{\Gamma}_2^\text{ind}\right)^{-1} \right) \\
		\mathbf{\hat{x}}_j &= \mathbf{\Gamma}_2 \left( \left(\mathbf{\Gamma}_2^\text{ind}\right)^{-1}  \boldsymbol{\gamma}_2^\text{ind} + \mathbf{\Gamma}^{-1}_m\boldsymbol{\gamma}_m \right) \label{eq_corr_struct_j}
	\end{align}
	If the mutual information component $\mathbf{\Gamma}_m, \boldsymbol{\gamma}_m $ is known, then the optimal fusion is simply the ratio of the product of local densities and the mutual information, as in eqn. \eqref{exactBayesian}. The result is,
	\begin{subequations}\label{eq_fusion_corr_struct}
		\begin{align}
			\mathbf{\Gamma}^f &= \left( \left( \mathbf{\Gamma}_1\right)^{-1} + \left( \mathbf{\Gamma}_2\right)^{-1} - \mathbf{\Gamma}^{-1}_m\right) \\
			\mathbf{\hat{x}}^f &= \left( \left( \mathbf{\Gamma}_1\right)^{-1}\mathbf{\hat{x}}_1 + \left( \mathbf{\Gamma}_2\right)^{-1}\mathbf{\hat{x}}_2 - \mathbf{\Gamma}^{-1}_m\boldsymbol{\gamma}_m \right)
		\end{align}
	\end{subequations}
	where $\mathbf{\Gamma}^f, \mathbf{\hat{x}}^f$ are the parameter of fused (Gaussian) density. 
	
	It is easy to prove that the resulting cross-correlation matrix according to the proposed correlation structure is,
	\begin{align}
		\mathbb{E}\left[\tilde{\mathbf{x}}_1\left(\tilde{\mathbf{x}}_2\right)^T \right] =  \mathbf{\Gamma}_1\mathbf{\Gamma}^{-1}_m\mathbf{\Gamma}_2\label{eq_crossCorr_struct}
	\end{align}	
	
	\subsection{Contradictions in the design of ICI}
	There are minor flaws in the design of ICI which are also evident from the aforementioned correlation structure it uses. In relation to eqn. \eqref{eq_fusion_corr_struct}, the mutual information component ICI uses are,
	\begin{align}
		\mathbf{\Gamma}_m^\text{ICI} = \left(\omega_1\mathbf{\Gamma}_1 +\omega_2\mathbf{\Gamma}_2 \right);\quad\boldsymbol{\gamma}_m^\text{ICI} = \left(\omega_1\mathbf{\hat{x}}_1 + \omega_2\mathbf{\hat{x}}_2\right) \label{eq_mutualInfo_compo_ici}
	\end{align}
	which do not form a consistent pair for the estimate of $\mathbf{x}$ as a mutual component like in the case of HMD-GA. In relation to the correlation structure presented below, this flaw is akin to using a Kalman filter with optimistic sensor/process noise covariance. 
	
	In such a case, the bound derived in eqn. \eqref{eqn_consis_cond} is no longer valid, and inconsistency of $(\boldsymbol{\gamma}_m^\text{ICI}, \mathbf{\Gamma}^\text{ICI}_m)$ needs to be additionally accounted for. Thus, in case of ICI, eqn. \eqref{eq_cond_consist_1} becomes,
	\begin{align}
		&\left(\omega_2\mathbf{\Gamma}_1^{{-1}}\mathbf{\Gamma}_{12}\mathbf{\Gamma}_m^{\text{ICI}^{-1}} + \omega_1\mathbf{\Gamma}_m^{\text{ICI}^{-1}}\mathbf{\Gamma}_{12}\mathbf{\Gamma}_2^{{-1}} - \mathbf{\Gamma}_1^{{-1}}\mathbf{\Gamma}_{12}\mathbf{\Gamma}_2^{{-1}}\right) + \notag \\
		& \quad \left( \omega_1\mathbf{\Gamma}_2^{{-1}}\mathbf{\Gamma}_{21}\mathbf{\Gamma}_m^{\text{ICI}^{-1}} + \omega_2\mathbf{\Gamma}_m^{\text{ICI}^{-1}}\mathbf{\Gamma}_{21}\mathbf{\Gamma}_1^{{-1}} - \mathbf{\Gamma}_2^{{-1}}\mathbf{\Gamma}_{21}\mathbf{\Gamma}_1^{{-1}}\right) \notag \\
					& \quad - \mathbf{\Gamma}_m^{\text{ICI}^{-1}}\tilde{\mathbf{\Gamma}}\mathbf{\Gamma}_m^{\text{ICI}^{-1}} \succeq \mathbf{0} \label{eq_cond_consist_ici}
	\end{align}
	 Where $\tilde{\mathbf{\Gamma}}$ is the spread-of-means term since $\mathbb{E}\left[\tilde{\boldsymbol{\gamma}}_m^\text{ICI} \left(\tilde{\boldsymbol{\gamma}}_m^\text{ICI}\right)^T \right] = \mathbf{\Gamma}^\text{ICI}_m + \tilde{\mathbf{\Gamma}}$. Eqn. \eqref{eq_cond_consist_ici} reveals a tighter consistency bound on $\mathbf{\Gamma}_m^{\text{ICI}^{-1}}$ than in the case of HMD-GA. 
	 
	 In \cite{noack2017inverse}, the consistency of ICI was proved by comparing it with optimal track-fusion covariance instead of its own mean-squared error. It should be noted that the optimal track-fusion algorithm is only a local optima, and only centralized track fusion is known to be efficient (reaches Cram\'er-Rao bound \cite{bar2001estimation}), for linear case and full communication rate \cite{bar1995multitarget}. 
	 
	 The ICI was designed to satisfy the aforementioned correlation structure \cite{noack2017inverse}. With respect to eqn. \eqref{eq_corr_struct} and eqn. \eqref{eq_corr_struct_j}, the necessary conditions on the covariances involved are,
	 \begin{itemize}
	 	\item As $\mathbf{\mathbf{\Gamma}}_1^\text{ind}$ and $\mathbf{\mathbf{\Gamma}}_2^\text{ind}$ are the covariance of independent component of $\mathbf{x}$, they have to be at least  positive semi-definite. 
	 	\item Since, according to the correlation structure, $\mathbf{\Gamma}_1$ and $\mathbf{\Gamma}_2$ are the result of naive fusion of dependent and independent components, the resulting covariance should be smaller than or equal to those components. Thus, 
	 	\begin{align}
	 		\mathbf{\Gamma}_m \geq  \mathbf{\Gamma}_1; \quad \mathbf{\Gamma}_m \geq  \mathbf{\Gamma}_2 \label{eq_cond_on_Gamma_m}
	 	\end{align}
	 \end{itemize}
		
	Now consider the approximation of $\mathbf{\Gamma}_m$ employed by ICI in eqn. \eqref{eq_mutualInfo_compo_ici} which suggests a clear violation of eqn. \eqref{eq_cond_on_Gamma_m}, since $\mathbf{\Gamma}_m^\text{ICI} \geq \min\left(\mathbf{\Gamma}_1, \mathbf{\Gamma}_2\right)$ always hold. 
		 
	 For $\boldsymbol{\gamma}_m= \left(\omega_1\mathbf{\hat{x}}_1 + \omega_2\mathbf{\hat{x}}_2\right)$, it can be proved that the requirement for consistency of overall fused estimate is,
	 \begin{align}
	 	\mathbf{\Gamma}_m \leq \omega\mathbf{\Gamma}_1 + (1-\omega)\mathbf{\Gamma}_2 \label{eq_ici_cons_cond}
	 \end{align}
	ICI uses this exact upper bound for the correlation component, due to which it is always consistent. But eqns. \eqref{eq_ici_cons_cond} and \eqref{eq_cond_on_Gamma_m} are in direct contradiction of each other. Therefore, ICI tends to break the exact correlation structure it was derived from.

	\section{Generalized Mean Density Fusion using Importance Sampling.}\label{implementation}
	Unlike HMD, geometric mean density fusion doesn't enjoy applicability to Gaussian mixtures with just one approximation. Although generalized Chernoff fusion and sigma-point-based Chernoff fusion are viable alternatives to the fusion of Gaussian mixtures, sampling-based alternatives are needed in some situations, especially in the case of low-precision systems where matrix inversion is not accurate. The non-positive definiteness of fused covariance matrices is another problem in such scenarios. Sampling methods also provide us with the option of fusing Gaussian and non-Gaussian local tracks together with no approximations. Such methods would also be useful in cases where track densities have to undergo non-linear transformations, like in heterogeneous fusion. 
	
	In \cite{ahmed2015s}, the concern of Gaussian mixture-based fusion was discussed using importance sampling. The focus of the paper was on different choices of proposal density to sample from. Here we show that due to the nature of expressions of generalized mean density-based fusion, the requirement of proposal density is nullified.
	
	We explain the procedure of calculating a generalized expectation of a function using importance sampling in the case of GMD by approximating its expression with a Gaussian and then generalize it to HMD. This requires calculating its first- and second-order moments. Using importance sampling, the expectation of a function $\mathit{f}\left(\mathbf{x}\right)$ with respect to a density $p\left(\mathbf{x}\right)$ can be calculated as,
	\begin{align}
		\mathbb{E}\left[\mathit{f}\left(\mathbf{x}\right)\right]_{p(\mathbf{x})} &= \int \mathit{f}\left(\mathbf{x}\right) p(\mathbf{x}) d\mathbf{x} = \int \frac{\mathit{f}\left(\mathbf{x}\right) p(\mathbf{x})}{q(\mathbf{x})}q(\mathbf{x}) dx \notag \\
				&= \mathbb{E}\left[\frac{p(\mathbf{x})}{q(\mathbf{x})} \mathit{f}\left(\mathbf{x}\right) \right]_{q(\mathbf{x})}
	\end{align}
	where $q(\mathbf{x})$ is the proposal density which is relatively easy to sample from, in comparison to $p(\mathbf{x})$. The expectation can then be computed using samples $\mathbf{x}_n$ from $q(\mathbf{x})$ as,
	\begin{align}
		\mathbb{E}\left[\frac{p(\mathbf{x})}{q(\mathbf{x})} \mathit{f}\left(\mathbf{x}\right) \right]_{\mathrlap{q(\mathbf{x})} } \quad =  \frac{1}{N} \sum_{n=1}^{N} \frac{\mathit{f}\left(\mathbf{x}_n\right) p(\mathbf{x}_n)}{q(\mathbf{x}_n)}
	\end{align}
	where $N$ is the number of samples. 
	
	Now, let's observe the expression for geometric mean density fusion,
	\begin{align}
		p^g(\mathbf{x}) \propto \frac{p^1(\mathbf{x})p^2(\mathbf{x})}{ \left(p^1(\mathbf{x})\right)^{\omega_2} \left(p^2(\mathbf{x})\right)^{\omega_1} } \label{eq_gmd}
	\end{align}
	where $p^1(\mathbf{x}), p^2(\mathbf{x})$ are the local track densities and $\omega_1, \omega_2$ are the fusion weights. The proportionality constant is the normalization factor $\zeta^g$ such that the $p^g(\mathbf{x})$ is a valid density. For now, let's assume that $p^1(\mathbf{x})$ is Gaussian (or easy to sample from). 
	
	Due to the structure of GMD, a generalized expectation with respect to $p^g(\mathbf{x})$ can be expressed as,
	\begin{align}
		\mathbb{E}[\mathit{f}(\mathbf{x})]_{p^g(\mathbf{x})} &= \frac{1}{\zeta^g}\int \mathit{f}(\mathbf{x}) \frac{p^1(\mathbf{x})p^2(\mathbf{x})}{ \left(p^1(\mathbf{x})\right)^{\omega_2} \left(p^2(\mathbf{x})\right)^{\omega_1} } d\mathbf{x} \notag \\
				&= \frac{1}{\zeta^g} \int \mathit{f}(\mathbf{x}) \frac{p^2(\mathbf{x})}{ \left(p^1(\mathbf{x})\right)^{\omega_2} \left(p^2(\mathbf{x})\right)^{\omega_1} } p^1(\mathbf{x}) d\mathbf{x} \notag \\
				&= \mathbb{E}\left[\mathit{f}(\mathbf{x})\frac{p^2(\mathbf{x})}{ \left(p^1(\mathbf{x})\right)^{\omega_2} \left(p^2(\mathbf{x})\right)^{\omega_1} }\right]_{p^1(\mathbf{x})} \label{eq_gmdSample_p1}
	\end{align}
	The resulting expectation can be computed using the following summation,
	\begin{align}
		\mathbb{E}[\mathit{f}(\mathbf{x})]_{p^g(\mathbf{x})} = \frac{1}{\zeta^g} \sum_{n=1}^N \mathit{f}(\mathbf{x}_n)\frac{p^2(\mathbf{x}_n)}{ \left(p^1(\mathbf{x}_n)\right)^{\omega_2} \left(p^2(\mathbf{x}_n)\right)^{\omega_1} }
	\end{align}
	with $\mathbf{x}_n$ being samples from $p^1(\mathbf{x})$. Note that due to symmetry of the mean-density expressions, it is also possible to obtain the expectation in question using samples from $p^2(\mathbf{x})$
	\begin{align}
		\mathbb{E}[\mathit{f}(\mathbf{x})]_{p^g(\mathbf{x})} &= \mathbb{E}\left[\mathit{f}(\mathbf{x})\frac{p^1(\mathbf{x})}{ \left(p^1(\mathbf{x})\right)^{\omega_2} \left(p^2(\mathbf{x})\right)^{\omega_1} }\right]_{p^2(\mathbf{x})} \label{eq_gmdSample_p2}
	\end{align}
	In comparison to HMD, the expression for GMD in eqn. \eqref{eq_gmd} only differs in the approximation of mutual information component in the denominator. Hence, HMD also enjoys sampling based evaluation with some obvious changes in eqns. \eqref{eq_gmdSample_p1} and \eqref{eq_gmdSample_p2}. Note that the normalization constant also needs to be calculated, which is given by,
	\begin{align}
		\zeta^g = \mathbb{E}[1]_{p^g(\mathbf{x})} &=  \sum_{n=1}^N \frac{p^1(\mathbf{x}_n)}{ \left(p^1(\mathbf{x}_n)\right)^{\omega_2} \left(p^2(\mathbf{x}_n)\right)^{\omega_1} } \notag \\
					&= \sum_{n=1}^N \frac{p^2(\mathbf{x}_n)}{ \left(p^1(\mathbf{x}_n)\right)^{\omega_2} \left(p^2(\mathbf{x}_n)\right)^{\omega_1} } \label{eq_zeta_g_sample}
	\end{align}
	where, again, samples from either $p^1(\mathbf{x}_n)$ or $p^2(\mathbf{x}_n)$ can be used with relevant expression. The evaluation of expectations for HMD follow similarly,
	\begin{subequations} \label{eq_hmdSample}
		\begin{align}
			\mathbb{E}[\mathit{f}(\mathbf{x})]_{p^h(\mathbf{x})} &= \mathbb{E}\left[\mathit{f}(\mathbf{x})\frac{p^1(\mathbf{x})}{ {\omega_2}\left(p^1(\mathbf{x})\right) +  {\omega_1}\left(p^2(\mathbf{x})\right)}\right]_{p^2(\mathbf{x})} \label{eq_hmdSample_p2} \\
			&= \mathbb{E}\left[\mathit{f}(\mathbf{x})\frac{p^2(\mathbf{x})}{ {\omega_2}\left(p^1(\mathbf{x})\right) + {\omega_1}\left(p^2(\mathbf{x})\right)}\right]_{p^1(\mathbf{x})} \label{eq_hmdSample_p1},
		\end{align}
	\end{subequations}
	Again, the normalization constant can be calculated by substituting $\mathit{f}(\mathbf{x}) = 1$ in eqn. \eqref{eq_hmdSample}.
	
	\subsection{Sampling based HMD fusion of Gaussian Mixtures}
	Considering Gaussian mixture densities in eqn. \eqref{eq_gmd}, 
	\begin{align}
		p^1(\mathbf{x}) &= \sum_{m=1}^M \alpha_m \mathcal{N}\left(\mathbf{x};\hat{\mathbf{x}}_m, \mathbf{\Gamma}_m\right) \\
		p^2(\mathbf{x}) &= \sum_{m=1}^M \beta_n \mathcal{N}\left(\mathbf{x};\hat{\bm{\lambda}}_n, \bm{\Lambda}_n\right)
	\end{align}
	where the notations are self-explanatory. The resulting HMD is then,
	\begin{align}
		p^h(\mathbf{x}) \propto  \sum_{m=1}^M\sum_{n=1}^N \alpha_m \beta_n \frac{\mathcal{N}\left(\mathbf{x};\hat{\mathbf{x}}_m, \mathbf{\Gamma}_m\right) \mathcal{N}\left(\mathbf{x};\hat{\bm{\lambda}}_n, \boldsymbol{\Lambda}_n\right) }{\omega_2 p^1(\mathbf{x}) + \omega_1p^2(\mathbf{x})} \label{eq_hmd_GM}
	\end{align}
	The trick we are going to use here is to approximate the fraction in the R.H.S. of eqn. \eqref{eq_hmd_GM} as a Gaussian density using the sampling method proposed before. Thus, 
	\begin{align}
		\frac{\mathcal{N}\left(\mathbf{x};\hat{\mathbf{x}}_m, \mathbf{\Gamma}_m\right) \mathcal{N}\left(\mathbf{x};\hat{\bm{\lambda}}_n, \bm{\Lambda}_n\right) }{\omega_2 p^1(\mathbf{x}) + \omega_1p^2(\mathbf{x})} \approx \frac{1}{\zeta}\mathcal{N}\left(\mathbf{x}; \hat{\mathbf{x}}^{m,n}_s, \mathbf{\Gamma}^{m,n}_s\right)
	\end{align}
	where $\zeta$ is the scaling factor so that the ratio is a valid probability distribution. The mean, $\hat{\mathbf{x}}^{m,n}_s$ is given by,
	\begin{align}
		\hat{\mathbf{x}}^{m,n}_s &= \frac{1}{\zeta}\sum_{s = 1}^S \mathbf{x}_s \frac{\mathcal{N}\left(\mathbf{x}_s;\hat{\bm{\lambda}}_n, \boldsymbol{\Lambda}_n \right)}{\omega_2 p^1(\mathbf{x}_s) + \omega_1p^2(\mathbf{x}_s)}\notag \\ 
			&\qquad= \frac{1}{\zeta} \sum_{s = 1}^S \mathbf{x}_s \frac{\mathcal{N}\left(\mathbf{x}_s;\hat{\mathbf{x}}_m, \mathbf{\Gamma}_m\right)}{\omega_2 p^1(\mathbf{x}_s) + \omega_1p^2(\mathbf{x}_s)}
	\end{align}
	depending upon whether $\mathcal{N}\left(\mathbf{x};\hat{\mathbf{x}}_m, \mathbf{\Gamma}_m\right)$ or $\mathcal{N}\left(\mathbf{x}_s;\hat{\mathbf{\lambda}}_n, \boldsymbol{\Lambda}_n \right)$ is used for sampling. Calculation of $\mathbf{\Gamma}^{m,n}_s$ follows similarly.
	\begin{align}
		\mathbf{\Gamma}^{m,n}_s = \frac{1}{\zeta} \sum_{s = 1}^S (\mathbf{x}_s)(\mathbf{x}_s)^T &\frac{\mathcal{N}\left(\mathbf{x}_s;\hat{\mathbf{\lambda}}_n, \boldsymbol{\Lambda}_n \right)}{\omega_2 p^1(\mathbf{x}_s) + \omega_1p^2(\mathbf{x}_s)} \notag \\
				&- (\hat{\mathbf{x}}^{m,n}_s)(\hat{\mathbf{x}}^{m,n}_s)^T,
	\end{align}
	when $\mathcal{N}\left(\mathbf{x}_s;\hat{\mathbf{x}}_m, \mathbf{\Gamma}_m \right)$ is used for sampling. Note that calculation of $\zeta$ is required which is straightforward as in eqn. \eqref{eq_zeta_g_sample}. The resulting fused Gaussian mixture is then, 
	\begin{align}
		p^h(\mathbf{x}) =  \sum_{m=1}^M\sum_{n=1}^N \frac{\alpha_m \beta_n}{\zeta} \mathcal{N}\left(\mathbf{x}; \hat{\mathbf{x}}^{m,n}_s, \mathbf{\Gamma}^{m,n}_s \right)
	\end{align}
	
	\subsection{Which Density to Sample From ?}
	An intuition for fused density is that it lies at the intersection of local track densities. Depending on the number of samples, sampling from a single distribution might lead the fused density to orient  towards the only distribution sampled from. Thus, if the problem enjoys the flexibility of sampling from both distributions, one can opt to utilize samples from one distribution over the other. This depends on the normalization weights in the importance sampling method.
	
	Considering eqns. \eqref{eq_gmdSample_p1} and \eqref{eq_gmdSample_p2}, suppose the user samples $\mathbf{x}^1_s$ and $\mathbf{x}^2_s$ from pdfs $p^1(\mathbf{x})$ and $p^2(\mathbf{x})$ respectively. Then, one should choose the sample which gives a larger value of normalization weight. Therefore, choose $\mathbf{x}^1_s$ if,
	\begin{align}
		\frac{p^2(\mathbf{x}^1_s)}{ \left(p^1(\mathbf{x}^1_s)\right)^{\omega_2} \left(p^2(\mathbf{x}^1_s)\right)^{\omega_1} } \geq \frac{p^1(\mathbf{x}^2_s)}{ \left(p^1(\mathbf{x}^2_s)\right)^{\omega_2} \left(p^2(\mathbf{x}^2_s)\right)^{\omega_1} } \label{eq_sampleCond}
	\end{align}
	else, choose $\mathbf{x}^2_s$. Note that in contemporary systems, sampling from normal distribution can be easily parallelized, resulting in much faster processing using a GPU.
	
\begin{figure}[t]
		\centering
		\begin{subfigure}[t]{0.48\columnwidth}
			\centering
			\includegraphics[width=\columnwidth]{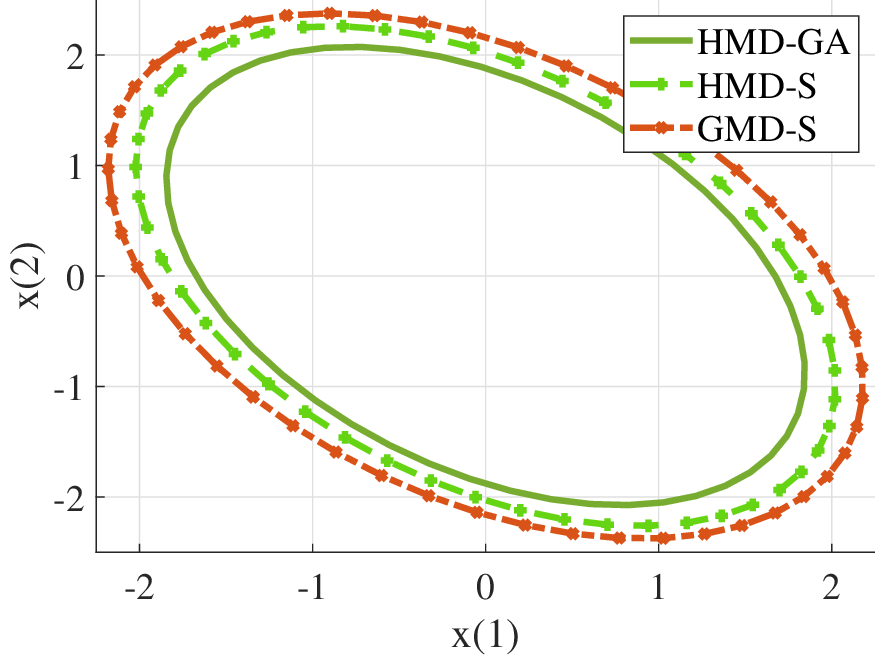}
			\caption{HMD-S v/s GMD-S.}
			\label{fig_hmd_s_vs_hmd_ga}
		\end{subfigure}
		\centering
		\begin{subfigure}[t]{0.48\columnwidth}
			\includegraphics[width=\columnwidth]{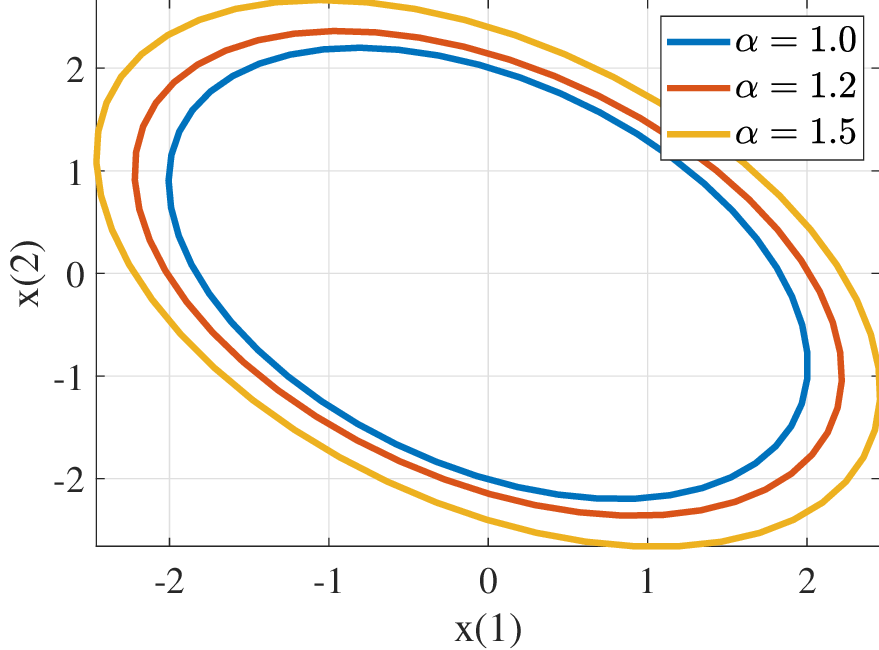}
			\caption{Effect of $\alpha$ on HMD-S.}
			\label{fig_hmdS_alpha}
		\end{subfigure}
		\caption{Uncertainty region (86.5\%) for HMD and GMD fusion of Gaussian densities using proposed sampling technique.}
		\label{fig_hmd_sample_Gauss}
	\end{figure}

	Finally, for demonstrating the case of fusing two Gaussian densities using the proposed sampling strategy, an example is set up, the results of which are shown in Fig. \ref{fig_hmd_sample_Gauss}. The local track densities are,
	\begin{align}
		p^1(\mathbf{x}) &= \mathcal{N}\bigg(\mathbf{x}; \begin{bmatrix} 	0.5\\ 1\end{bmatrix}, \begin{bmatrix} 2.5 & -1 \\ -1 & 1.2\end{bmatrix}\bigg), \\
		p^2(\mathbf{x}) &= \mathcal{N}\bigg(\mathbf{x}; \begin{bmatrix} 2\\ 1\end{bmatrix}, \begin{bmatrix} 0.8 & -0.5 \\ -0.5 & 4\end{bmatrix}\bigg)
	\end{align}
	
	where $5000$ samples were used from both the distributions based on eqn. \eqref{eq_sampleCond}. As expected, the geometric mean-density fusion, which is an analogue of CI, has the largest uncertainty region. Note that the covariance of HMD-GA is less than that of HMD-S, which is obvious since the mutual information component in the former is approximated by a Gaussian distribution, which has the largest entropy and hence the highest covariance. Therefore, the resulting fused covariance is minimum.
	
	The sampling method allows us to vary covariance-inflation with ease. To demonstrate this, we use an inflation factor $\alpha \geq 1$, and sample from the same densities but inflated covariance, $\mathbf{\Gamma}' = \alpha \mathbf{\Gamma}$. The resulting uncertainty regions are shown in Fig. \ref{fig_hmdS_alpha}. Small variations in $\alpha$ lead to a resulting covariance even greater than GMD without any changes in computation. Such results would be helpful in track fusion cases where the local densities suffer from inconsistency, track bias, anomalies, or adversarial attacks. 
	
\begin{figure}[t]
		\centering
		\begin{subfigure}{0.48\columnwidth}
			\centering
			\includegraphics[width = \linewidth]{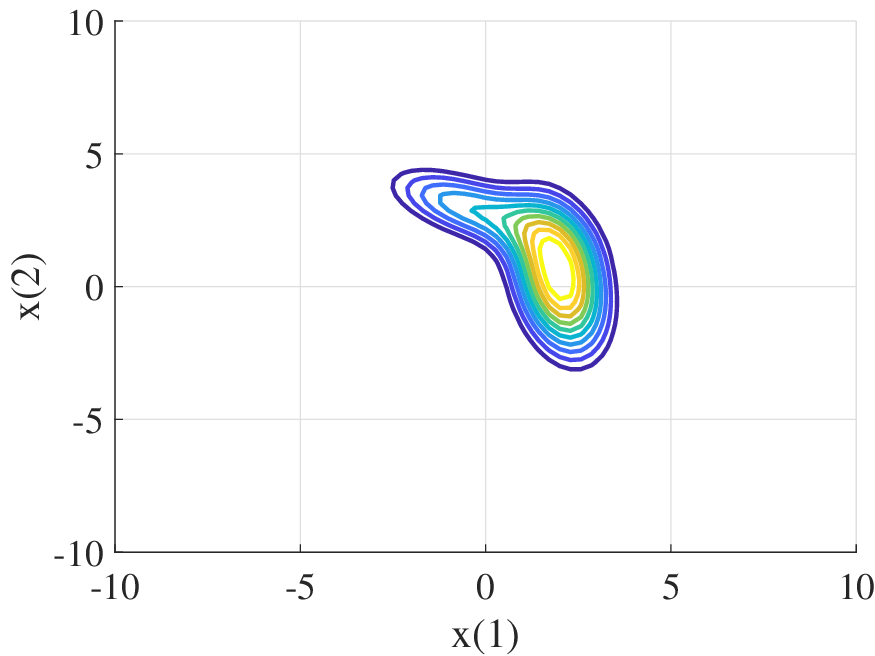}
			\caption{Local mixture density 1.}
			\label{fig_mixture_1}
		\end{subfigure}
		\begin{subfigure}{0.48\columnwidth}
			\centering
			\includegraphics[width = \linewidth]{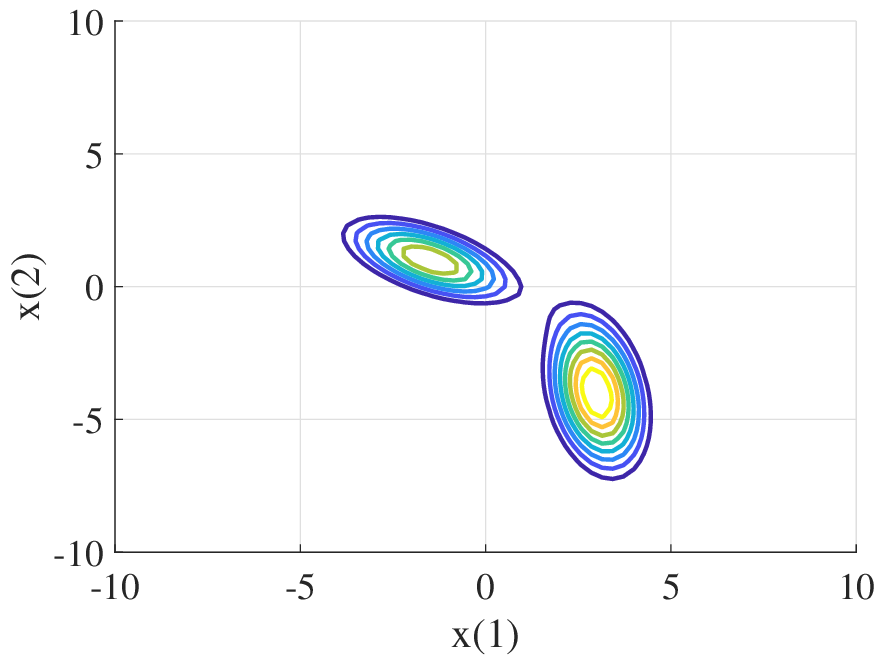}
			\caption{Local mixture density 2.}
			\label{fig_mixture_2}
		\end{subfigure}
		\begin{subfigure}{0.48\columnwidth}
			\centering
			\includegraphics[width = \linewidth]{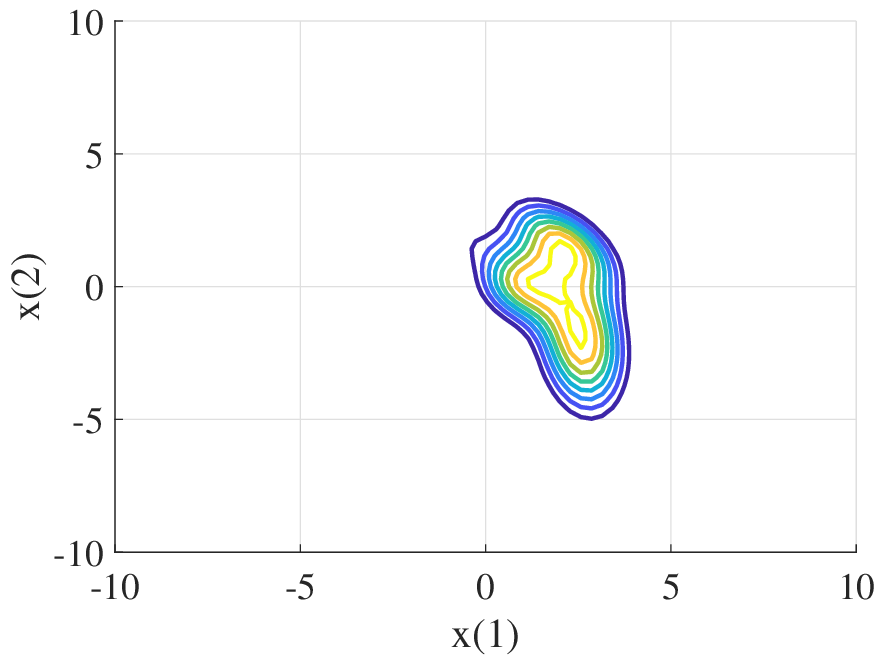}
			\caption{Fused mixture density using GMD.}
			\label{fig_mixture_g}
		\end{subfigure}
		\begin{subfigure}{0.48\columnwidth}
			\centering
			\includegraphics[width = \linewidth]{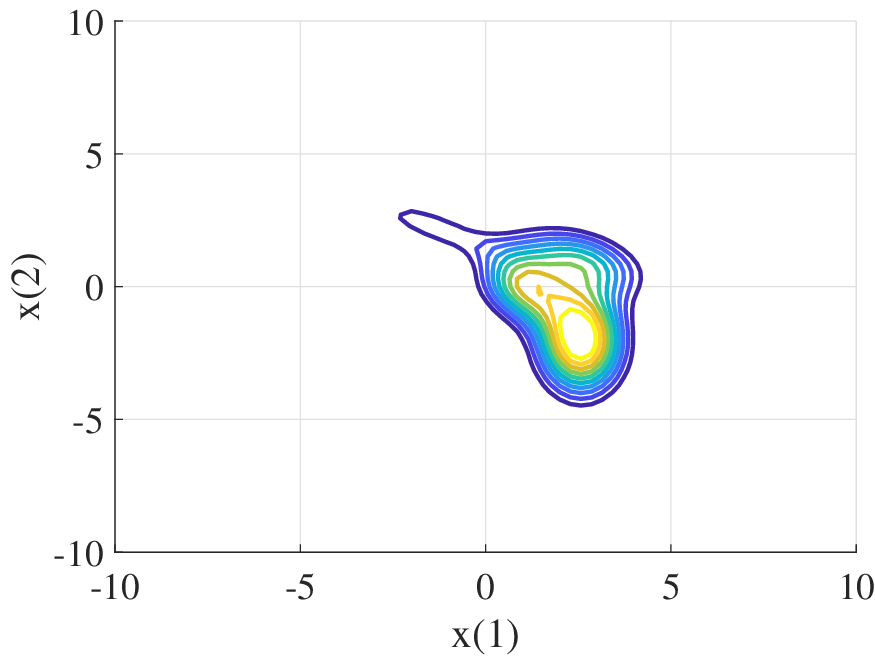}
			\caption{Fused mixture density using HMD.}
			\label{fig_mixture_h}
		\end{subfigure}
		\caption{Fusion of Gaussian mixtures using proposed sampling technique.}
	\end{figure}

	For the case of Gaussian mixtures, we consider the following track densities,
	\begin{align}
		p^1(\mathbf{x}) &= 0.3 \mathcal{N}\left(\mathbf{x}; \begin{bmatrix} -0.5\\3\end{bmatrix}, \begin{bmatrix} 2.5 &-1 \\ -1 & 1.2\end{bmatrix}\right) \notag \\ 
						&\quad + 0.7 \mathcal{N}\left(\mathbf{x}; \begin{bmatrix} 2\\0.3\end{bmatrix}, \begin{bmatrix} 0.8 &-0.5\\ -0.5& 4\end{bmatrix}\right) \notag \\						
		p^1(\mathbf{x}) &= 0.4 \mathcal{N}\left(\mathbf{x}; \begin{bmatrix} -1.5\\1\end{bmatrix}, \begin{bmatrix} 2.5 &-1 \\ -1 & 1.2\end{bmatrix}\right) \notag \\ 
		&\quad + 0.6 \mathcal{N}\left(\mathbf{x}; \begin{bmatrix} 3\\-4\end{bmatrix}, \begin{bmatrix} 0.8 &-0.5\\ -0.5& 4\end{bmatrix}\right) \notag					
	\end{align}
	The corresponding mode covariance is kept the same in both the densities without any loss of generalization. The local densities are plotted in Figs. \ref{fig_mixture_1} and \ref{fig_mixture_2}, and the corresponding fused mixtures are plotted in Figs. \ref{fig_mixture_g} and \ref{fig_mixture_h} for the cases of GMD and HMD, respectively. Interestingly, both the HMD and GMD mixtures lie at the intersection of local uncertainty regions, as in the case of Gaussian densities, with the uncertainty in HMD-S being less than that in GMD-S, as expected. Again, 5000 samples were used in the simulation.
	
	\section{Simulations} \label{sim}
	
	To assess the performance of HMD, especially its consistency, we performed a rigorous evaluation based on three different simulation scenarios. Note that HMD-GA was shown to produce the least covariance, which is also shown in Fig. \ref{fig_hmd_s_vs_hmd_ga}. Therefore, we preferred using HMD-GA for comparison in order to establish empirical consistency. In sampling-based methods, the main question of consistency can be tackled easily, which is evident in Fig. \ref{fig_hmdS_alpha}, which is why we kept it out of this section. Another reason for not testing sampling-based methods is the computational time involved in sampling in addition to Monte-Carlo (MC) runs.
	
	We first discuss the evaluation of fusion weights for the proposed HMD-GA fusion. The findings are interesting in the case of scalar estimates. Among the simulation scenarios, the first two cases are aimed at only consistency, and no error-based performance metrics are employed. The third simulation is aimed at evaluating the performance of HMD-GA in a real-life multi-target, multi-sensor tracking scenario wherein the fusion center is equipped with memory. Due to the huge sensor coverage of 300 km, the local tracks become inconsistent due to the curvature of cross-range uncertainty, and thus the consistency of the fusion methodology cannot be guaranteed among any of the fusion methodologies discussed in this article. 
	
	The aim of this section is to deem HMD, especially HMD-GA, a viable candidate for generalized fusion of processed densities, which is divided into two phases based on the aim, viz., consistency-based and performance-based scenarios. 
	
	\subsection{Fusion Weights}
	
	The fundamental basis of computing fusion weight $\omega$ in all conservative fusion techniques is to minimize some scalar function of output (reported) covariance. Throughout the literature, this problem has been perceived from the angles of minimizing the entropy of the fused estimate, or in Mahler's words, to increase the peakiness of the fused distribution \cite{julier2006empirical}, both of which tend to incline towards minimizing the fused covariance in the Gaussian case.
	
	Based on such heuristics, we try to optimize $\omega$ such that output covariance is minimized, but from a different view. We attempt to maximize the mutual information component, which also results in the minimization of fused covariance. The optimization becomes,
	\begin{align}
		\omega^* = \argmin_{\omega} \mathit{f}\left[\left((1-\omega) \mathbf{\Gamma_1} + \omega\mathbf{\Gamma_2} + \tilde{\mathbf{\Gamma}}\right)^{-1}\right] \label{eq_optim_omega}
	\end{align}
	where $\tilde{\mathbf{\Gamma}}$ which also depends on $\omega$, is defined in eqn. \eqref{eq_spread_of_means_alter}; and $\mathit{f}[.]$ is any scalar representation of a matrix like a determinant or trace. The reason for a different optimization approach is that covariance inflation is not a problem in HMD-GA, so the user should not be concerned about minimizing it.  Note that this optimization is slightly faster to compute as compared to that in CI and ICI since it requires only one matrix inversion compared to 2 and 3 in CI and ICI, respectively. 
	
	Table \ref{tab_omega_time} reports the computation times for implementing different fusion algorithms relative to naive fusion. All algorithms calculate fusion weights to minimize their respective covariances, whereas eqn. \eqref{eq_optim_omega} is employed in the case of HMD-GA. Ellipsoidal intersection (EI) \cite{sijs2010state}, which does not require calculating fusion weights, is also included for comparison.
	
	\begin{table}[h!]
		\caption{ Relative computational time of fusion algorithms.}
		\label{tab_omega_time}\centering
		\begin{tabular}{cccccc}
			\toprule
				\rowcolor{gray!10}
				& \textbf{Naive} & \textbf{CI} & \textbf{ICI} & \textbf{EI} & \textbf{HMD-GA} \\
			\midrule
			\textbf{Relative Time} & 1 & 3.74 & 4.43 & 4.09 & 3.68\\
			\bottomrule
		\end{tabular}
	\end{table}
	
	An interesting point can be observed in the case of scalar estimates. The fusion weight in the case of CI is calculated as ($\Gamma_1, \Gamma_2$ are scalars),
	\begin{align}
		\omega^* &=  \argmin_{\omega} \left[\omega {\Gamma}_1^{-1} + (1-\omega){\Gamma}_2^{-1}\right]^{-1} \notag \\
				&= \argmin_{\omega}\left[\Gamma_1 + \omega\left(\Gamma_2 - \Gamma_1\right)  \right]^{-1} \notag \\
				&= \begin{cases}
					0, \quad  \text{if }\Gamma_1 > \Gamma_2\\
					1,	\quad \text{if }\Gamma_2 > \Gamma_1	
				\end{cases}
	\end{align} 
	Similarly in the case of ICI, the given optimization problem becomes,
	\begin{align}
		\omega^* &=  \argmin_{\omega} \left[\frac{1}{\Gamma_1} + \frac{1}{\Gamma_2} - \frac{1}{\omega\Gamma_1 + (1-\omega)\Gamma_2}\right]^{-1} \notag \\
				&= \argmax_{\omega}\left[\omega\Gamma_1 + (1-\omega)\Gamma_2\right] \notag \\
				&= \begin{cases}
					1, \quad  \text{if }\Gamma_1 > \Gamma_2\\
					0,	\quad \text{if }\Gamma_2 > \Gamma_1	
				\end{cases}
	\end{align}
	Thus, the result in both cases is the local track estimate corresponding to the least covariance. Therefore, the fusion weights should not be optimized in CI and ICI algorithms for the case of scalar estimates, even though the results are consistent (assuming participating densities are consistent). Due to the presence of the spread-of-means term, HMD-GA does not suffer from any such problem.
	
\begin{figure}
		\centering
		\includegraphics[width = \columnwidth]{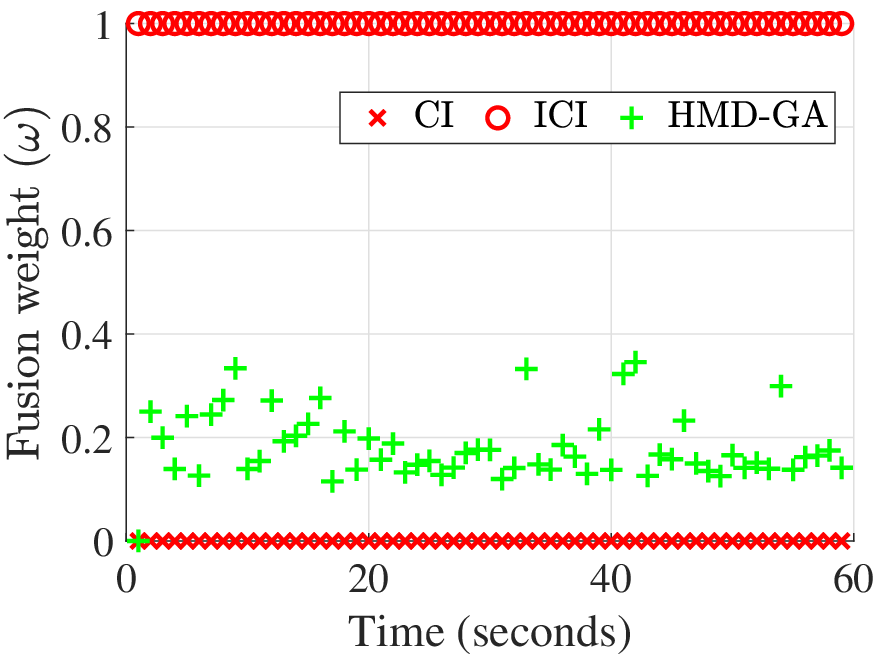}
		\caption{Comparison of fusion weights in case of scalar estimates.}
		\label{fig_omega_compare}
		
	\end{figure}

	To support this claim, a scalar parameter estimation problem was set up wherein the state $x_k$ and measurements $z_k$ at instant $k$ are described by the following equations,
	\begin{align}
		x_{k} &= x_{k-1} + w_{k-1} \\
		z_k &= x_k + v_k
	\end{align}
	where the noise parameters $w_{k-1}$ and $v_k$ are zero-mean Gaussian and mutually correlated with a correlation coefficient $\rho$. The correlation is assumed to be unknown, and at each instant, the prior distribution is attempted to fuse with the measurement distribution using conservative techniques. The fusion weight $\omega$, averaged over 500 Monte-Carlo runs, is shown in Fig. \ref{fig_omega_compare}. As observed, the fusion weight in the case of CI is always 0, and in the case of ICI, it is always 1. Thus, the output of these algorithms is the measurement itself. The figure proves that the only fusion performed here is in the case of HMD-GA, with an average fusion weight value 0.2. Thus, measurement is given more weight, but prior information is also fused.
	
	\subsection{Consistency-Based Test 1}
\begin{figure}[t]
		\centering
		\includegraphics[width=0.9\columnwidth]{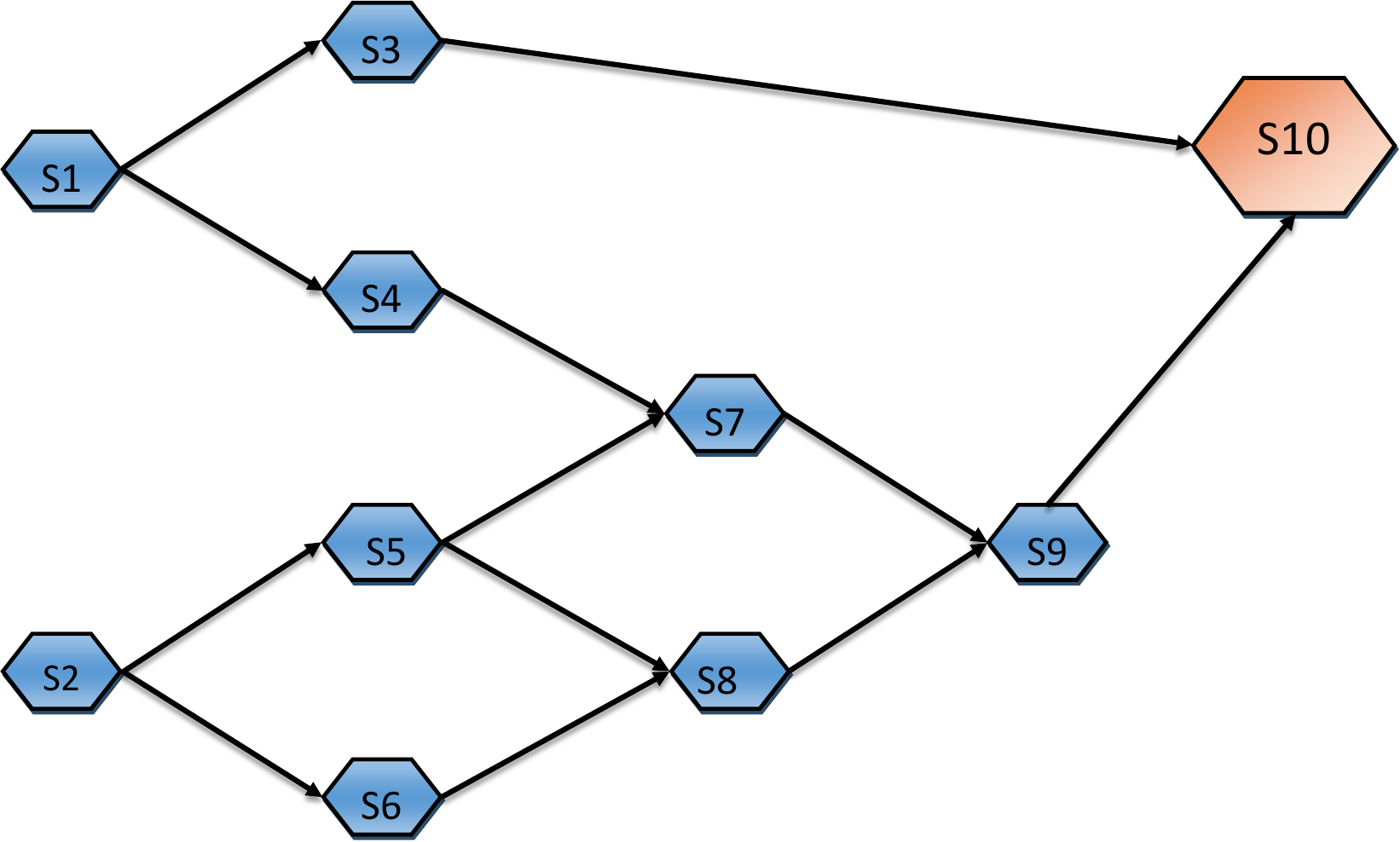}
		\caption{Consistency-based test scenario 1.}
		\label{sim_consBased_10Node_1}
	\end{figure}

	For consistency-based tests, we re-developed the scenarios employed in \cite{noack2017decentralized} and \cite{noack2017inverse}. The tests are aimed at testing consistency by visualizing covariance ellipsoids after averaging them over numerous Monte-Carlo runs. The setup for the first test consists of 10 nodes in a configuration as shown in Fig. \ref{sim_consBased_10Node_1}. The transmission of information follows the direction of the arrows in the figure. For example, node S7 receives estimates from nodes S4 and S5, which also gather estimates from nodes S1 and S2, respectively. The network flow for the rest of the nodes follows similarly. Each node is capable of processing the local measurements using a local Kalman filter (KF) before transmitting or fusing them with incoming tracks as per the network. No ambiguity on time synchronization or sensor bias has been included in the problem. 
	
	During an MC run, each node is initialized by sampling from the distribution, $\mathcal{N}\left(\begin{bmatrix} 0 \\ 0 \end{bmatrix}, \begin{bmatrix}2&0\\0&2  \end{bmatrix} \right) $. The measurement matrix at each node $i$ is,
	\begin{align}
		\mathbf{H}_i = \begin{bmatrix} 
			\sin\left(\frac{\pi}{2}.\frac{i}{10}\right) &  \cos\left(\frac{\pi}{2}.\frac{i}{10}\right) \\
			\cos\left(\frac{\pi}{2}.\frac{i}{10}\right) & 	\sin\left(\frac{\pi}{2}.\frac{i}{10}\right)
		\end{bmatrix},
	\end{align} 
	in addition to the zero-mean Gaussian measurement noise with variance $R_i = 0.2\mathbf{I}_2$. The initial estimate at each node is updated using the local measurement using a standard KF and then fused with other nodes along the path. At the end of transmission, fused estimates are received at node S10, where we compare the reported covariance ellipsoids for each fusion methodology with the sample covariance ellipsoids, averaged over 50,000 MC runs. The algorithms used for comparison are covariance intersection, inverse-covariance intersection, and naive fusion, among which the latter is proven to be inconsistent. Also, as a baseline, centralized fusion is employed, which is optimal in the sense of a minimum mean square error. 
	
	Among the covariance ellipsoids plotted in Fig. \ref{sim_consBased_10Node_1}, it can be seen that centralized fusion is consistent with respect to the sample covariance, $\mathbb{E}\left[\left(\hat{\mathbf{x}}_i - \mathbf{x}\right) \left(\hat{\mathbf{x}}_i - \mathbf{x}\right)^T\right]$. The naive fusion produces a highly inconsistent ellipsoid, as the reported covariance overestimates the actual error-covariance by a large margin. This is obviously due to the double counting of common information. As expected, the ICI and CI produce conservative estimates, with the CI producing a heavily bloated covariance compared to the sample covariance. Not surprisingly, the least conservative result is produced by HMD-GA, with an actual error even tighter than ICI.
	
	\subsection{Consistency-Based Test 2}
\begin{figure}[t]
		\centering
		\includegraphics[width=0.8\columnwidth]{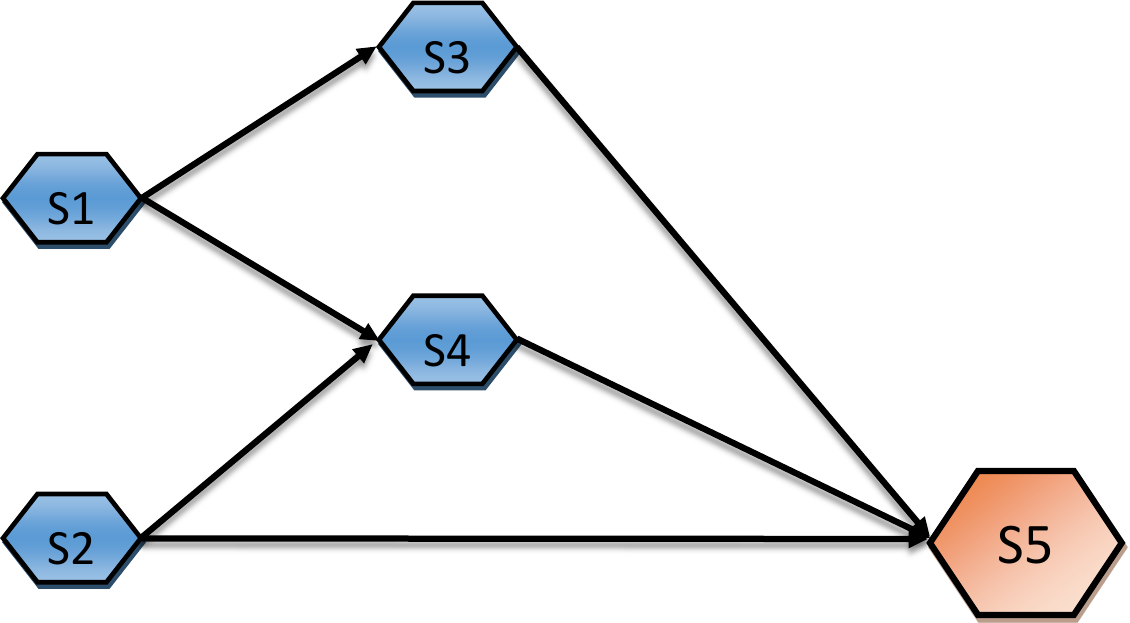}
		\caption{Consistency-based test scenario 2.}
		\label{sim_consBased_10Node_2}
	\end{figure}

	The source of correlation in the first scenario was only due to process noise, since there was no measurement history. To check the performance of the proposed fusion algorithm due to the presence of both correlations, we evaluated the consistency of HMD-GA on another scenario presented in Fig. \ref{sim_consBased_10Node_2}. The difference here is that the local nodes are allowed to process measurements over five time steps before the estimates are sent to node S5, as per the network configuration. Again, the reported covariance is compared with the sampled covariance averaged over 50,000 MC runs. Since the two-dimensional state is now evolving with time, the following linear model is used for propagation,
	\begin{align}
		\mathbf{x}_{k+1} = \mathbf{F}_k\mathbf{x}_k + \mathbf{w}_k
	\end{align} 
	where the state-transition matrix $\mathbf{F}_k = \begin{bmatrix} 1&0.5\\0&1\end{bmatrix}$, the zero-mean process noise $\mathbf{w}_k$ follows the distribution $\mathcal{N}\left(\mathbf{0}, 0.5\mathbf{I}_2\right)$ with $\mathbf{I}_n$ being an n-dimensional identity matrix. The observation model at node $i$ is,
	\begin{align}
		\mathbf{z}^i_k = \mathbf{H}_k\mathbf{x}_k + \mathbf{v}^i_k
	\end{align}
	where $\mathbf{H}_k = \mathbf{I}_2$ and $\mathbf{v}^i_k$ is sampled from a zero-mean Gaussian distribution with covariance $\begin{bmatrix} 0.5&0\\0&0.2\end{bmatrix}$ for nodes S1, S3 and S5; and $\begin{bmatrix} 0.1&0\\0&0.5\end{bmatrix}$ for nodes S2, S4. The initial estimate at each node is sampled from the distribution $\mathcal{N}\left(\begin{bmatrix}0\\0 \end{bmatrix}, \begin{bmatrix} 2&1\\1&2\end{bmatrix}\right)$. 
	
\begin{figure*}[h!]
		\centering
		\begin{subfigure}{0.17\linewidth}
			\centering
			\includegraphics[width=\textwidth]{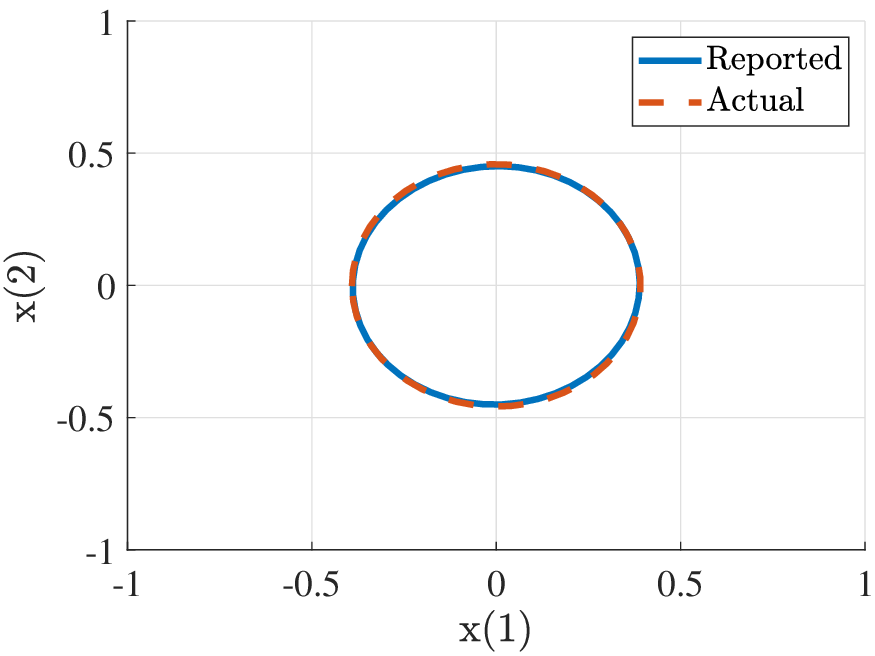}
			\caption{Centralized.}
			\label{fig_cent_sim2}
		\end{subfigure}
		\begin{subfigure}{0.17\linewidth}
			\centering
			\includegraphics[width=\textwidth]{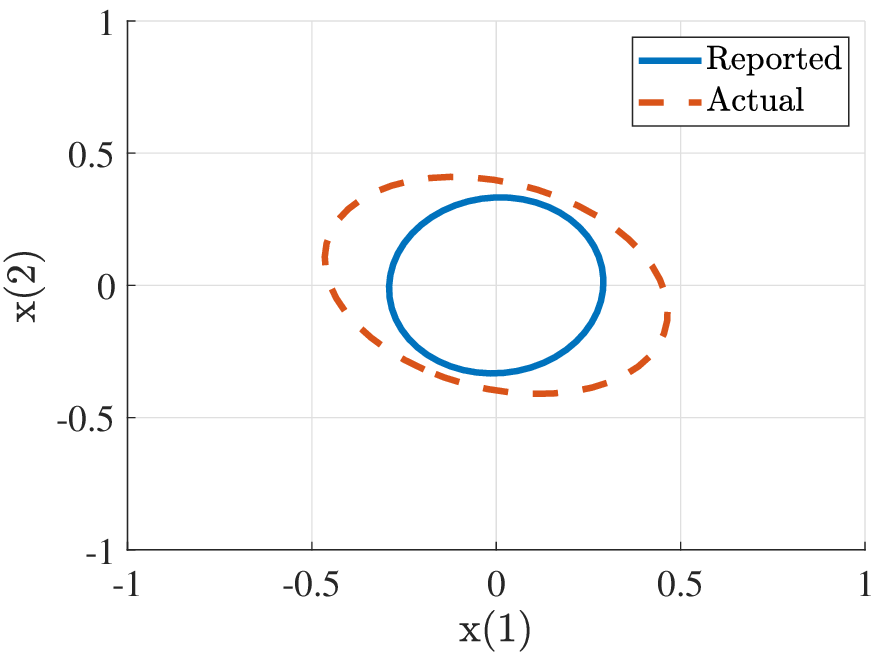}
			\caption{Naive.}
			\label{fig_naive_sim2}
		\end{subfigure}
		\begin{subfigure}{0.17\linewidth}
			\centering
			\includegraphics[width=\textwidth]{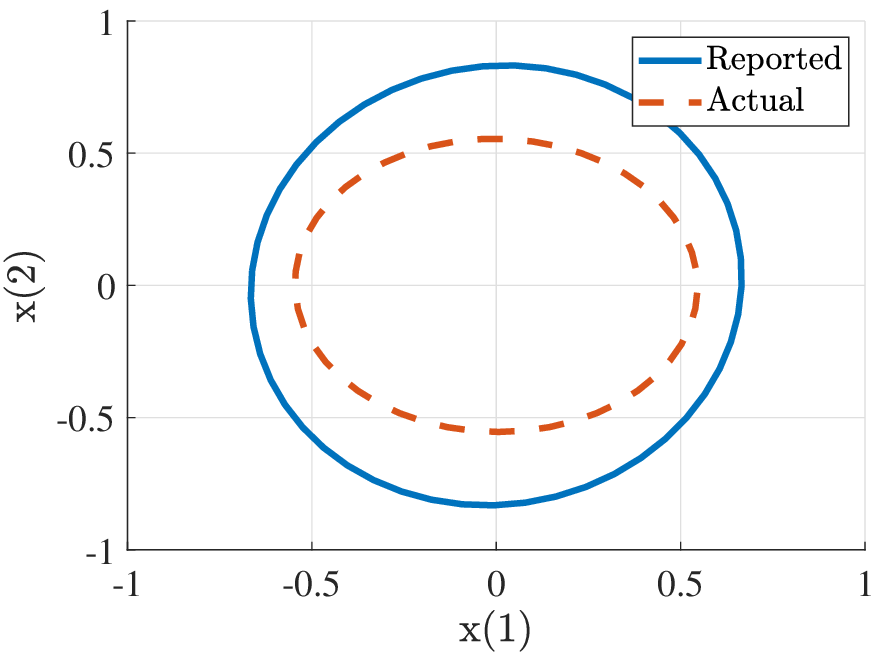}
			\caption{ICI}
			\label{fig_ici_sim2}
		\end{subfigure}
		\begin{subfigure}{0.17\linewidth}
			\centering
			\includegraphics[width=\textwidth]{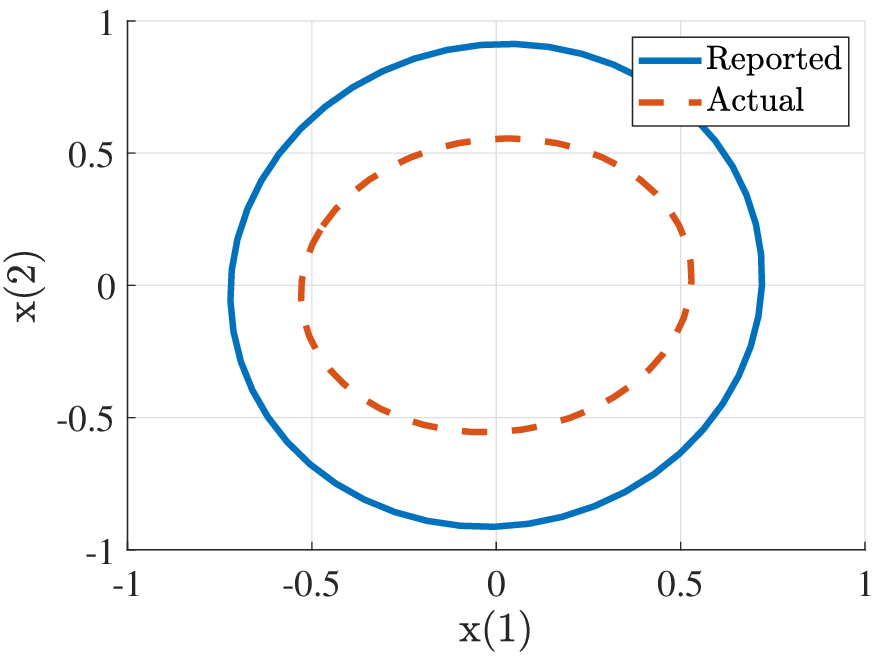}
			\caption{CI.}
			\label{fig_ci_sim2}
		\end{subfigure}
		\begin{subfigure}{0.17\linewidth}
			\centering
			\includegraphics[width=\textwidth]{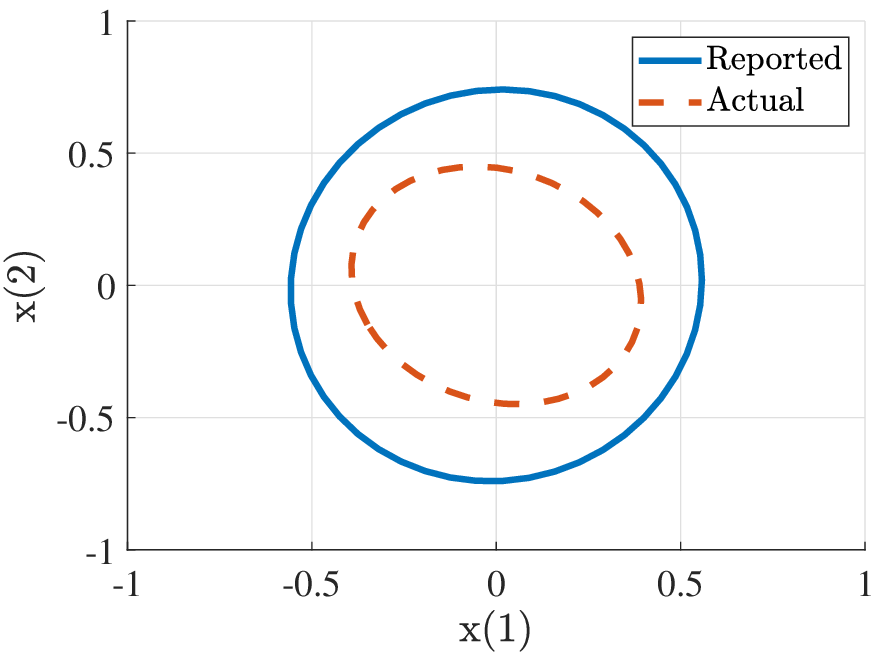}
			\caption{HMD-GA}
			\label{fig_hmd_sim2}
		\end{subfigure}
		\caption{Fusion result for consistency-based test 2.}
		\label{fig_sim1b}
	\end{figure*}

	The results are shown in Fig. \ref{fig_sim1b}, wherein again, the centralized fusion is completely consistent, whereas the naive result severely fails to produce a consistent estimate. In this scenario as well, HMD-GA produces the least conservative result, followed by ICI, and then CI, which produces the largest covariance. Also note that along with the reported covariance, the actual error in the case of HMD-GA is also less compared to other conservative fusion methods, with the size of the covariance ellipsoid almost equaling the centralized fusion result. In both of these cases, HMD-GA has been shown to perform well in scenarios with both sources of correlation.  
	
	\begin{figure*}[h!]
		\centering
		\begin{subfigure}{0.17\linewidth}
			\centering
			\includegraphics[width=\textwidth]{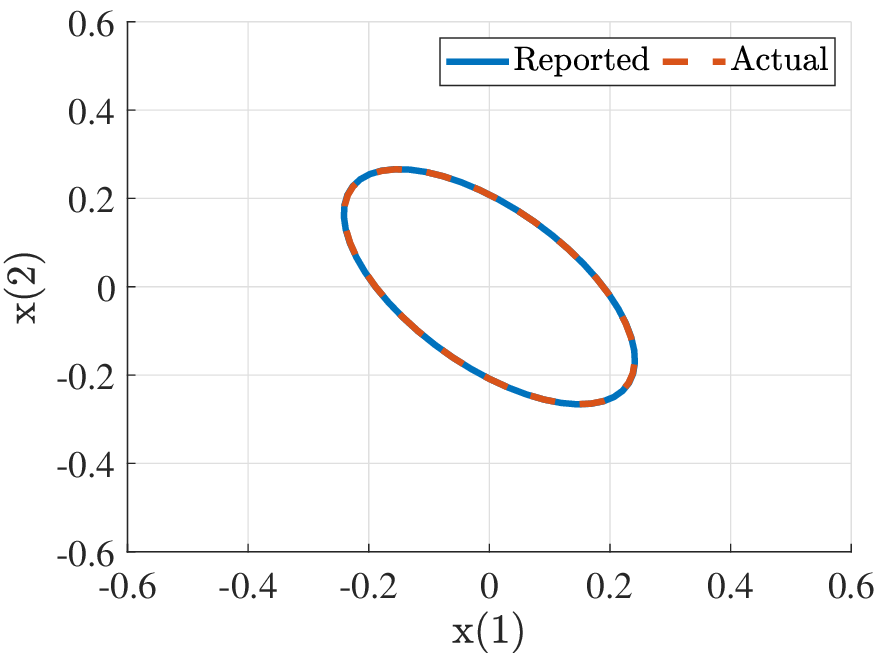}
			\caption{Centralized.}
			\label{fig_cent_sim1}
		\end{subfigure}
		\begin{subfigure}{0.17\linewidth}
			\centering
			\includegraphics[width=\textwidth]{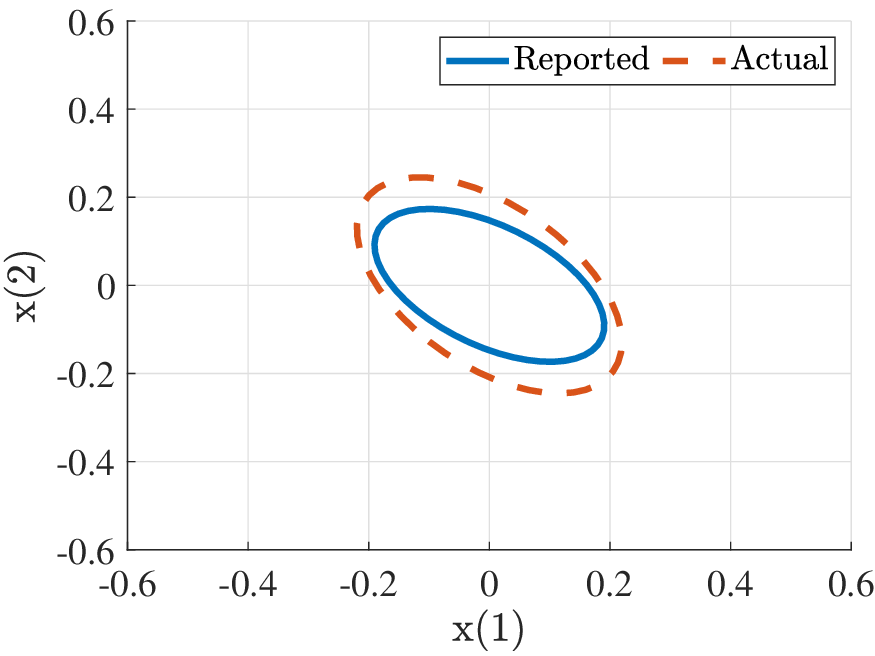}
			\caption{Naive.}
			\label{fig_naive_sim1}
		\end{subfigure}
		\begin{subfigure}{0.17\linewidth}
			\centering
			\includegraphics[width=\textwidth]{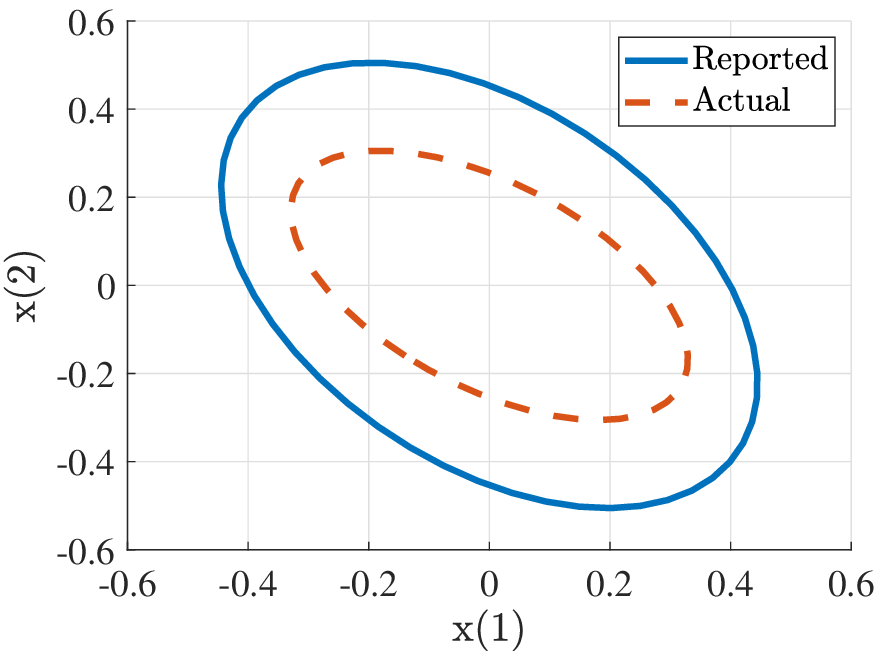}
			\caption{ICI}
			\label{fig_ici_sim1}
		\end{subfigure}
		\begin{subfigure}{0.17\linewidth}
			\centering
			\includegraphics[width=\textwidth]{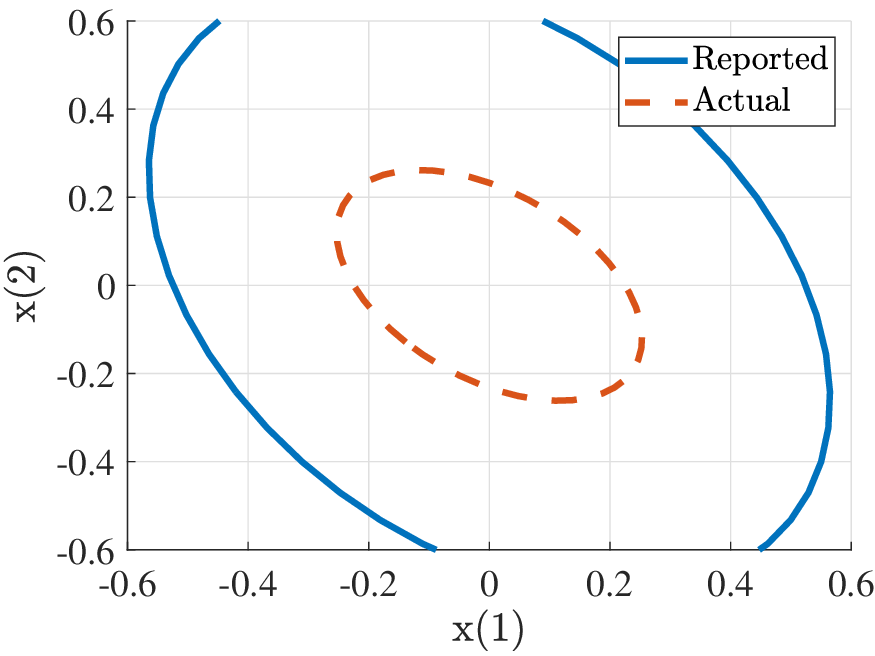}
			\caption{CI.}
			\label{fig_ci_sim1}
		\end{subfigure}
		\begin{subfigure}{0.17\linewidth}
			\centering
			\includegraphics[width=\textwidth]{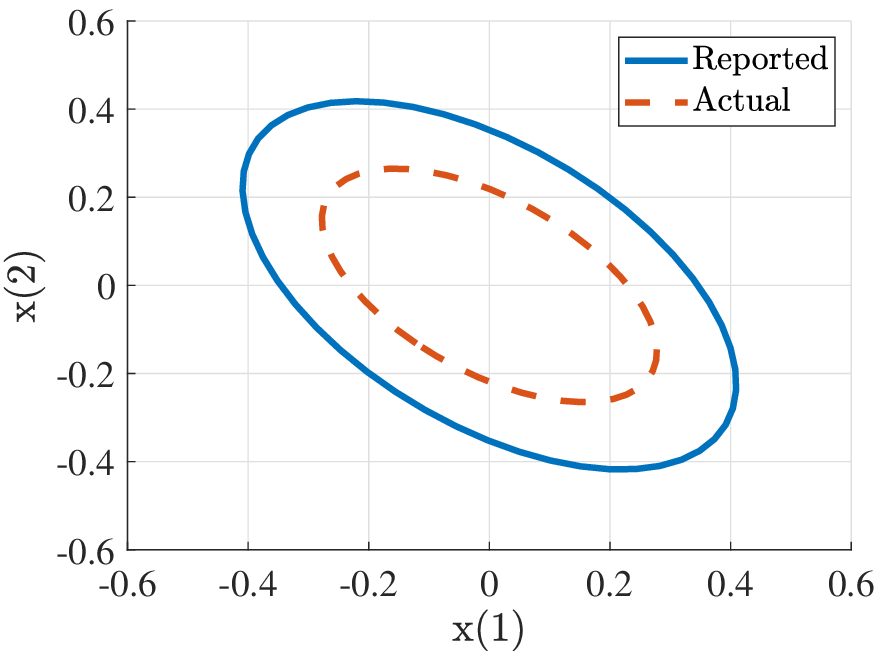}
			\caption{HMD-GA}
			\label{fig_hmd_sim1}
		\end{subfigure}
		\caption{Fusion result for consistency-based test 1.}
	\end{figure*}
	
	\subsection{A Practical Multi-Target Tracking Scenario.} \label{subsec_sim2}
	While, the tests in the previous subsections prove the consistency of the proposed algorithm, they lack practicality in the sense that they are oversimplified. For example, the linear measurement assumption guarantees that the local Kalman filter is consistent. In addition, the total number of measurements was low in comparison to a real-life scenario to prove that the repeated fusion of the same track using the proposed method doesn't hamper consistency. An anticipated metric such as root-mean-square error (RMSE) is also missing in the earlier simulations.
	
	This scenario consists of three radars surrounded by 20 targets within a significant amount of clutter, as shown in Fig. \ref{fig_sim_multiSen}. The sensors have a limited field of view (FOV), as can be seen with the circular markers around the radar in the same figure. All the radars are equipped with local processors such that only updated tracks are sent to the fusion center (FC) every $T_\mathit{f}$ seconds. The FC is assumed to be collocated with Radar 1. As in the previous simulations, ambiguity in time synchronizations and sensor biases have been ignored in this scenario. The scenario parameters are listed in Table \ref{tab_param_scen2}. The targets move in an area spanning nearly 3600 $\text{km}^2$ whereas the sensors are static at coordinates mentioned in Table \ref{tab_loc_param_scen2}. Geodetic coordinates of targets according to the WGS84 datum are also provided for replication. Note that a nearly constant velocity (NCV) model is used in tracking, which the target doesn't adhere to, resulting in spikes in the RMSE plots. We tackle this phenomenon by using a relatively high process noise intensity. The scenario is considered sufficiently pragmatic due to the following features :
	
	\begin{enumerate}[label= (\roman*)]
		\item On average, the targets are at least a hundred kilometers far from a sensor. The curvature of cross-range uncertainty is a known phenomenon while tracking at such distances, due to which local consistency is impacted. For more details on this topic, readers are referred to \cite{tian2014consistency}.
		\item Local trackers do not enjoy a clear view. Due to the high volume of clutter, the probability data association (PDA) tracker \cite{kirubarajan2017target} is employed.
		\item The targets do not necessarily adhere to the state-space motion model employed by local trackers.
		\item Track association at the fusion center is performed without knowledge of the correlation between the tracks.
		\item The target position is captured in geodetic coordinates, which are converted to local east-north-up (ENU) for tracking.
	\end{enumerate}
	\subsubsection{Model Equations}
	Although the coordinates mentioned represent a spherical system, the trackers are implemented in a standard mixed coordinate system, wherein the process model is Cartesian and the sensor model is spherical in two dimensions. For conversion between the geodetic and local tangent plane coordinates, readers are referred to \cite{civicioglu2012transforming}. 
	
	The target motion and the measurements, respectively, are modeled by the following equations at time $k$ (dropping target and measurement indices),
	\begin{subequations} \label{eq_model_scen2}
		\begin{align}
			\mathbf{x}_{k+1} &= \mathbf{F}\mathbf{x}_k + \mathbf{w}_k, \\
			\mathbf{z}_k &= \mathbf{h}(x_k,y_k) + \mathbf{v}_k,
		\end{align}
	\end{subequations}
	where (using $\mathbf{I}_n$ for n dimensional identity matrix),
	\begin{align*}
		\mathbf{F} &= \begin{bmatrix}
			\mathbf{I}_2 & \Delta T\mathbf{I}_2 \\
			\mathbf{0}_2 & \mathbf{I}_2
		\end{bmatrix} \quad \text{and,} \\
		\mathbf{h}(x_k,y_k,z_k) &= 	\begin{bmatrix} r_k & \theta_k \end{bmatrix}^T,				
	\end{align*}
	where,
	\begin{align*}
		r_k &= \sqrt{x_k^2 + y_k^2};\quad \theta_k = \tan^{-1}\left(\frac{y_k}{x_k}\right) 
	\end{align*}
	$\Delta T$ is the local sampling time, and $\mathbf{w}_k$, $\mathbf{v}_k$ are uncorrelated, zero-mean Gaussian distributed process noise vector and measurement noise vector respectively, such that $\forall \{k,j\}$,
	\begin{align*}
		E[\mathbf{w}_k\mathbf{w}_j] = \delta_{kj}\mathbf{Q};\quad E[\mathbf{w}_k\mathbf{v}_j] &= \mathbf{0};\quad E[\mathbf{v}_k\mathbf{v}_j] = \delta_{kj}\mathbf{R}, 
	\end{align*}
	where $\mathbf{Q}$ and $\mathbf{R}$ are the process noise and measurement noise covariance matrix respectively, 
	\begin{align*}
		\mathbf{Q} &= \tilde{q}\begin{bmatrix}
			\frac{\Delta T^3}{3} \mathbf{I}_2 & \frac{\Delta T^2}{2} \mathbf{I}_2 \\
			\frac{\Delta T^2}{2} \mathbf{I}_2 & \Delta T \mathbf{I}_2
		\end{bmatrix}, \\
		\mathbf{R}	&= \text{diag}(\sigma_r^2, \sigma_\theta^2),			
	\end{align*}
	and $\tilde{q}$ is the process noise intensity in $\text{meter}^2/\text{second}^3$.
	
	Due to high clutter volume, the local tracking algorithm employs PDA, thereby, indirectly assuming wide separation of targets, which is the case here. The PDA algorithm assigns weights to all available measurements in a single scan, and performs measurement update by taking a weighted average of all innovations. Since the targets are widely separated, there can only be one target-originated measurement in a gate. After local processing, the updated tracks are sent to FC (Radar 1), every $T_\mathit{f} = 10$ seconds. 
	
	The fusion center is equipped with memory. Thus, a global track is always maintained for each fused track at least until the next fusion step. Track-management at FC is similar to that at local node. When a track-list arrives at FC, it is associated with existing global track lists by solving the following 2-D optimization based cost, 
	\begin{align}
		J &= \sum_{i = 1}^{N_g}\sum_{j= 1}^{N_l}C_{ij}z_{ij}, \quad \text{subjected to constraints,}  \\
		\sum_{i=1}^{N_g}z_{ij} &= 1; \quad \sum_{j=1}^{N_l}z_{ij} = 1, \quad \forall i,j
	\end{align}
	where $z_{ij}$ is the assignment between $i^\text{th}$ global track and $j^\text{th}$ local track. The corresponding cost $C_{ij}$ is defined as,  
	  \begin{align}
	  	C_{ij} = \left(\hat{\mathbf{x}}^i - \hat{\mathbf{x}}^j\right)^T\left[\mathbf{\Gamma}^i + \mathbf{\Gamma}^j - \mathbf{\Gamma}^{ij} - \mathbf{\Gamma}^{ji}\right]\left(\hat{\mathbf{x}}^i - \hat{\mathbf{x}}^j\right) \notag \\
	  		\approx \left(\hat{\mathbf{x}}^i - \hat{\mathbf{x}}^j\right)^T\left[\mathbf{\Gamma}^i + \mathbf{\Gamma}^j\right]\left(\hat{\mathbf{x}}^i - \hat{\mathbf{x}}^j\right)
	  \end{align}
	 due to the unavailability of $\mathbf{\Gamma}^{ij} = \left(\mathbf{\Gamma}^{ji}\right)^T$. Next step is the fusion of associated tracks wherein we used ICI, CI and the proposed HMD-GA. This procedure is repeated until track-lists from all local nodes have been fused.

	 \subsubsection{Performance Metrics} \label{subsec_perfMetric}
	 The simulation is evaluated on the following metrics :
	 \begin{enumerate}[label=(\roman*)]
	 	\item \textbf{ Root mean-square error (RMSE)} : 
	 	RMSE evaluates the absolute filtering performance in simulation where the true system trajectory is known. At any time-step $k$, the RMSE is given by the equation,
	 	\begin{align}
	 		\text{RMSE}_k = \sqrt{\frac{1}{M} \sum_{m=1}^M \lvert\lvert \hat{x}_{m,k} - x_k \rvert\rvert ^2  },  \label{eq_meas_upd_times}
	 	\end{align}
	 	where $M$ is the number of Monte-Carlo (MC) iterations over which the root-mean is evaluated. $\hat{x}_{m,k}$ is the filter estimate at $k^\text{th}$ time-step of $m^\text{th}$ MC run. Note that the truth $x_k$ is the same throughout all MC iterations.
	 	
	 	\item\textbf{ Normalized estimation error squared (NEES)}:
	 	NEES is a measure of estimator consistency, which is based on the fact that a $\chi^2$ distribution has a mean equal to its degrees of freedom. Thus the following expression
	 	\begin{align}
	 		\text{NEES}_k = \frac{1}{M} \sum_{m=1}^M \left[\left( \hat{\mathbf{x}}_{m,k} - \mathbf{x}_k\right)^T\mathbf{P}_{m,k|k}^{-1}\left( \hat{\mathbf{x}}_{m,k} - \mathbf{x}_k\right) \right]
	 	\end{align}
	 	should have $n_x$ degrees of freedom (asymptotically). Since the number of MC runs are finite, the NEES for a consistent estimator should lie between the appropriate confidence bounds given by the tail probabilities of respective $\chi^2$ density \cite{bar2001estimation}.		
	\end{enumerate}
	Due to the long simulation time and large number of targets, both the metrics are evaluated over 200 MC runs.
	
	\subsubsection{Remarks}
\begin{figure*}[h!]
		\centering
		\begin{subfigure}[t]{0.48\linewidth}
			\centering
		\includegraphics[width=\linewidth]{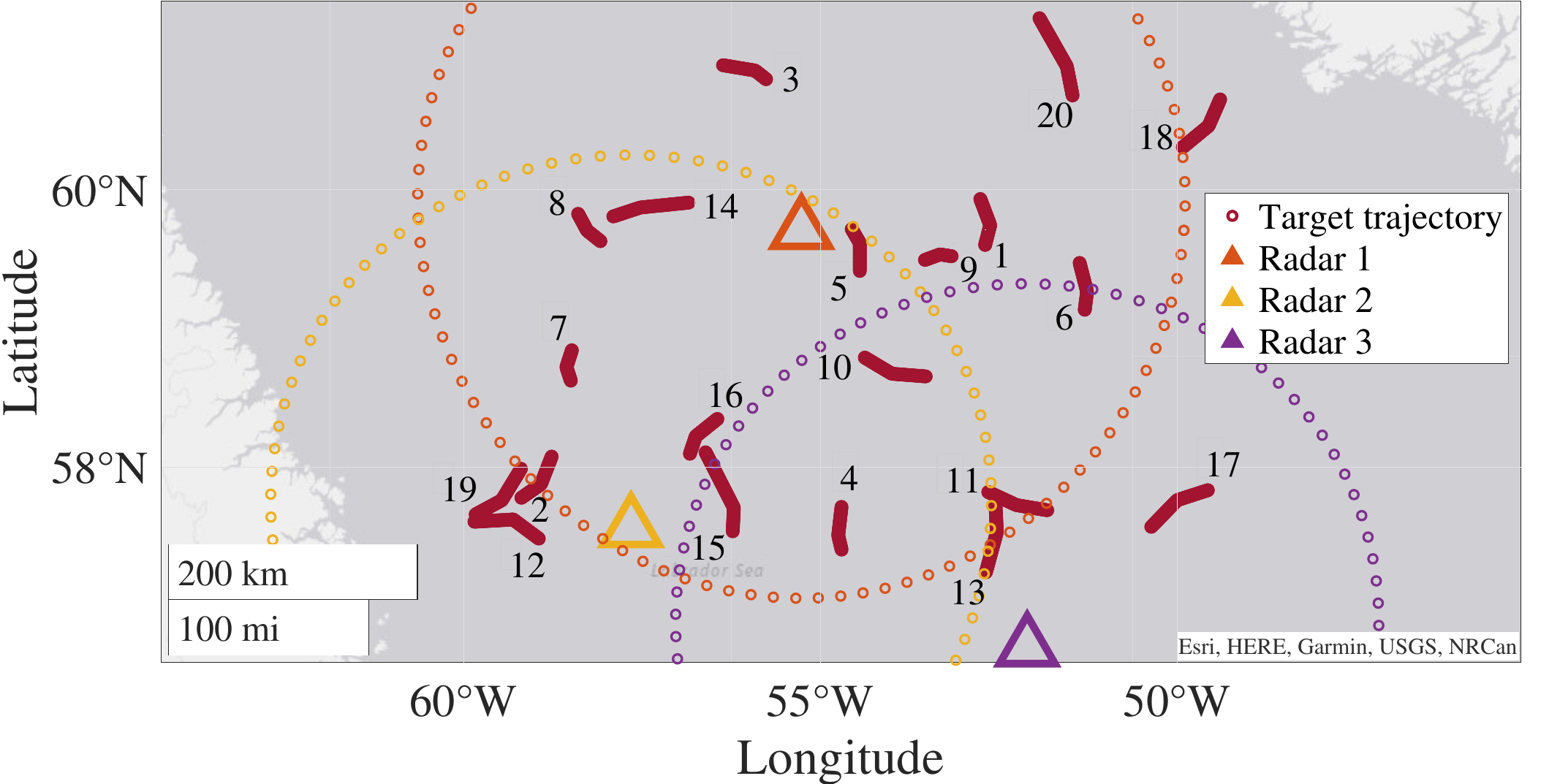}
		\caption{Multi-target multi-sensor scenario used in section \ref{subsec_sim2}.}
		\label{fig_sim_multiSen}
		\end{subfigure}
		\begin{subfigure}[t]{0.48\linewidth}
			\centering	
			\includegraphics[width=\linewidth]{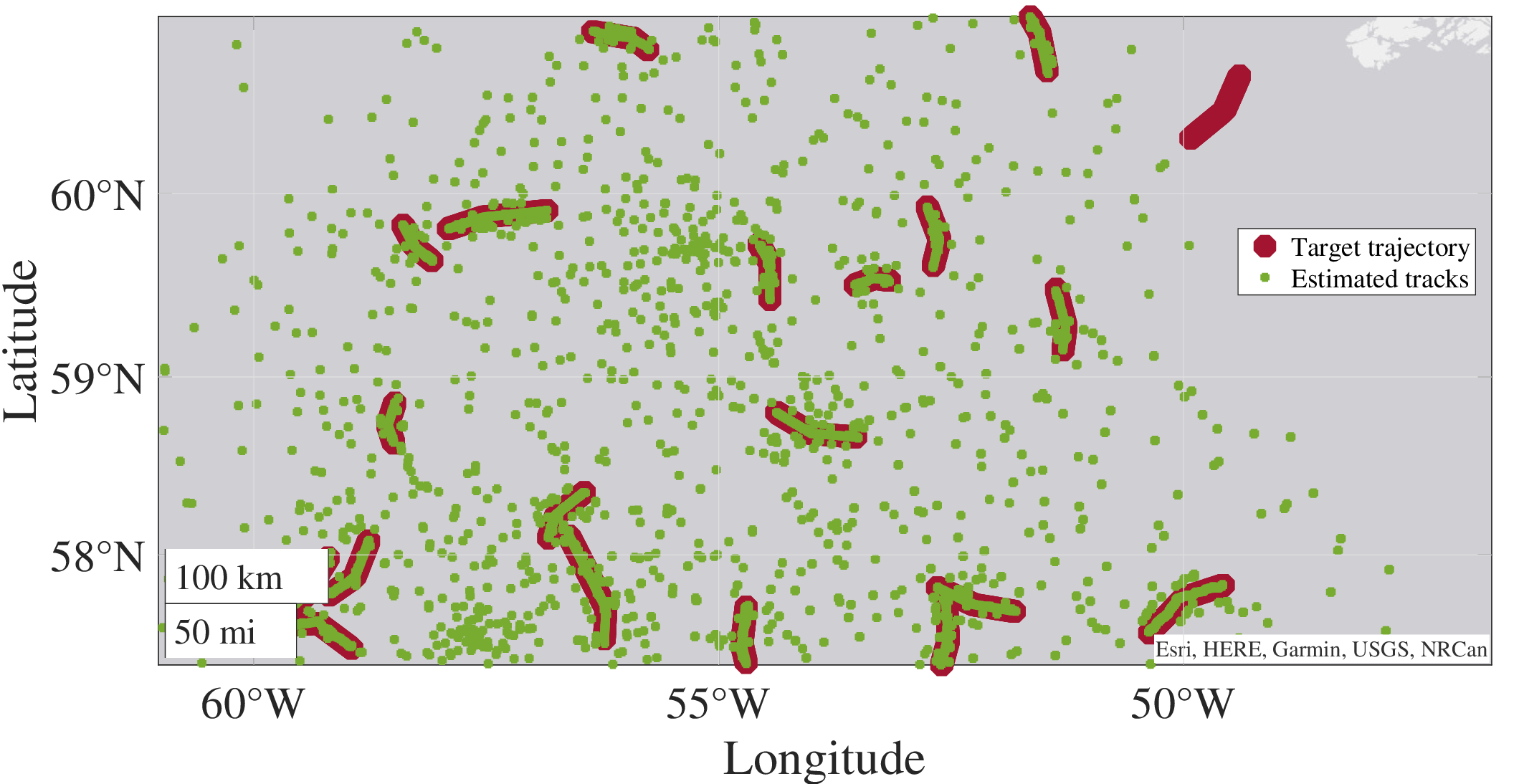}
			\caption{Finalized tracks (among clutter) using HMD-GA.}
			\label{fig_sim_multiSen_tracks}
		\end{subfigure}
	\end{figure*}

\begin{figure}
		\centering
		\begin{subfigure}{0.48\columnwidth}
			\centering
			\includegraphics[width=\linewidth]{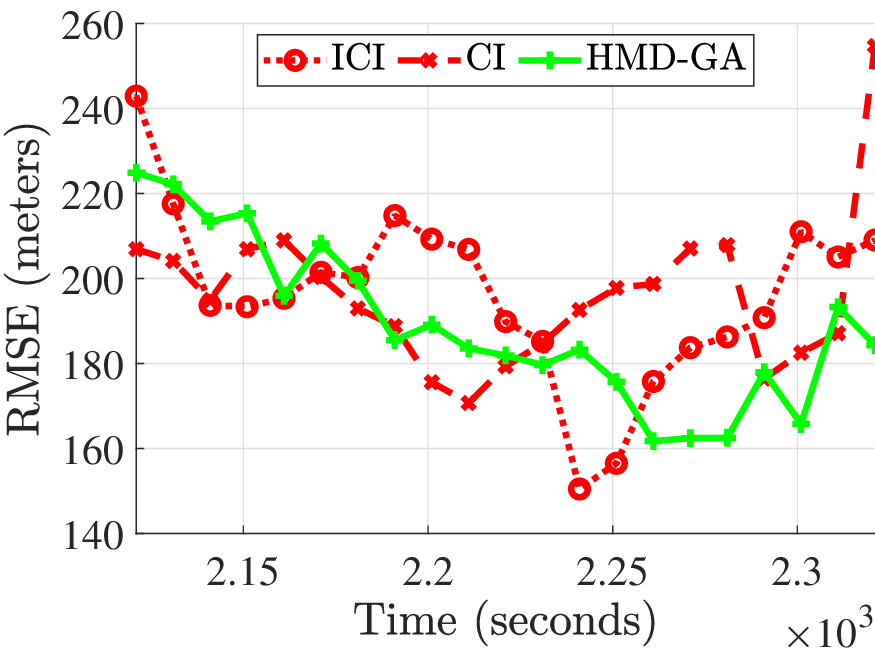}
		\end{subfigure}	
		\begin{subfigure}{0.48\columnwidth}
			\centering
			\includegraphics[width=\linewidth]{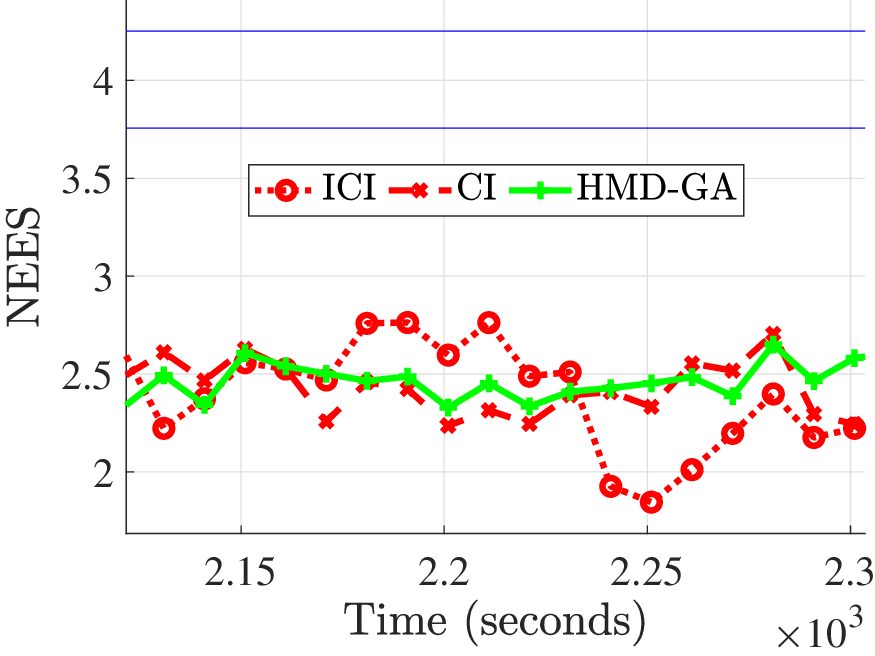}
		\end{subfigure}
		\caption{Target 1.}
		\label{sim_scene3R_tar1}
	\end{figure}

\begin{figure}
		\centering	
		\begin{subfigure}{0.48\columnwidth}
			\centering
			\includegraphics[width=\linewidth]{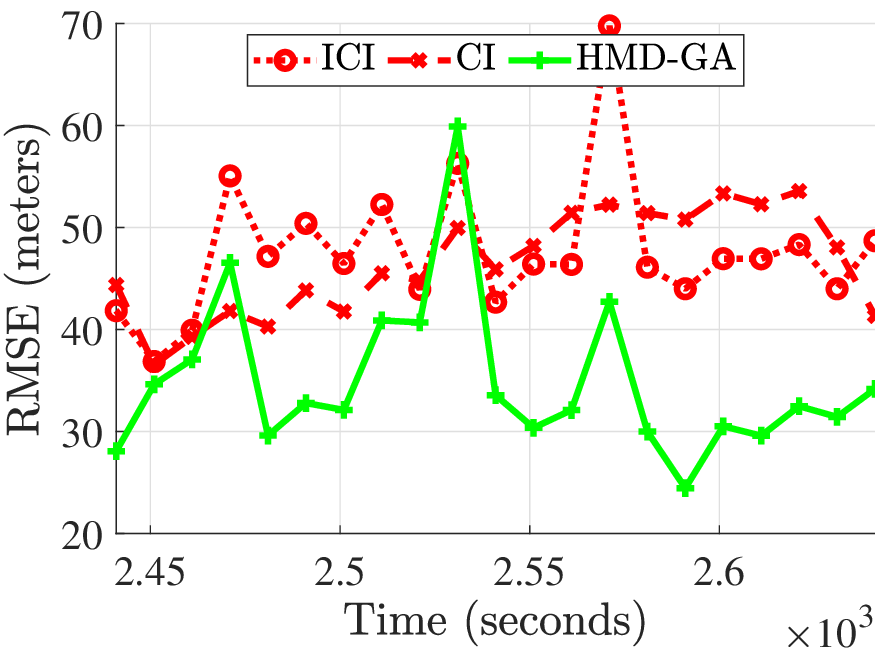}
		\end{subfigure}	
		\begin{subfigure}{0.48\columnwidth}
			\centering
			\includegraphics[width=\linewidth]{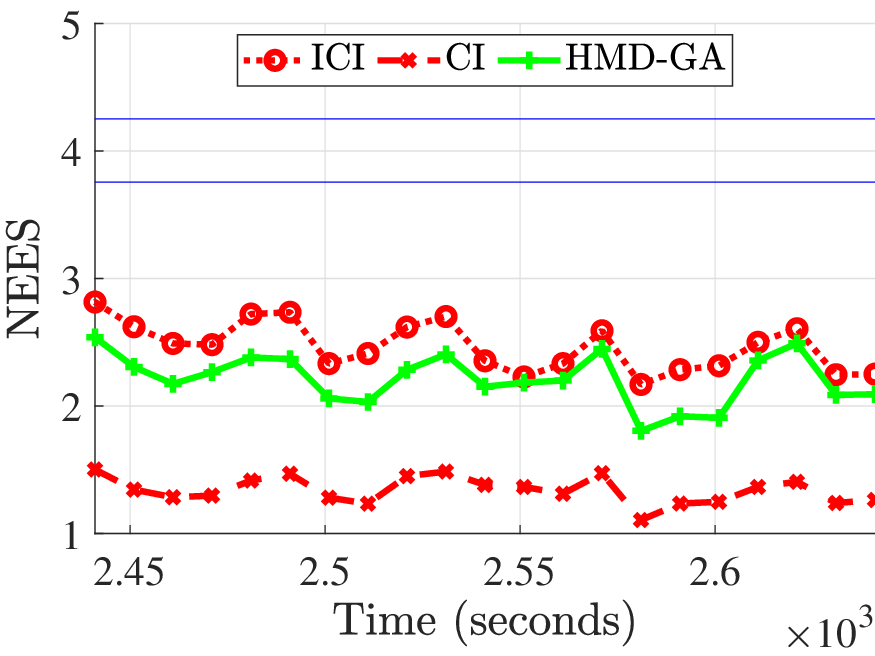}
		\end{subfigure}
		\caption{Target 2.}
		\label{sim_scene3R_tar2}
	\end{figure}

\begin{figure}
		\begin{subfigure}{0.48\columnwidth}
			\centering
			\includegraphics[width=\linewidth]{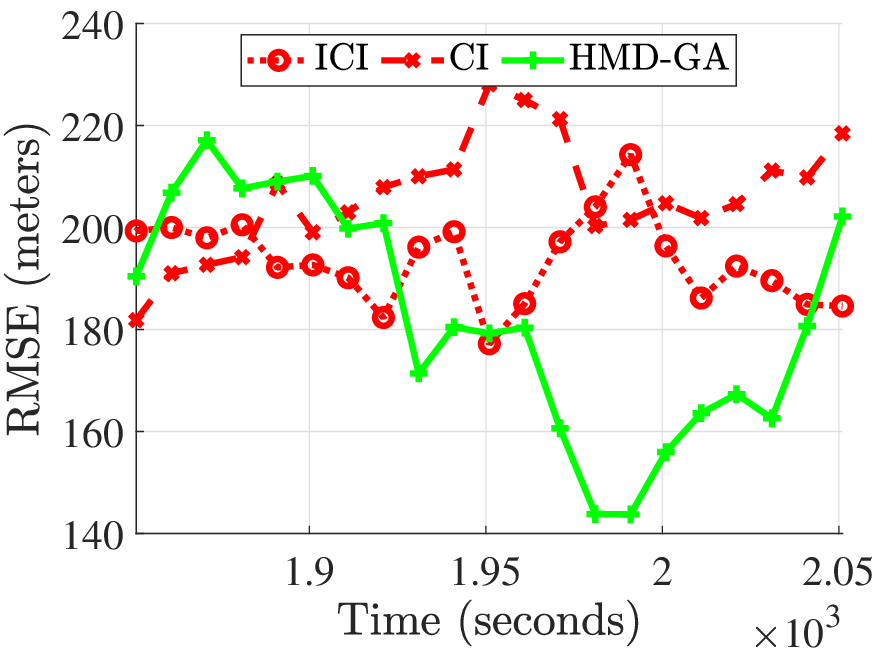}
		\end{subfigure}	
		\begin{subfigure}{0.48\columnwidth}
			\centering
			\includegraphics[width=\linewidth]{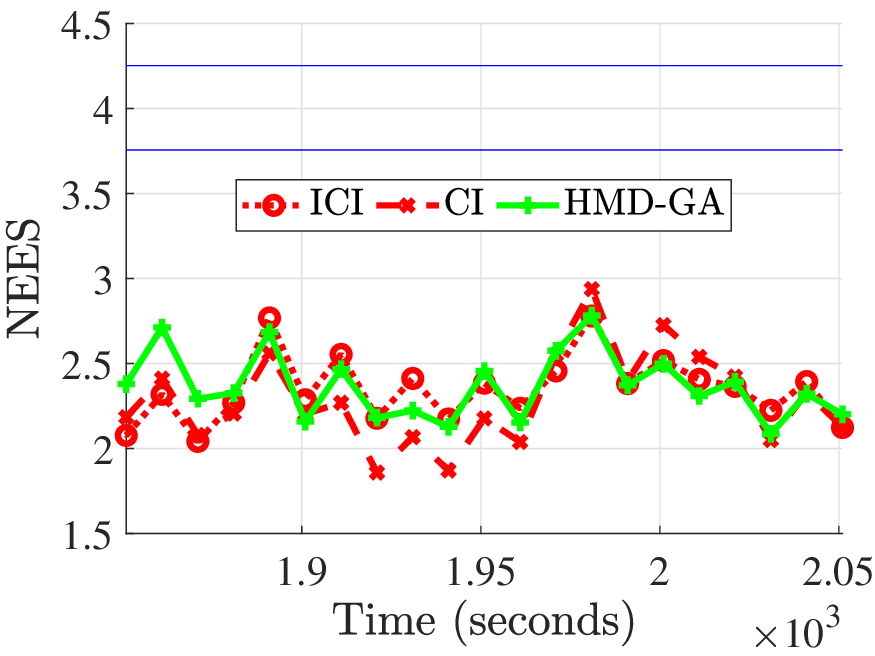}
		\end{subfigure}
		\caption{Target 3.}
		\label{sim_scene3R_tar3}
	\end{figure}

\begin{figure}
		\centering	
		\begin{subfigure}{0.48\columnwidth}
			\centering
			\includegraphics[width=\linewidth]{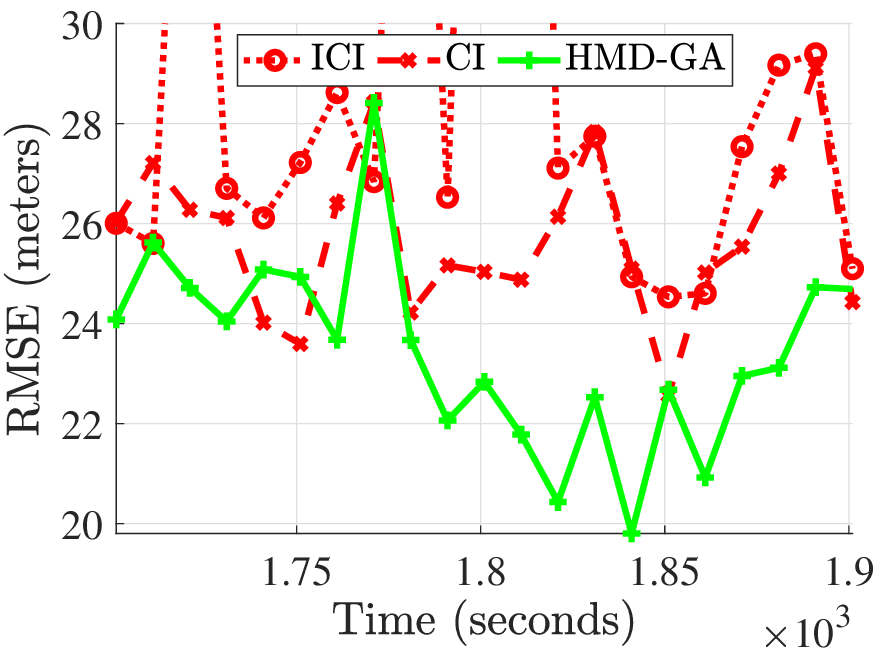}
		\end{subfigure}	
		\begin{subfigure}{0.48\columnwidth}
			\centering
			\includegraphics[width=\linewidth]{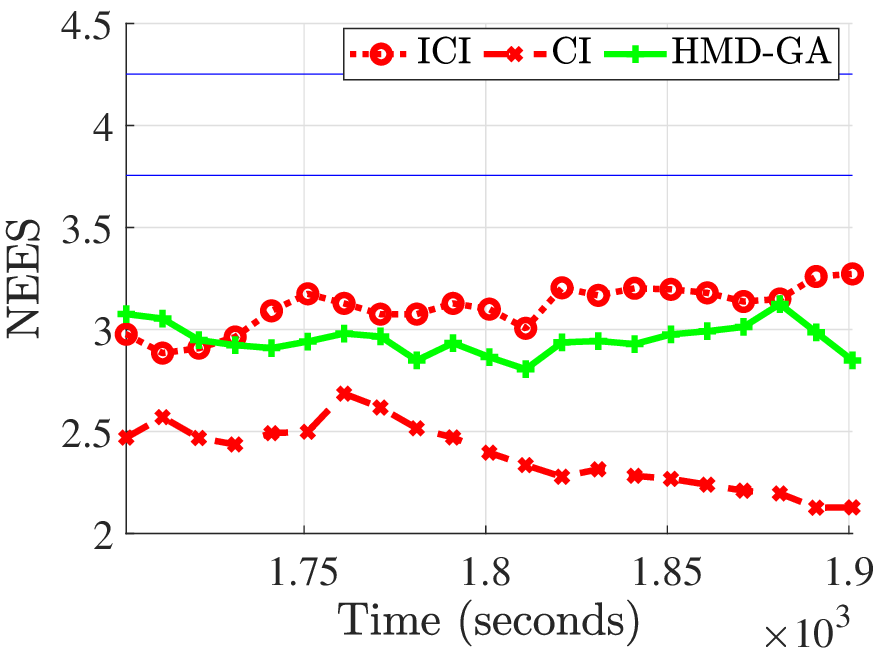}
		\end{subfigure}
		\caption{Target 4.}
		\label{sim_scene3R_tar4}
	\end{figure}

\begin{figure}	
		\begin{subfigure}{0.48\columnwidth}
			\centering
			\includegraphics[width=\linewidth]{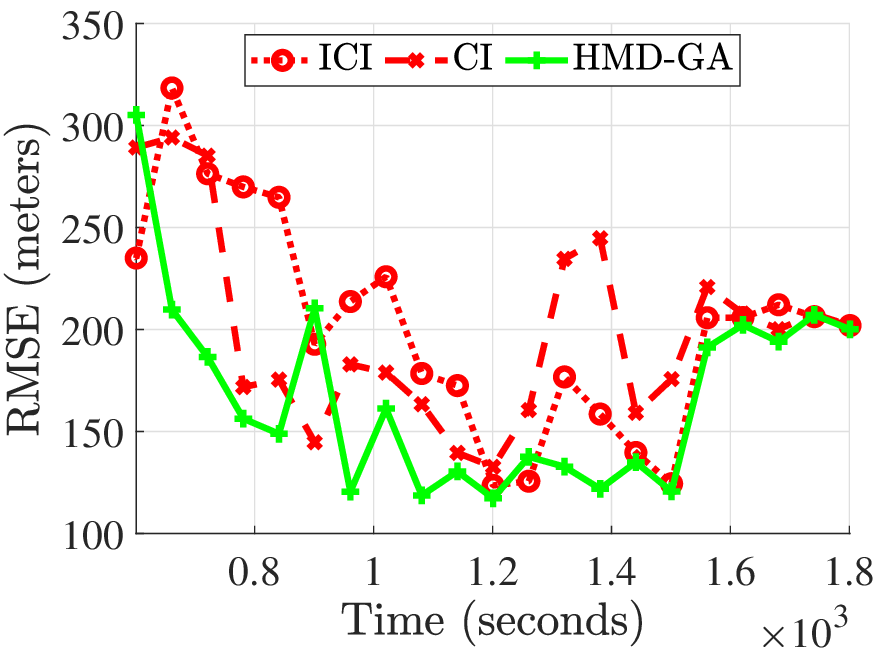}
		\end{subfigure}	
		\begin{subfigure}{0.48\columnwidth}
			\centering
			\includegraphics[width=\linewidth]{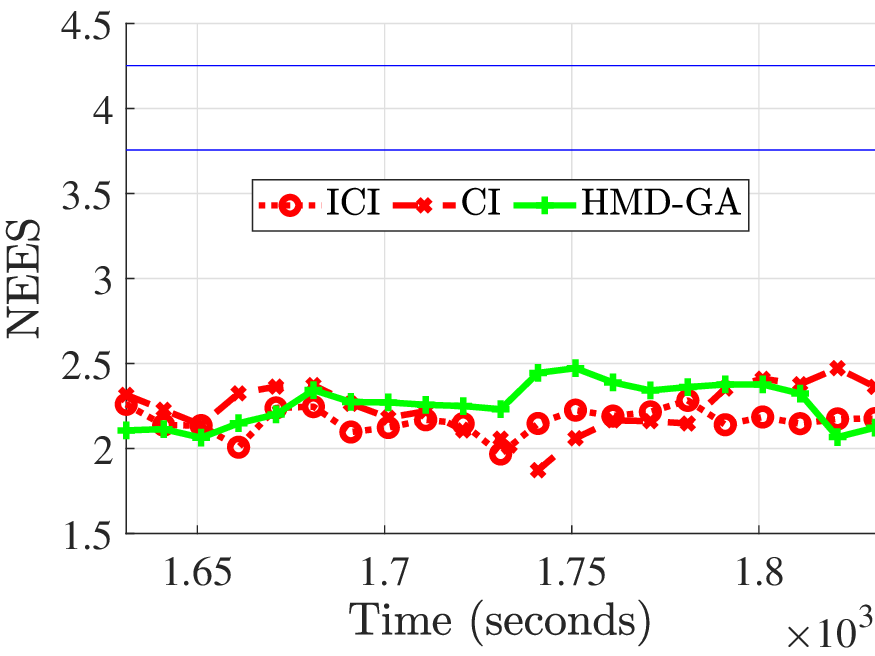}
		\end{subfigure}
		\caption{Target 5.}
		\label{sim_scene3R_tar5}
	\end{figure}

\begin{figure}	
		\begin{subfigure}{0.48\columnwidth}
			\centering
			\includegraphics[width=\linewidth]{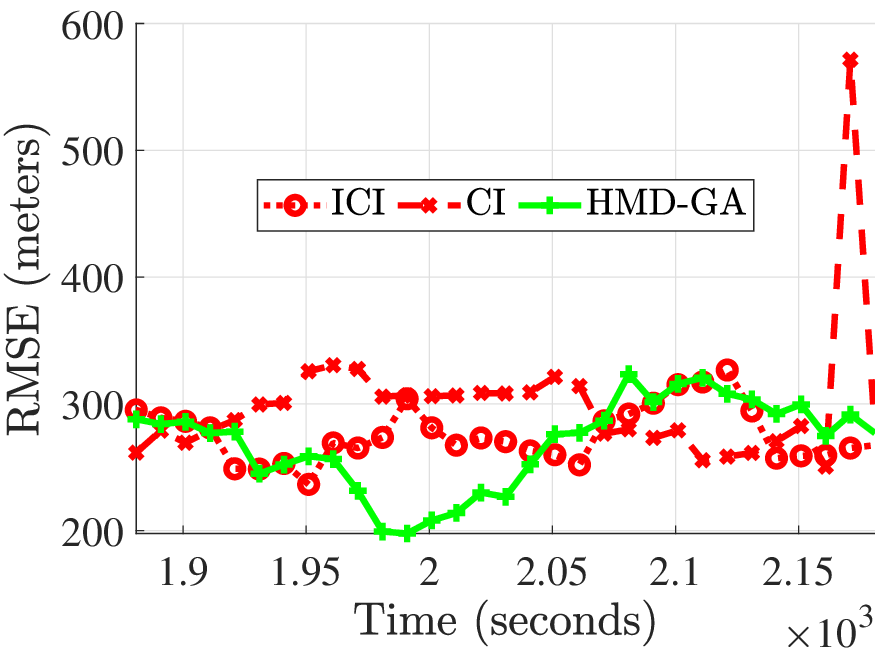}
		\end{subfigure}	
		\begin{subfigure}{0.48\columnwidth}
			\centering
			\includegraphics[width=\linewidth]{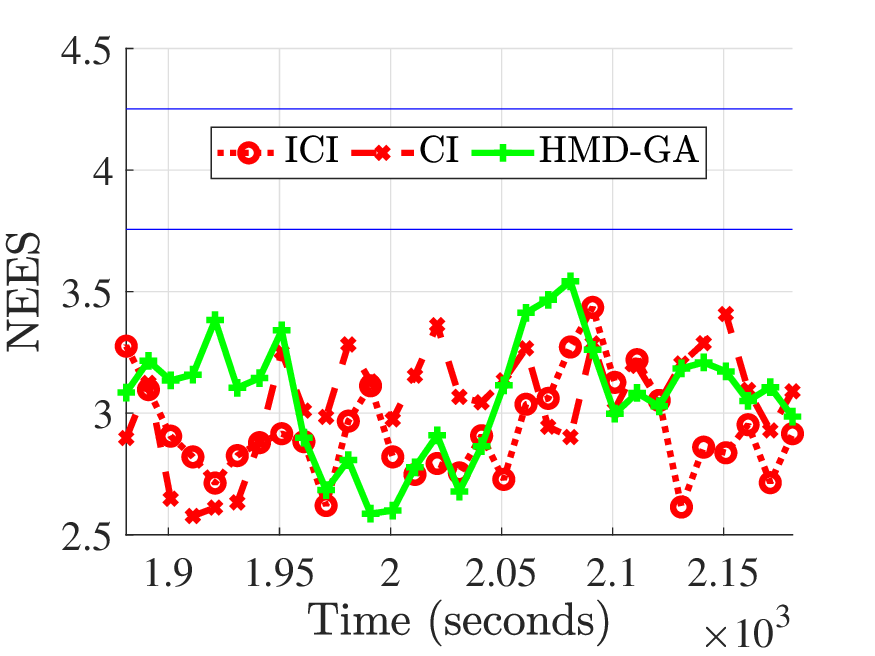}
		\end{subfigure}
		\caption{Target 6.}
		\label{sim_scene3R_tar6}
	\end{figure}

\begin{figure}
		\centering
		\begin{subfigure}{0.48\columnwidth}
			\centering
			\includegraphics[width=\linewidth]{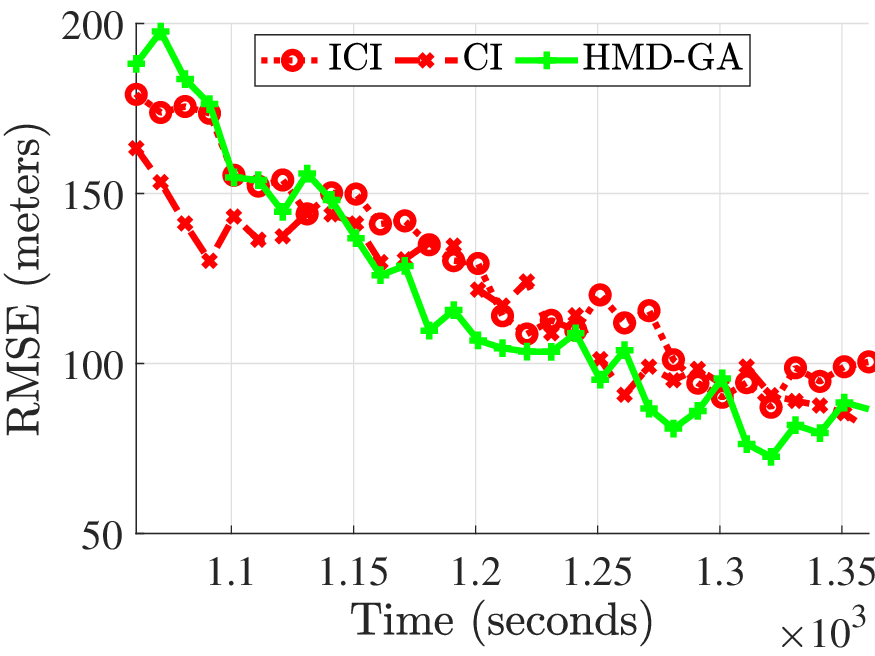}
		\end{subfigure}	
		\begin{subfigure}{0.48\columnwidth}
			\centering
			\includegraphics[width=\linewidth]{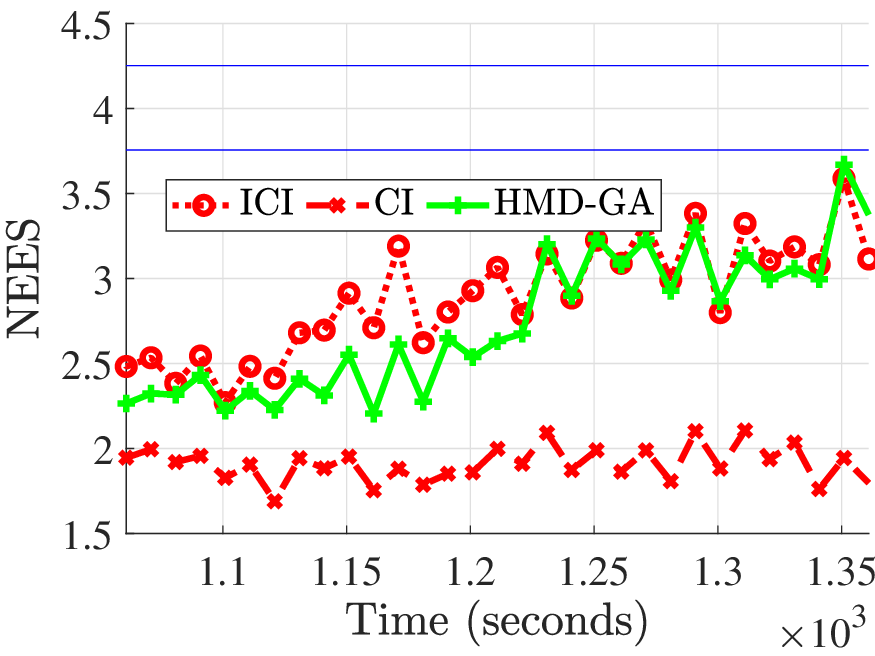}
		\end{subfigure}
		\caption{Target 7.}
		\label{sim_scene3R_tar7}
	\end{figure}

\begin{figure}	
		\begin{subfigure}{0.48\columnwidth}
			\centering
			\includegraphics[width=\linewidth]{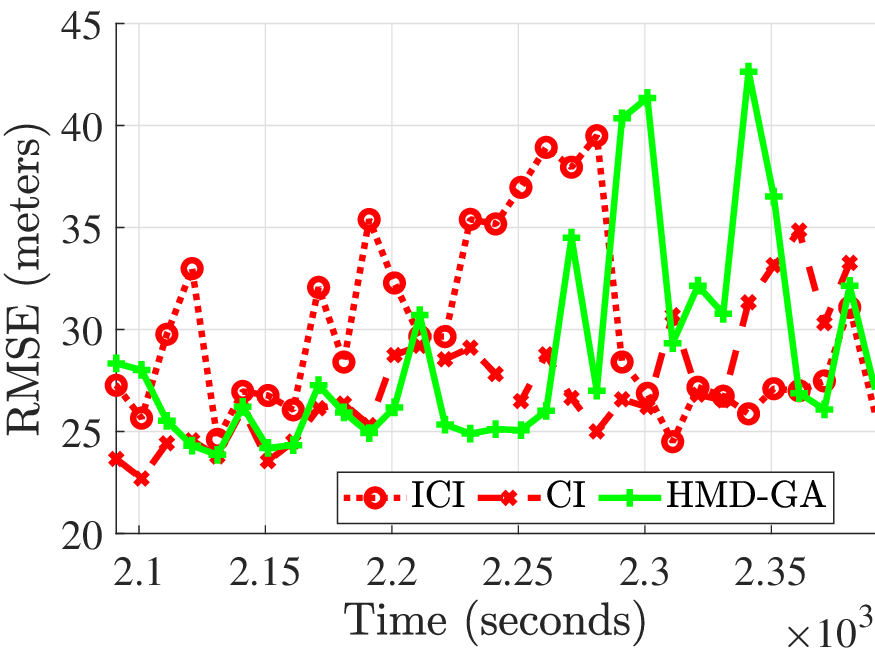}
		\end{subfigure}	
		\begin{subfigure}{0.48\columnwidth}
			\centering
			\includegraphics[width=\linewidth]{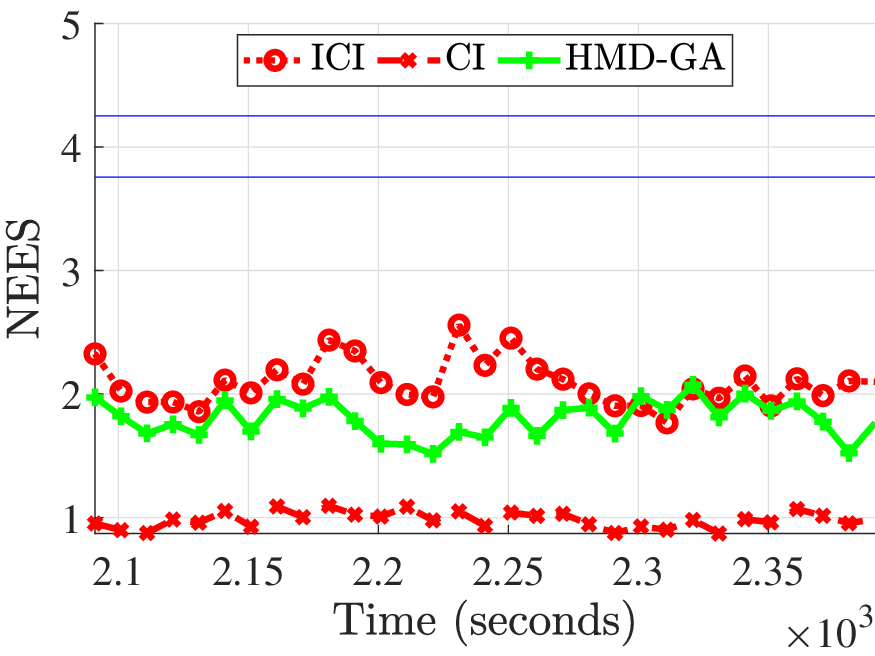}
		\end{subfigure}
		\caption{Target 8.}
		\label{sim_scene3R_tar8}
	\end{figure}

\begin{figure}	
		\begin{subfigure}{0.48\columnwidth}
			\centering
			\includegraphics[width=\linewidth]{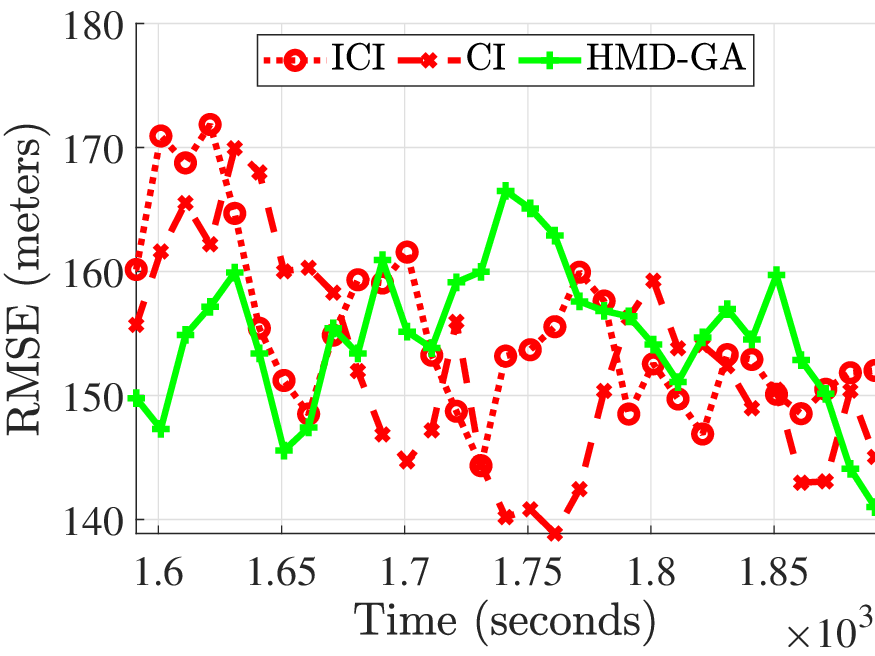}
		\end{subfigure}	
		\begin{subfigure}{0.48\columnwidth}
			\centering
			\includegraphics[width=\linewidth]{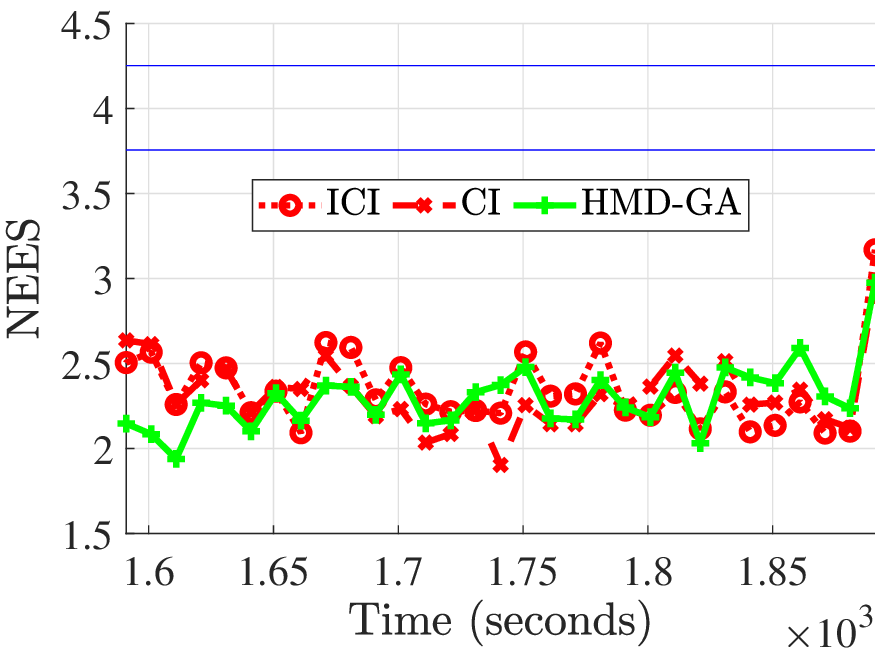}
		\end{subfigure}
		\caption{Target 9.}
		\label{sim_scene3R_tar9}
	\end{figure}

\begin{figure}	
		\begin{subfigure}{0.48\columnwidth}
			\centering
			\includegraphics[width=\linewidth]{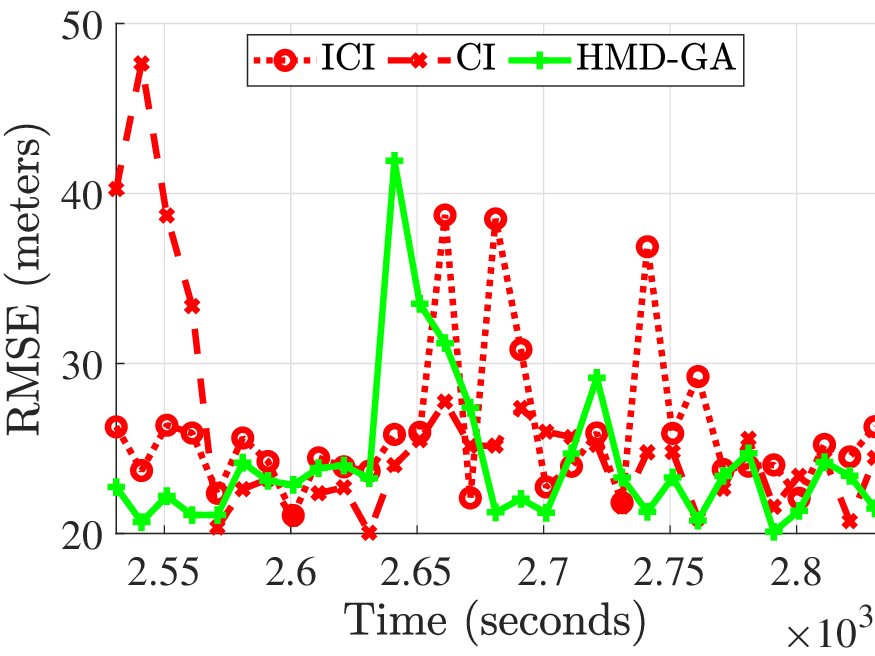}
		\end{subfigure}	
		\begin{subfigure}{0.48\columnwidth}
			\centering
			\includegraphics[width=\linewidth]{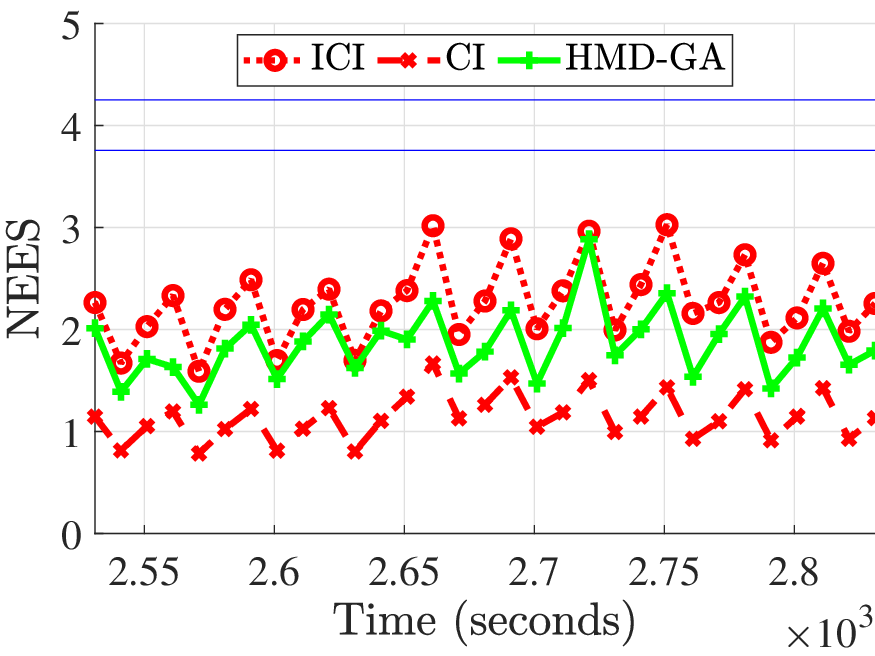}
		\end{subfigure}
		\caption{Target 10.}
		\label{sim_scene3R_tar10}
	\end{figure}

\begin{figure}	
		\begin{subfigure}{0.48\columnwidth}
			\centering
			\includegraphics[width=\linewidth]{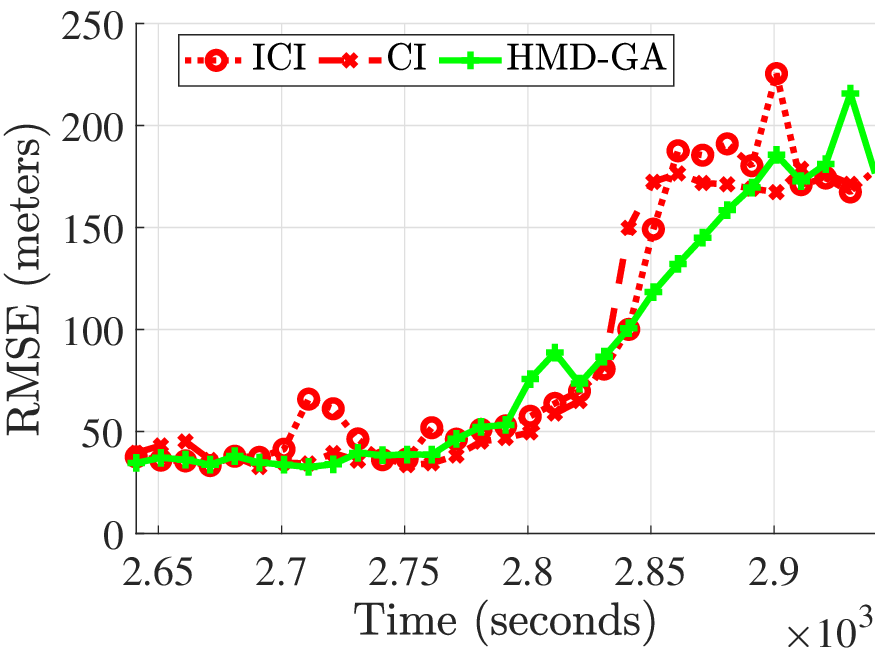}
		\end{subfigure}	
		\begin{subfigure}{0.48\columnwidth}
			\centering
			\includegraphics[width=\linewidth]{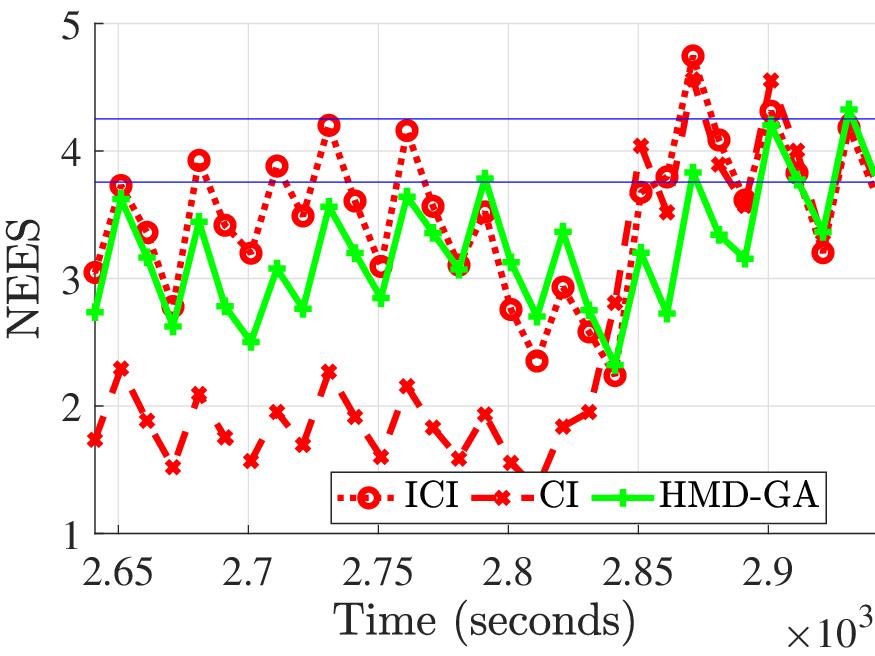}
		\end{subfigure}
		\caption{Target 11.}
		\label{sim_scene3R_tar11}
	\end{figure}

\begin{figure}
		\begin{subfigure}{0.48\columnwidth}
			\centering
			\includegraphics[width=\linewidth]{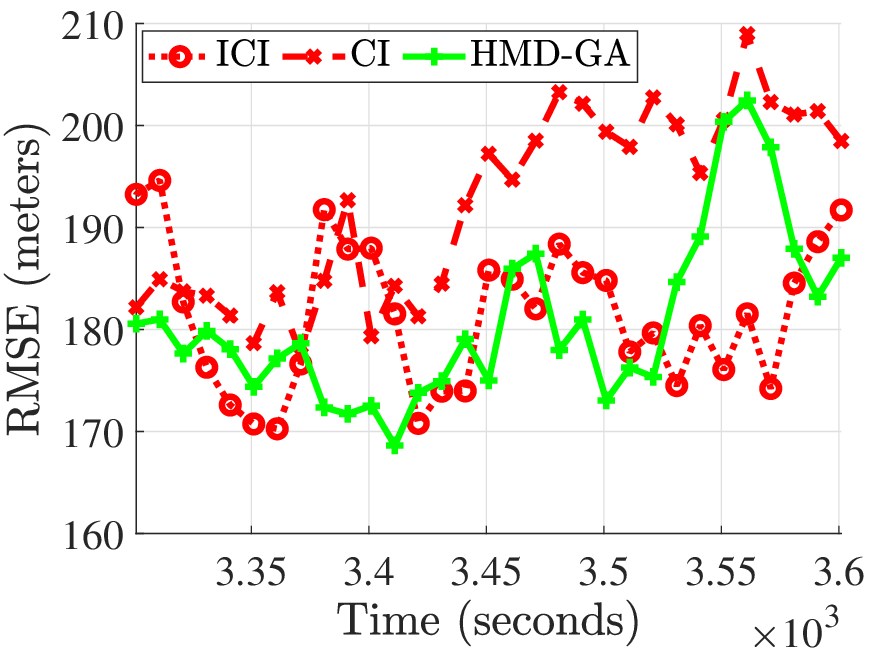}
		\end{subfigure}	
		\begin{subfigure}{0.48\columnwidth}
			\centering
			\includegraphics[width=\linewidth]{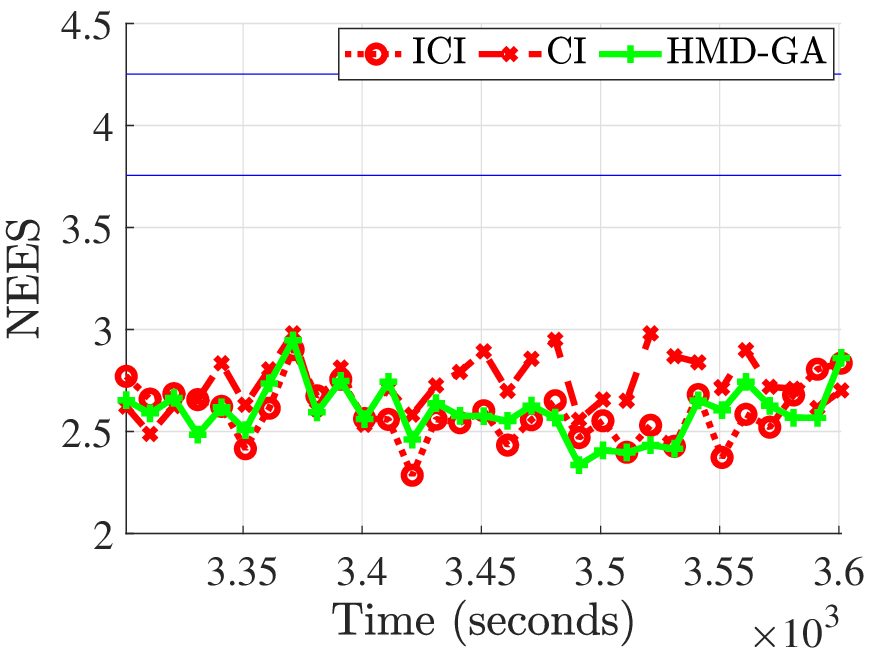}
		\end{subfigure}
		\caption{Target 12.}
		\label{sim_scene3R_tar12}
	\end{figure}

\begin{figure}
		\centering
		\begin{subfigure}{0.48\columnwidth}
			\centering
			\includegraphics[width=\linewidth]{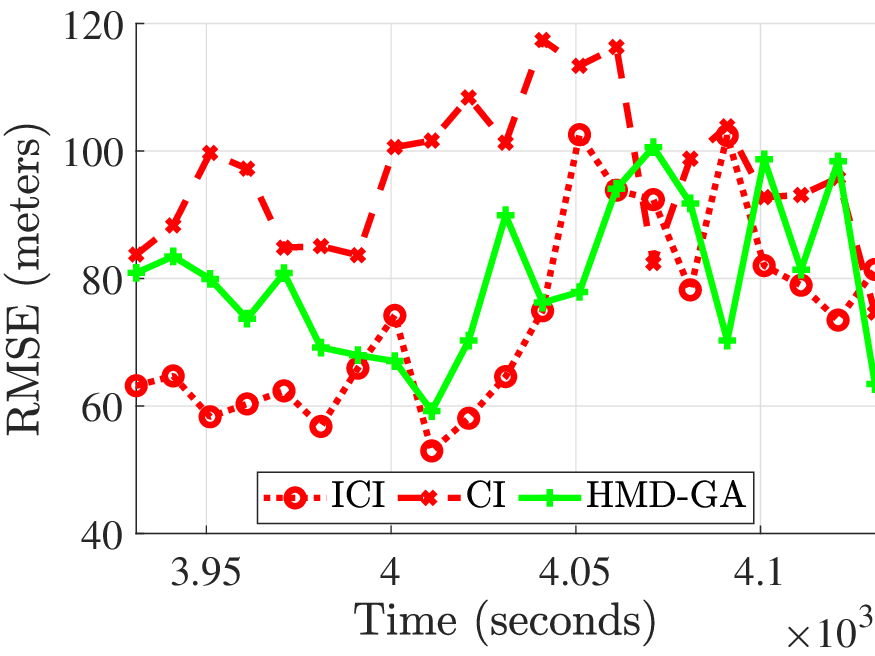}
		\end{subfigure}	
		\begin{subfigure}{0.48\columnwidth}
			\centering
			\includegraphics[width=\linewidth]{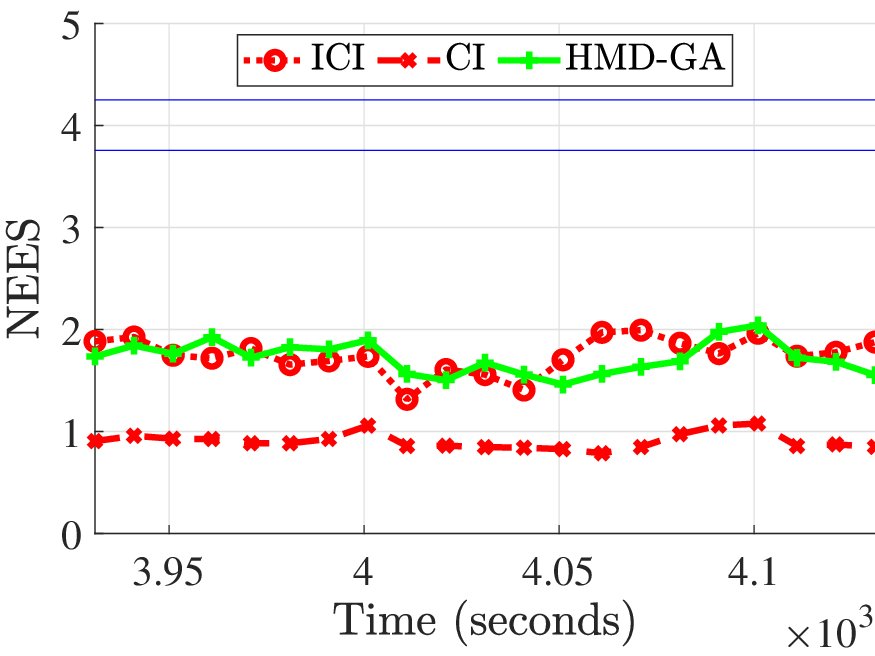}
		\end{subfigure}
		\caption{Target 13.}
		\label{sim_scene3R_tar13}
	\end{figure}

\begin{figure}	
		\begin{subfigure}{0.48\columnwidth}
			\centering
			\includegraphics[width=\linewidth]{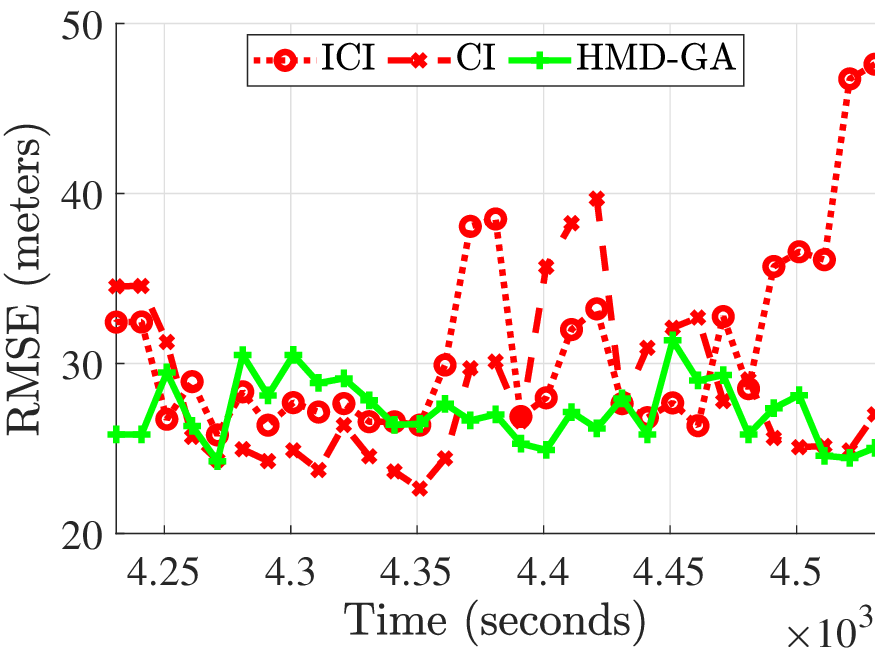}
		\end{subfigure}	
		\begin{subfigure}{0.48\columnwidth}
			\centering
			\includegraphics[width=\linewidth]{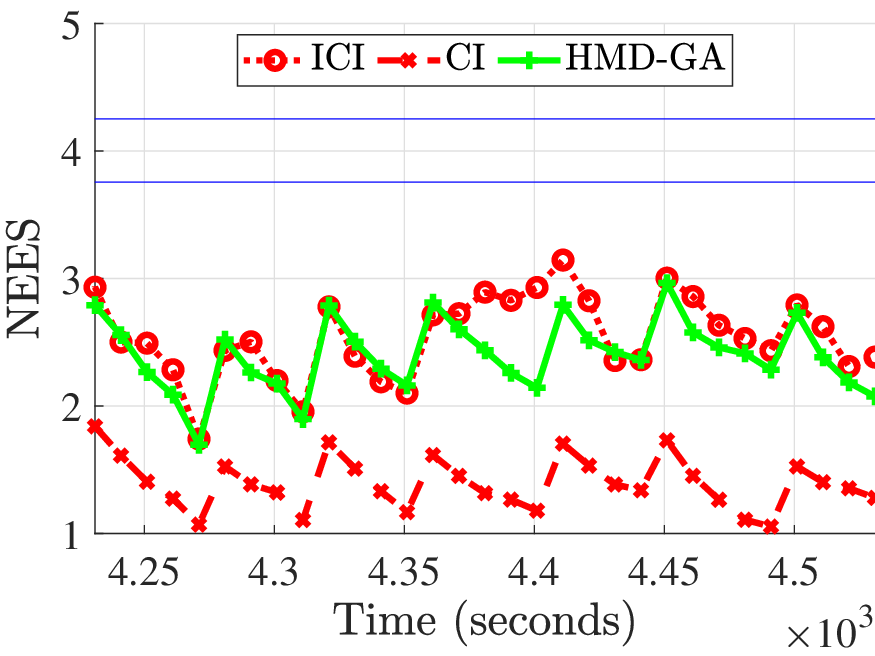}
		\end{subfigure}
		\caption{Target 14.}
		\label{sim_scene3R_tar14}
	\end{figure}

\begin{figure}	
		\begin{subfigure}{0.48\columnwidth}
			\centering
			\includegraphics[width=\linewidth]{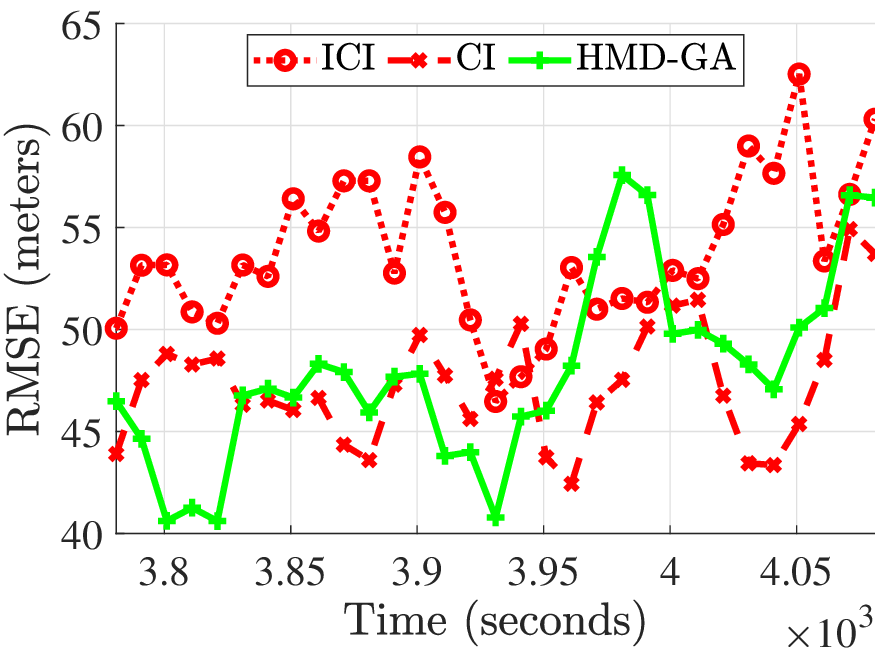}
		\end{subfigure}	
		\begin{subfigure}{0.48\columnwidth}
			\centering
			\includegraphics[width=\linewidth]{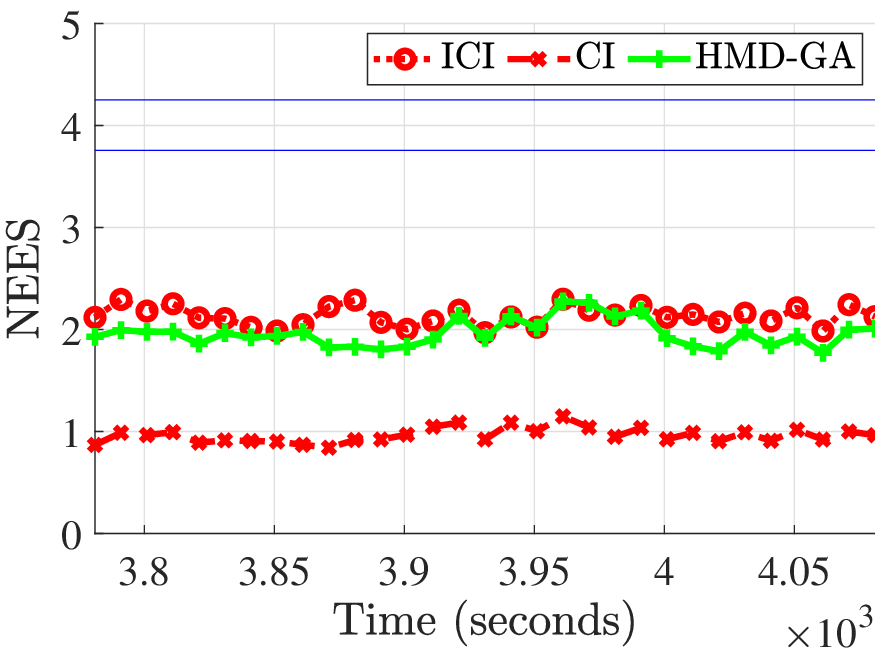}
		\end{subfigure}
		\caption{Target 15.}
		\label{sim_scene3R_tar15}
	\end{figure}

\begin{figure}	
		\begin{subfigure}{0.48\columnwidth}
			\centering
			\includegraphics[width=\linewidth]{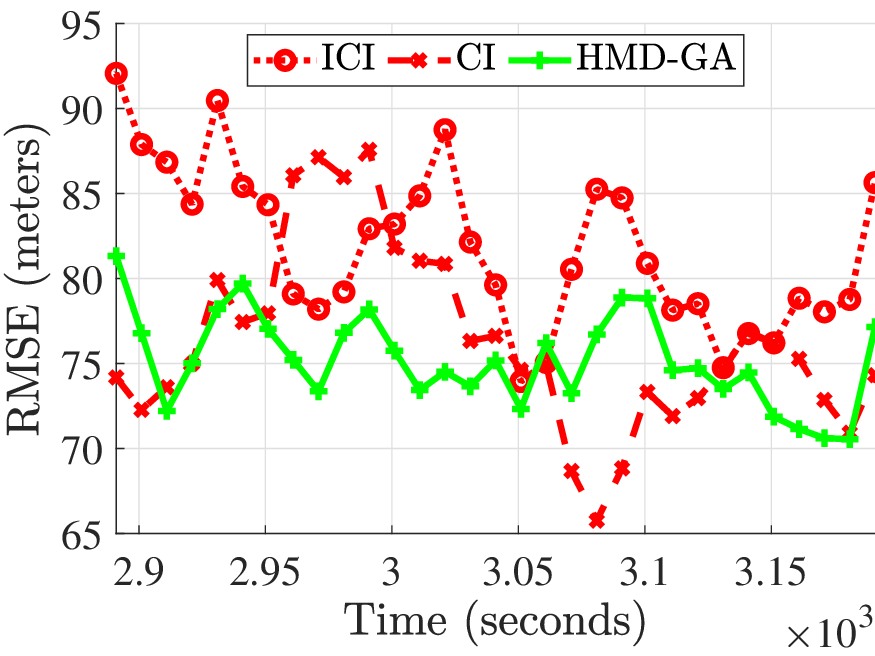}
		\end{subfigure}	
		\begin{subfigure}{0.48\columnwidth}
			\centering
			\includegraphics[width=\linewidth]{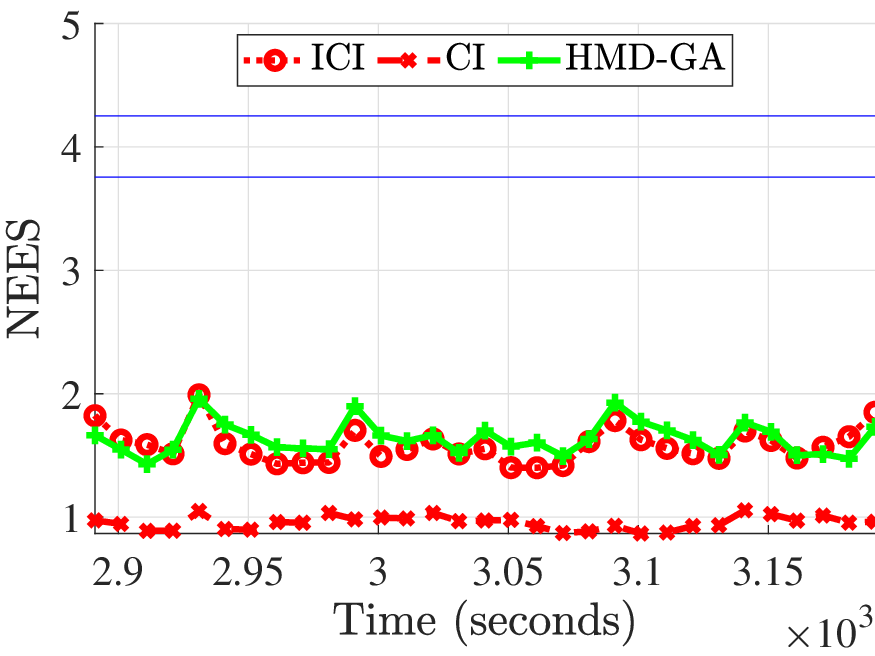}
		\end{subfigure}
		\caption{Target 16.}
		\label{sim_scene3R_tar16}
	\end{figure}

\begin{figure}	
		\begin{subfigure}{0.48\columnwidth}
			\centering
			\includegraphics[width=\linewidth]{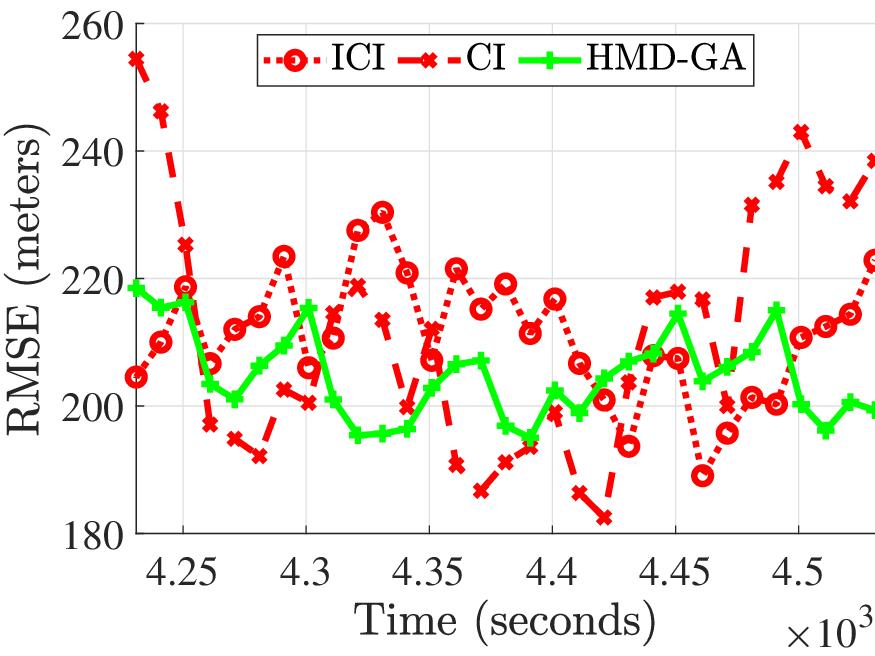}
		\end{subfigure}	
		\begin{subfigure}{0.48\columnwidth}
			\centering
			\includegraphics[width=\linewidth]{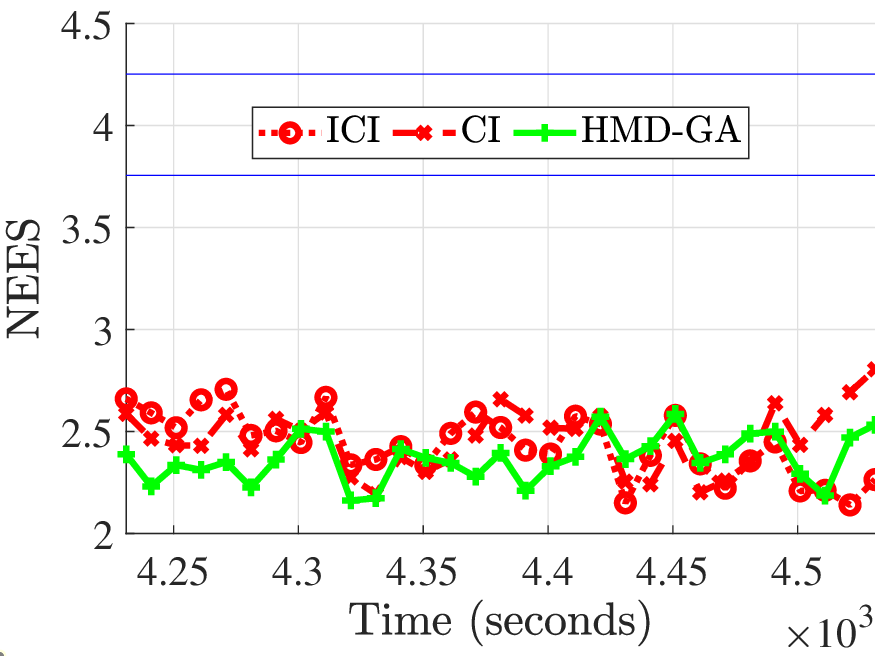}
		\end{subfigure}
		\caption{Target 17.}
		\label{sim_scene3R_tar17}
	\end{figure}

\begin{figure}	
		\begin{subfigure}{0.48\columnwidth}
			\centering
			\includegraphics[width=\linewidth]{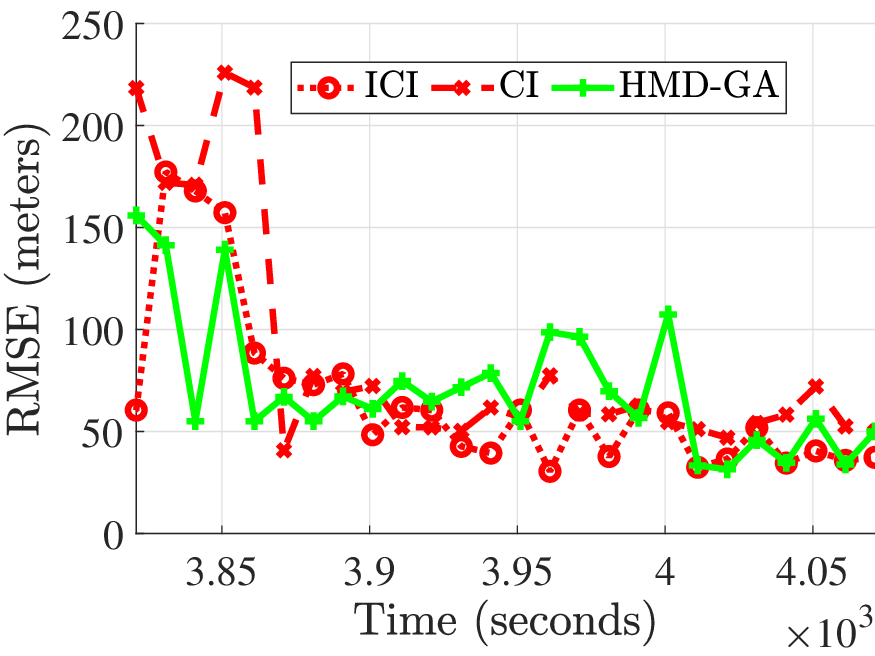}
		\end{subfigure}	
		\begin{subfigure}{0.48\columnwidth}
			\centering
			\includegraphics[width=\linewidth]{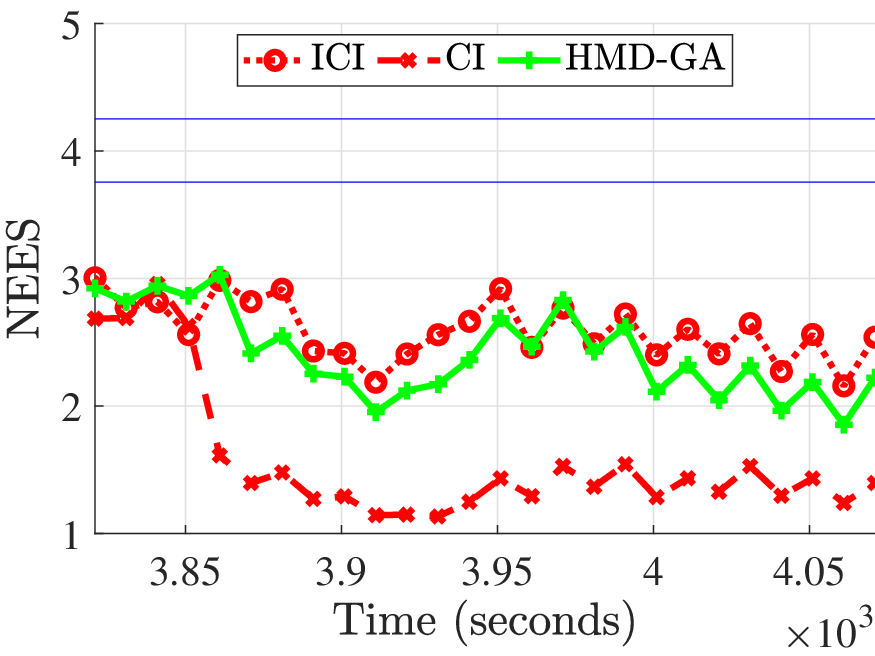}
		\end{subfigure}
		\caption{Target 19.}
		\label{sim_scene3R_tar19}
	\end{figure}

\begin{figure}[h!]
		\begin{subfigure}{0.48\columnwidth}
			\centering
			\includegraphics[width=\linewidth]{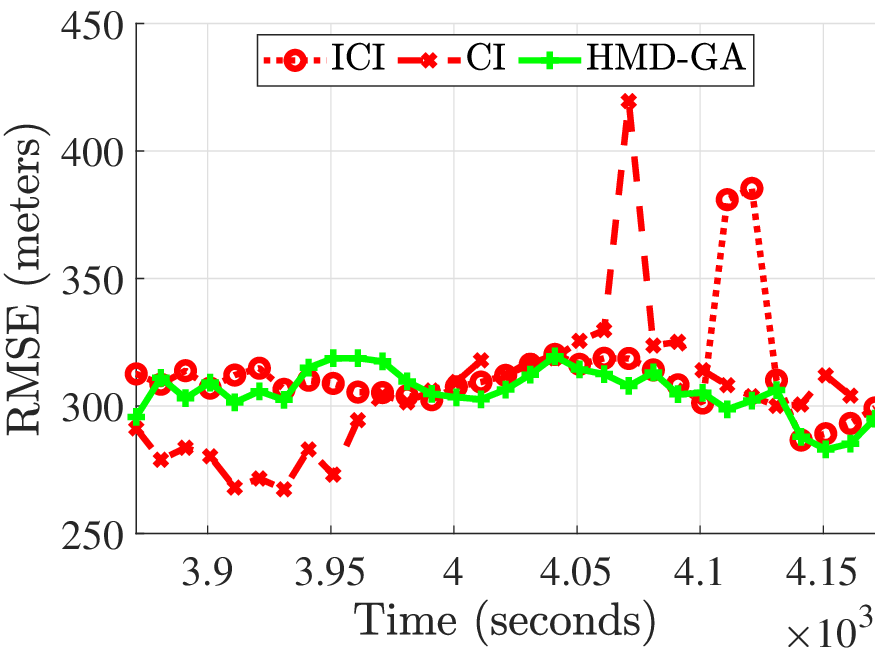}
		\end{subfigure}	
		\begin{subfigure}{0.48\columnwidth}
			\centering
			\includegraphics[width=\linewidth]{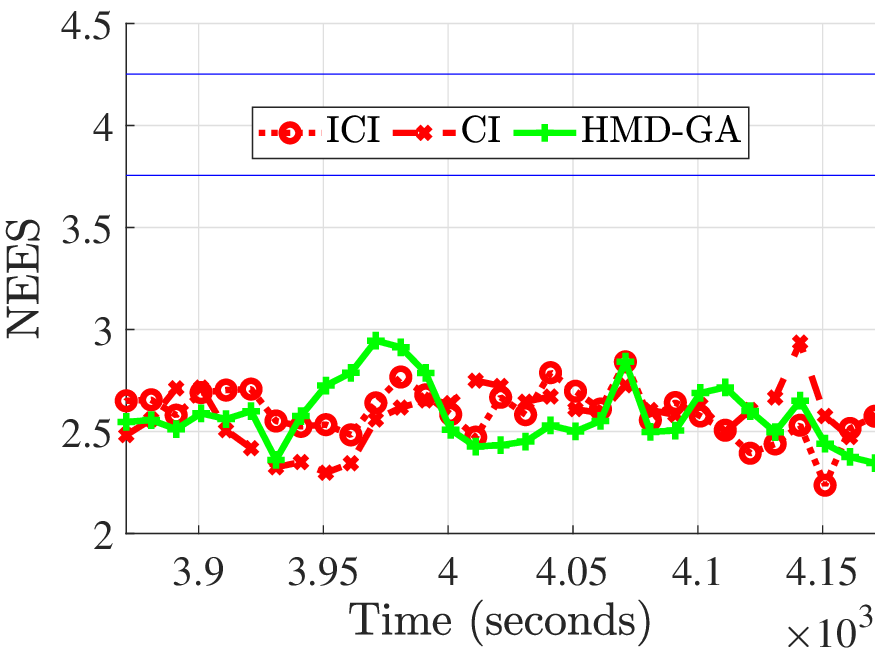}
		\end{subfigure}
		\caption{Target 20.}
		\label{sim_scene3R_tar20}
	\end{figure}

	The results in the form of RMSE and NEES plots are presented in Figs. \ref{sim_scene3R_tar1} through Fig. \ref{sim_scene3R_tar20}. Note that target 18 does not lie within the FOV of either sensor, which is why only 19 sets of plots are present (see Fig. \ref{fig_sim_multiSen}). The estimated tracks for a single run using HMD-GA are shown in Fig. \ref{fig_sim_multiSen_tracks}. Due to high clutter, the number of tentative tracks initialized at the fusion center is clearly evident.

	Based on the varying dynamics of the scenario, we have categorized the targets into three categories :
	\begin{itemize}
		\item Category A : Targets that lie within the intersection of all three sensor nodes for most of their lifetime.
		\item Category B : Targets that lie within the intersection of two sensor nodes.
		\item Category C : Targets that lie in the FOV of only one sensor. Due to memory, such tracks will be maximally correlated. 
	\end{itemize}
	
	The tracks in Category A are fused by all three sensors and consist of Targets 4, 10, and 15. Due to maximum information availability, these targets have the least error relative to the error in estimation of other targets. This is also evident in the relevant plots in Figs. \ref{sim_scene3R_tar4}, \ref{sim_scene3R_tar10} and \ref{sim_scene3R_tar15}. The NEES plots for ICI and HMD-GA show no difference, while CI, proven to be highly conservative, lies at the bottom of all NEES plots. 
	
	The distance from the node plays an important role in the error performance, as the cross-range error increases with range. Thus, as observed, Target 4 has the highest RMSE, whereas Target 10 has the least. Target 15 moves to Category B towards the end of its life, due to which the error increases. The expected error in CI, however, remains close to ICI throughout the simulation. HMD-GA has the least estimation error for most of the sample  time, as evident in the plots.
	
	The tracks in Category B consist of Targets 2, 5, 7, 8, 11, 12, 13, 14, and 16. The RMSE is moderately higher than Category A due to information availability from only two sensor nodes. Due to the upper bound on the employed track-fusion algorithms, the NEES plot also suggests a higher reported covariance than in Category A. It is obvious that recurring fusion should decrease fused covariance, due to which NEES plots are comparatively higher in the case of Category A.
	
	The third category is interesting since these targets lie in the FOV of only one sensor. Due to memory, the fusion track contains a predicted estimate of the last track sent from the same node. If the sensor node has reached a steady state such that the track covariance is not varying enough with time, the correlation between the global track and local track will be the highest among all three categories. The correlation also strongly depends on the maneuvering index, which is a function of sample time. Thus, with decreasing values of $T_\mathit{f}$ and local sample-time, the correlation increases.
	
	It is to be noted that in the case of unbiased and consistent local estimates, the correlation coefficient increases as the local pdf become similar to each other. The expressions for HMD and CI become, in this case,
	\begin{align}
		\frac{p^1(\mathbf{x})p^2(\mathbf{x})}{ \left(p^1(\mathbf{x})\right)^{\omega_2} \left(p^2(\mathbf{x})\right)^{\omega_1} }_{p_1 \rightarrow p_2} = \frac{p_1(\mathbf{x})p_2(\mathbf{x})}{\omega_2p_1(\mathbf{x}) + \omega_1p_2(\mathbf{x})}_{p_1 \rightarrow p_2}
	\end{align} 
	 and similarly in the case of ICI.
	 \begin{align}
	 	\left.\mathbf{\Gamma}_\text{ICI}\right\vert_{\mathbf{\Gamma}_1 \rightarrow \mathbf{\Gamma}_2} \rightarrow \mathbf{\Gamma}_1
	 \end{align}

	 Thus, the result is the local track density itself. Therefore, in the case of perfect correlation, almost no fusion takes place with conservative fusion techniques. This is shown in the NEES plot for Category C targets, where the relative conservativeness of CI in comparison to ICI and HMD-GA vanishes. All fusion techniques tend to produce a similar fused covariance, due to which the NEES plot looks similar in all three cases. Such objects are targets 3, 6, 9, 17, and 20. The corresponding RMSE plot shows the highest error among all categories (almost equal to the local estimate covariance).
	 
	 The reported covariance in the case of HMD-GA is always less than ICI (see appendix for proof), while being overall conservative with respect to true error. The same is reflected in the NEES result, with exceptions in the case of high cross-correlation, wherein the fused covariance is almost the same for all implemented fusers. In some cases, the NEES graph suggests higher covariance for HMD-GA, but this is not true. The result is only due to the use of finite MC samples.

	\begin{table}[t]
		\centering
		\caption{Simulation parameters -- Scenario 2}
		\label{tab_param_scen2}
		\begin{tabular}{l l}
			\toprule
			Parameter & Value \\
			\midrule
			Process noise intensity ($\tilde{q}$) & 0.15 $\text{m}^2/\text{s}^3$ \\
			Number of targets & 20 \\
			Detection probability & 0.99 \\
			False alarm probability $(P_{fa})$ & $10^{-6}$ \\
			Clutter density ($\lambda$)& $10^{-6}$ \\
			Maximum target speed & 30 $\text{m}/\text{s}$ \\
			Number of sensors & 3 \\
			Sensor type & Monostatic 2-D Radar \\
			Sensor coverage & $300,000 \text{ m}\times 360^\circ $ \\
			Sensor std. deviation - range ($\sigma_r$) & 50 m \\ 
			Sensor std. deviation - azimuth ($\sigma_\theta$) & $2^\circ$ \\
			Gating threshold ($\mu$) & 0.95 \\
			Track deletion threshold (num. of misses) & 6 \\
			Track loss threshold & 500 m \\
			Fusion sample period ($T_f$) & 10 seconds \\
			Total simulation time & 4537 seconds \\
			Sensor sample time & 2 seconds \\
			\bottomrule 
		\end{tabular}
	\end{table}
	
	\begin{table}
		\centering
		\caption{Locative Parameters -- Scenario 2}
		\label{tab_loc_param_scen2}
		\begin{tabular}{l l l}
			\toprule
			Parameter & Initial coordinates ($^\circ\text{N}, ^\circ\text{E}$) & Final coordinates ($^\circ\text{N}, ^\circ\text{E}$)  \\
			\midrule
			Target 1 & 59.6104,	-52.6939 & 59.9352,	-52.7611 \\
			Target 2 & 57.7699,	-59.1922 & 	58.0822, -58.7635 \\
			Target 3 & 60.7648, -55.762 & 60.8607,	-56.3676 \\
			Target 4 & 57.7025, -54.7059 & 57.3769,	-54.7036 \\
			Target 5 & 59.4254, -54.4498 & 59.7284,	-54.5691 \\
			Target 6 & 59.1483, -51.2977 & 59.4869,	-51.3718 \\
			Target 7 & 58.8561, -58.484 & 58.6288,	-58.495 \\
			Target 8 & 59.831, -58.3974 & 59.6367, -58.0819 \\
			Target 9 & 59.5306, -53.1629  & 59.5036, -53.5418 \\
			Target 10 & 58.8028, -54.3803  & 58.6656, -53.5222\\
			Target 11 & 57.8132, -52.6504 & 57.676, -51.8147 \\
			Target 12 & 57.4636, -58.9469 & 57.5906, -59.8504\\
			Target 13 &  57.2096, -52.6824 & 57.7325, -52.5499\\
			Target 14 & 59.9093, -56.8477 & 59.8108, -57.9123 \\
			Target 15 & 57.5181, -56.2302 & 58.1104, -56.6123 \\
			Target 16 & 58.3548, -56.4447 & 58.0948, -56.8321 \\
			Target 17 & 57.8268, -49.5721 & 57.55, -50.3705\\
			Target 18 & 60.2969, -49.9242 & 60.6296, -49.3978 \\
			Target 19 & 57.6451, -59.8337 & 57.988, -59.1956\\
			Target 20 & 60.6533, -51.4683 & 61.1832, -51.9363 \\
			Radar 1 & 59.7138694,   -55.2676093   & - \\
			Radar 2 & 57.5399008,   -57.6551522 & - \\
			Radar 3 & 56.6320564,   -52.104272 & - \\
			\bottomrule
		\end{tabular}	
	\end{table}
	
	
	
	
		%
		%

	\section{Conclusion} \label{conclusion}
	A deeper investigation of harmonic mean density based fusion was performed in this article, wherein we showed that HMD is a naturalized consensus-based fusion algorithm that minimizes a weighted $\chi^2$ divergence between the local densities. Thus, the fusion is an agreement rather than a distributed Bayes' type result. The agreement occurs by optimizing a fused density such that it is closer to local track densities in the sense of a statistical distance ($\chi^2$ divergence). Two versions of HMD are investigated. The first one, which was first proposed in \cite{nikhil_journal_1} by approximating the mixture in the denominator by a Gaussian, is termed HMD-GA, and the second is HMD-S, which uses samples from either of the participating local densities, thus providing versatility in fusion.
	
	A consistency analysis of HMD-GA has been performed and a proof of its consistency has been provided. In extreme cases of correlation, the HMD-GA was proven to be consistent. It is clear that the consistency of HMD-GA strongly depends on the consistency of local estimates. Due to similarity of expression, HMD-GA is closely linked with the inverse covariance intersection, which has been closely investigated here. It was found that though ICI is considered for a specific correlation structure, it does tend to contradict the same.
	
	Sampling-based HMD implementation possesses the desirable quality of being as conservative as possible by inflating the local covariance. This might have appreciable utility in situations with anomalies and track biases. Future research will be based on such problems. Also, tests on three different simulation scenarios reveal the consistent nature of HMD-GA, with the tightest covariance among CI and ICI. The results also show that HMD-GA performs best in terms of error metrics. 
	
	The latest research on multi-sensor, multi-target tracking promotes employing extended-target mechanisms, random finite set (RFS) based filtering, multi-hypothesis trackers, and anomaly-tolerant filtering on local nodes. Flexible and easy implementation of HMD-GA will definitely serve as a promising fusion methodology in all of these cases.

	\appendix
	\section{Proof that reported covariance in ICI is greater than in HMD-GA}
	Indirectly, it is sufficient to prove that all eigen values of the covariance in ICI are greater than that in HMD-GA. For notation, the eigen values of an n-dimensional matrix $\mathbf{A}$ are represented by $\lambda^1_A, \lambda^2_A, \dots, \lambda^n_A$ with $\lambda^1_A \geq \lambda^2_A, \cdots \geq \lambda^n_A $.
	
	It's easy to see that interchange of fusion weight $\omega$ in the formulation of both HMD-GA and ICI has no significant effect since it hovers around $0.3-0.6$ in most pragmatic cases. Thus, we can perceive $\mathbf{\Gamma}_m^\text{HMD} = \mathbf{\Gamma}_m^\text{ICI} + \omega(1-\omega)\tilde{\mathbf{\Gamma}}$, where $\tilde{\mathbf{\Gamma}}$ is the spread-of-means term. The inverse of mutual information component in HMD-GA is (using Sherman-Morrison Formula \cite{hager1989updating}),
	\begin{align}
	\left(\mathbf{\Gamma}_m^\text{HMD}\right)^{-1} = \left(\mathbf{\Gamma}_m^\text{ICI}\right)^{-1} - \omega(1-\omega) \frac{\left(\mathbf{\Gamma}_m^\text{ICI}\right)^{-1}\tilde{\mathbf{\Gamma}}\left(\mathbf{\Gamma}_m^\text{ICI}\right)^{-1}}{1 + \bar{\mathbf{x}}^T\left(\mathbf{\Gamma}_m^\text{ICI}\right)^{-1}\bar{\mathbf{x}}}
	\end{align}
	
	where $\bar{x} = \left(\hat{\mathbf{x}}^i - \hat{\mathbf{x}}^j\right) $. This leads to following relation between fused covariance in the case of HMD-GA and ICI,
	\begin{align}
		\mathbf{\Gamma}^\mathit{f}_\text{HMD} = \left[ \left(\mathbf{\Gamma}^\mathit{f}_\text{ICI}\right)^{-1}  + \frac{\left(\mathbf{\Gamma}_m^\text{ICI}\right)^{-1}\tilde{\mathbf{\Gamma}}\left(\mathbf{\Gamma}_m^\text{ICI}\right)^{-1}}{1 + \bar{\mathbf{x}}^T\left(\mathbf{\Gamma}_m^\text{ICI}\right)^{-1}\bar{\mathbf{x}}}\right]^{-1}
	\end{align}
	
	Again, using rank-one update of an inverse,
	\begin{align}
		\mathbf{\Gamma}^\mathit{f}_\text{HMD} &= \mathbf{\Gamma}^\mathit{f}_\text{ICI} - \omega(1-\omega) \frac{1}{1 + \bar{\mathbf{x}}^T\left(\mathbf{\Gamma}_m^\text{ICI}\right)^{-1}\bar{\mathbf{x}}}\notag \\
		&\quad \quad \times \frac{\mathbf{\Gamma}^\mathit{f}_\text{ICI}\left(\mathbf{\Gamma}_m^\text{ICI}\right)^{-1}\tilde{\mathbf{\Gamma}}\left(\mathbf{\Gamma}_m^\text{ICI}\right)^{-1}\mathbf{\Gamma}^\mathit{f}_\text{ICI}}{1 + \bar{\mathbf{y}}^T\mathbf{\Gamma}^\mathit{f}_\text{ICI}\bar{\mathbf{y}}} \label{eq_compareCov_ici_hmd}
	\end{align}
	where $\bar{\mathbf{y}} = \left(\mathbf{\Gamma}_m^\text{ICI}\right)^{-1}\bar{\mathbf{x}}$. Note that the last term on LHS is still a rank one matrix with at most one eigen value $\lambda^\text{SOM}$, since it's still related to spread-of-means (SOM) term. Using eqn. \eqref{eq_compareCov_ici_hmd}, the eigen values should follow the relation,
	\begin{align}
		\lambda_1^\text{HMD} +  \lambda_2^\text{HMD} + \dots = \lambda_1^\text{ICI} +  \lambda_2^\text{ICI} + \dots + \lambda^\text{SOM}.
	\end{align}
	Relations between individual eigen values are not available in literature as per author's knowledge. The best result is Cauchy's interlacing theorem \cite{mercer2000cauchy}, which when applied to eqn. \eqref{eq_compareCov_ici_hmd}, results in,
	\begin{align}
		\lambda_n^\text{HMD} \leq \lambda_n^\text{ICI} \leq \lambda_{n-1}^\text{HMD} \leq \lambda_{n-1}^\text{ICI} \cdots \lambda_1^\text{HMD} \leq \lambda_1^\text{ICI} 
	\end{align}
	where $n$ is the dimension of the matrices. 
	
	\bibliographystyle{elsarticle-num}
	\bibliography{IEEEabrv,main_journal_4.bib}

\begin{thebibliography}{10}
\expandafter\ifx\csname url\endcsname\relax
  \def\url#1{\texttt{#1}}\fi
\expandafter\ifx\csname urlprefix\endcsname\relax\def\urlprefix{URL }\fi
\expandafter\ifx\csname href\endcsname\relax
  \def\href#1#2{#2} \def\path#1{#1}\fi

\bibitem{bar1981track}
Y.~Bar-Shalom, On the track-to-track correlation problem, IEEE Transactions on
  Automatic control 26~(2) (1981) 571--572.

\bibitem{nikhil_journal_1}
N.~Sharma, R.~Tharmarasa, S.~Bhaumik, T.~Kirubarajan, On pooling-based track
  fusion strategies : Harmonic mean density (submitted), Signal Processing
  (2024).

\bibitem{julier1997non}
S.~J. Julier, J.~K. Uhlmann, A non-divergent estimation algorithm in the
  presence of unknown correlations, in: Proceedings of the 1997 American
  Control Conference (Cat. No. 97CH36041), Vol.~4, IEEE, 1997, pp. 2369--2373.

\bibitem{chen2002fusion}
L.~Chen, P.~O. Arambel, R.~K. Mehra, Fusion under unknown
  correlation-covariance intersection as a special case, in: Proceedings of the
  Fifth International Conference on Information Fusion. FUSION 2002.(IEEE Cat.
  No. 02EX5997), Vol.~2, IEEE, 2002, pp. 905--912.

\bibitem{sijs2010state}
J.~Sijs, M.~Lazar, P.~Bosch, State fusion with unknown correlation: Ellipsoidal
  intersection, in: Proceedings of the 2010 American Control Conference, IEEE,
  2010, pp. 3992--3997.

\bibitem{noack2017decentralized}
B.~Noack, J.~Sijs, M.~Reinhardt, U.~D. Hanebeck, Decentralized data fusion with
  inverse covariance intersection, Automatica 79 (2017) 35--41.

\bibitem{noack2017inverse}
B.~Noack, J.~Sijs, U.~D. Hanebeck, Inverse covariance intersection: New
  insights and properties, in: 2017 20th International Conference on
  Information Fusion (Fusion), IEEE, 2017, pp. 1--8.

\bibitem{mahler2000optimal}
R.~P. Mahler, Optimal/robust distributed data fusion: a unified approach, in:
  Signal Processing, Sensor Fusion, and Target Recognition IX, Vol. 4052,
  International Society for Optics and Photonics, 2000, pp. 128--138.

\bibitem{hurley2002information}
M.~B. Hurley, An information theoretic justification for covariance
  intersection and its generalization, in: Proceedings of the Fifth
  International Conference on Information Fusion. FUSION 2002.(IEEE Cat. No.
  02EX5997), Vol.~1, IEEE, 2002, pp. 505--511.

\bibitem{battistelli2014kullback}
G.~Battistelli, L.~Chisci, \uppercase{K}ullback--\uppercase{L}eibler average,
  consensus on probability densities, and distributed state estimation with
  guaranteed stability, Automatica 50~(3) (2014) 707--718.

\bibitem{upcroft2005rich}
B.~Upcroft, L.~L. Ong, S.~Kumar, M.~Ridley, T.~Bailey, S.~Sukkarieh,
  H.~Durrant-Whyte, Rich probabilistic representations for bearing only
  decentralised data fusion, in: 2005 7th International Conference on
  Information Fusion, Vol.~2, IEEE, 2005, pp. 8--pp.

\bibitem{julier2006empirical}
S.~J. Julier, An empirical study into the use of \uppercase{C}hernoff
  information for robust, distributed fusion of \uppercase{G}aussian mixture
  models, in: 2006 9th International Conference on Information Fusion, IEEE,
  2006, pp. 1--8.

\bibitem{gunay2016chernoff}
M.~Gunay, U.~Orguner, M.~Demirekler, Chernoff fusion of \uppercase{G}aussian
  mixtures based on sigma-point approximation, IEEE Transactions on Aerospace
  and Electronic Systems 52~(6) (2016) 2732--2746.

\bibitem{ajgl2015approximation}
J.~Ajgl, M.~{\v{S}}imandl, J.~Dun{\'\i}k, Approximation of powers of
  \uppercase{G}aussian mixtures, in: 2015 18th International Conference on
  Information Fusion (Fusion), IEEE, 2015, pp. 878--885.

\bibitem{lu2019distributed}
K.~Lu, C.~Sun, Q.~Fu, Q.~Zhu, Distributed track-to-track fusion for non-linear
  systems with gaussian mixture noise, IET Radar, Sonar \& Navigation 13~(5)
  (2019) 740--749.

\bibitem{ahmed2015s}
N.~R. Ahmed, What's one mixture divided by another?: A unified approach to
  high-fidelity distributed data fusion with mixture models, in: 2015 IEEE
  International Conference on Multisensor Fusion and Integration for
  Intelligent Systems (MFI), IEEE, 2015, pp. 289--296.

\bibitem{koliander2022fusion}
G.~Koliander, Y.~El-Laham, P.~M. Djuri{\'c}, F.~Hlawatsch, Fusion of
  probability density functions, Proceedings of the IEEE 110~(4) (2022)
  404--453.

\bibitem{yury}
P.~Yury, Y.~Wu, Information Theory : From Coding to Learning, 1st Edition,
  Cambridge University Press, 2022.

\bibitem{csiszar2004information}
I.~Csisz{\'a}r, P.~C. Shields, et~al., Information theory and statistics: A
  tutorial, Foundations and Trends{\textregistered} in Communications and
  Information Theory 1~(4) (2004) 417--528.

\bibitem{julier2008fusion}
S.~Julier,
  \href{https://digital-library.theiet.org/content/conferences/10.1049/ic_20080050}{Fusion
  without independence}, IET Conference Proceedings (2008) 1--4(3).
\newline\urlprefix\url{https://digital-library.theiet.org/content/conferences/10.1049/ic_20080050}

\bibitem{li2020arithmetic}
T.~Li, X.~Wang, Y.~Liang, Q.~Pan, On arithmetic average fusion and its
  application for distributed multi-bernoulli multitarget tracking, IEEE
  Transactions on Signal Processing 68 (2020) 2883--2896.

\bibitem{gul2016robust}
G.~G{\"u}l, A.~M. Zoubir, Robust hypothesis testing with alpha-divergence, IEEE
  Transactions on Signal Processing 64~(18) (2016) 4737--4750.

\bibitem{ghosal1997review}
S.~Ghosal, A review of consistency and convergence of posterior distribution,
  in: Varanashi Symposium in Bayesian Inference, Banaras Hindu University,
  Citeseer, 1997.

\bibitem{ghosal1995convergence}
S.~Ghosal, J.~K. Ghosh, T.~Samanta, On convergence of posterior distributions,
  The Annals of Statistics 23~(6) (1995) 2145--2152.

\bibitem{bar2001estimation}
Y.~Bar-Shalom, X.~R. Li, T.~Kirubarajan, Estimation with applications to
  tracking and navigation: theory algorithms and software, John Wiley \& Sons,
  2001.

\bibitem{bhatia1997and}
R.~Bhatia, P.~Rosenthal, How and why to solve the operator equation ax- xb= y,
  Bulletin of the London Mathematical Society 29~(1) (1997) 1--21.

\bibitem{bar1995multitarget}
Y.~Bar-Shalom, X.-R. Li, Multitarget-multisensor tracking: principles and
  techniques, Vol.~19, YBs Storrs, CT, 1995.

\bibitem{tian2014consistency}
X.~Tian, Y.~Bar-Shalom, A consistency-based gaussian mixture filtering approach
  for the contact lens problem, IEEE Transactions on Aerospace and Electronic
  Systems 50~(3) (2014) 1636--1646.

\bibitem{kirubarajan2017target}
T.~Kirubarajan, Y.~Bar-Shalom, Target tracking using probabilistic data
  association-based techniques with applications to sonar, radar, and eo
  sensors, in: Handbook of multisensor data fusion, CRC Press, 2017, pp.
  223--262.

\bibitem{civicioglu2012transforming}
P.~Civicioglu, Transforming geocentric cartesian coordinates to geodetic
  coordinates by using differential search algorithm, Computers \& Geosciences
  46 (2012) 229--247.

\bibitem{hager1989updating}
W.~W. Hager, Updating the inverse of a matrix, SIAM review 31~(2) (1989)
  221--239.

\bibitem{mercer2000cauchy}
A.~M. Mercer, P.~R. Mercer, et~al., Cauchy's interlace theorem and lower bounds
  for the spectral radius, International Journal of Mathematics and
  Mathematical Sciences 23 (2000) 563--566.

\end{thebibliography}

\end{document}